\documentclass[10pt]{article}

\usepackage{graphicx} % For including graphics
\usepackage[hidelinks]{hyperref} % For hyperlinks without borders
\usepackage{url} % For handling URLs
\usepackage{amsmath,amssymb} % For mathematical symbols and formulas
\usepackage[utf8]{inputenc} % For UTF-8 input encoding
\usepackage{natbib} % For bibliography and citations
\usepackage{float} % For forcing table/figure positions
\usepackage{array} % For advanced table formatting
\usepackage{tabularx} % For full-width tables
\usepackage{longtable} % For tables spanning multiple pages
\usepackage{booktabs} % For professional-quality tables
\usepackage{multirow} % For merging rows in tables
\usepackage{ragged2e} % For text alignment and justification
\usepackage{geometry} % For controlling page margins
\usepackage{setspace} % For line spacing
\usepackage{titlesec} % For controlling section titles
\usepackage{changepage} % For adjusting text margins locally

\justifying % Apply full justification for the entire document

\graphicspath{{figures/}} % Specify the path for graphics

\newcolumntype{P}[1]{>{\raggedright\arraybackslash}p{#1}} % Define raggedright column type

\usepackage{parskip}  % For paragraph spacing management
\titleformat{\section}{\bfseries\large}{\thesection}{1em}{}
\titleformat{\subsection}{\bfseries\normalsize}{\thesubsection}{1em}{}
\titleformat{\subsubsection}{\bfseries\small}{\thesubsubsection}{1em}{}

\title{Title of Your Manuscript}
\author{Author Name}
\date{}

\usepackage{ragged2e}

\title{Cloud Platforms for Developing Generative AI Solutions:\\A Scoping Review of Tools and Services}
\date{}

\begin{document}
\maketitle

% Author list with affiliations
{\noindent\bfseries
Dhavalkumar Patel\textsuperscript{1}, 
Ganesh Raut\textsuperscript{1}, 
Satya Narayan Cheetirala\textsuperscript{1}, 
Girish N Nadkarni\textsuperscript{2}, 
Robert Freeman\textsuperscript{2}, 
Benjamin S. Glicksberg\textsuperscript{2}, 
Eyal Klang\textsuperscript{2}, 
Prem Timsina\textsuperscript{1}}

\begin{enumerate}
\def\labelenumi{\arabic{enumi}.}
\item
  \textbf{Institute for Healthcare Delivery Science, Icahn School of
  Medicine at Mount Sinai, New York, NY 10029, United States.}
\item
  \textbf{Department of Medicine, Division of Data-Driven and Digital
  Medicine, Icahn School of Medicine at Mount Sinai, New York, NY 10029,
  United States.}
\end{enumerate}

\noindent\textbf{*Corresponding author:} Dhavalkumar Patel\\
\noindent\textbf{Email:} Dhaval.Patel@mountsinai.org

\noindent\textbf{Abstract:}

Generative AI is transforming enterprise application development by
enabling machines to create content, code, and designs. These models,
however, demand substantial computational power and data management.
Cloud computing addresses these needs by offering infrastructure to
train, deploy, and scale generative AI models.~This review examines
cloud services for generative AI, focusing on key providers like Amazon
Web Services (AWS), Microsoft Azure, Google Cloud, IBM Cloud, Oracle
Cloud, and Alibaba Cloud. It compares their strengths, weaknesses, and
impact on enterprise growth.~We explore the role of high-performance
computing (HPC), serverless architectures, edge computing, and storage
in supporting generative AI. We also highlight the significance of data
management, networking, and AI-specific tools in building and deploying
these models.~Additionally, the review addresses security concerns,
including data privacy, compliance, and AI model protection. It assesses
the performance and cost efficiency of various cloud providers and
presents case studies from healthcare, finance, and entertainment.~We
conclude by discussing challenges and future directions, such as
technical hurdles, vendor lock-in, sustainability, and regulatory
issues.~ Put together, this work can serve as a guide for practitioners
and researchers looking to adopt cloud-based generative AI solutions,
serving as a valuable guide to navigating the intricacies of this
evolving field.


\section{Introduction}

The advent of generative artificial intelligence (AI) marks a
transformative era in technology, fundamentally reshaping industries
such as healthcare, finance, creative arts, and scientific research.
Generative AI models---capable of producing human-like text, images,
code, and multimedia content---have swiftly evolved from academic
concepts to powerful tools driving innovation across diverse
sector{[}1{]}. For instance, OpenAI\textquotesingle s GPT3.5{[}2{]},
GPT-4{[}3{]} and Google\textquotesingle s Gemini {[}4{]} have
demonstrated unprecedented capabilities in language understanding and
image generation, respectively. According to {[}5{]}, the global
generative AI market is projected to reach \$110.8 billion by 2030,
growing at a CAGR of 34.6\% from 2023. However, the development,
training, and deployment of these sophisticated models require immense
computational resources and specialized expertise, presenting
significant opportunities and challenges for organizations aiming to
leverage their potential.

Cloud computing and their Platform services has emerged as a critical
enabler in this AI revolution, providing scalable infrastructure,
specialized hardware (such as GPUs and TPUs), and managed services that
democratize access to generative AI technologies. The synergy between
cloud platforms and generative AI is not only accelerating technological
advancements but also lowering barriers to entry, allowing organizations
of all sizes to participate in this transformative movement.

\subsection{Objectives and Scope}

This comprehensive scoping review aims to critically analyze cloud
platforms specifically designed for developing generative AI solutions.
Our investigation encompasses several key objectives that address the
rapidly evolving landscape of cloud-based AI development. First, we
conduct a thorough assessment of state-of-the-art cloud services offered
by major providers---including Amazon Web Services (AWS), Microsoft
Azure, Google Cloud Platform (GCP), IBM Cloud, Oracle Cloud, and Alibaba
Cloud---examining their specific capabilities for generative AI
development. Through detailed comparative analysis, we evaluate the
strengths, weaknesses, and unique features of each platform, with
particular emphasis on their support for large language models (LLMs),
multimodal AI, and emerging AI paradigms. The review then delves into
the complex landscape of technical, ethical, and regulatory challenges
associated with leveraging cloud platforms for generative AI, while
simultaneously identifying promising opportunities for innovation and
improvement in this domain. We extend our analysis to examine future
directions in cloud-based AI development, exploring cutting-edge areas
such as quantum computing, edge AI, federated learning, and sustainable
AI practices. Throughout this investigation, we maintain a strong focus
on practical applications, offering actionable recommendations for
researchers, practitioners, and decision-makers who must navigate the
intricate ecosystem of cloud-based generative AI development. This
comprehensive approach ensures that our review not only advances
theoretical understanding but also provides concrete guidance for
implementing generative AI solutions in real-world cloud environments.

\subsection{Significance and Timeliness}

As organizations increasingly adopt generative AI to drive innovation,
enhance productivity, and create new value propositions, understanding
the intricacies of the cloud ecosystem is crucial for informed
decision-making. For example,{[}6{]} reports that over 60\% of
enterprises plan to integrate generative AI into their operations by
2025. This review also addresses the growing adoption of generative AI
by organizations to drive innovation, enhance productivity, and create
new value propositions. With over 60\% of enterprises expected to
integrate generative AI into their operations by 2025 {[}6{]},
understanding the complexities of the cloud ecosystem is essential for
making informed decisions.

This paper serves as a valuable resource for diverse stakeholders. For
researchers, it provides a comprehensive overview of the technological
landscape, highlighting key areas for further investigation and
fostering interdisciplinary collaboration. Practitioners benefit from
insights into the capabilities and limitations of various cloud
platforms and services, guiding the selection of appropriate tools for
AI projects. For policymakers, the paper elucidates the technical
foundations of generative AI and cloud computing, informing discussions
on governance, ethics, data privacy, and regulation in the AI domain.
Moreover, the rapid advancements in both generative AI and cloud
computing---such as the introduction of GPT-4{[}3{]}, the rise of
multimodal models like DALL·E 3{[}7{]} and Gemini{[}4{]}, and the
increasing emphasis on ethical AI and sustainability make this review
particularly timely and necessary.

\subsection{Methodology}

Our investigation employs a systematic and rigorous analytical approach,
following the established scoping review framework proposed by Arksey
and O\textquotesingle Malley{[}8{]}. The methodology begins with an
extensive literature search spanning multiple scholarly databases,
including IEEE Xplore, ACM Digital Library, and Google Scholar,
complemented by comprehensive reviews of official cloud service provider
documentation, websites, and white papers. This broad search strategy
ensures capture of both academic perspectives and industry developments.
In establishing our inclusion criteria, we focused specifically on
studies and documents addressing cloud platforms for generative AI,
encompassing technical capabilities, implementation case studies, and
emerging trends in the field. Our data sources represent a carefully
curated mix of academic literature, technical documentation from cloud
service providers, authoritative industry reports from organizations
such as Gartner and IDC, and detailed case studies of real-world
implementations. For data extraction and analysis, we developed and
implemented a comparative framework that evaluates cloud platforms
across several key dimensions, including computational capabilities,
AI-specific services, data management tools, pricing models, and support
for emerging technologies. To validate our findings and ensure
comprehensive coverage, we engaged in extensive consultation with
industry experts and academics, incorporating their insights and
feedback throughout the review process. This methodological approach
enables a thorough and balanced examination of the current state of
cloud platforms for generative AI development, while maintaining
academic rigor and practical relevance.

\subsection{Structure of the Review}

The paper is structured to guide the reader through the multifaceted
landscape of cloud-based generative AI development and \textbf{Figure 1}
provide all detail subsections of each section.

\textbf{Section 2: Background and Context} -- Provides an overview of
generative AI and cloud computing, tracing their evolution and
convergence.

\textbf{Section 3: Cloud Service Providers Overview} -- Offers a
detailed examination of major cloud providers, their AI strategies,
market positions, and strategic partnerships.

\textbf{Section 4: Cloud Services for Generative AI} -- Delves into
specific services crucial for AI development, including high-performance
computing, serverless architectures, edge computing, data management,
and AI development ecosystems.

\textbf{Section 5: Comparative Analysis} -- Presents a comprehensive
comparison of cloud platforms, highlighting strengths, weaknesses,
unique offerings, and providing SWOT analyses for each provider.

\textbf{Section 6: Challenges and Future Directions} -- Discusses
technical, strategic, and ethical challenges, including data bias,
explainability, regulatory compliance, and sustainability. Examines
future trends such as AI model efficiency, cloud-native innovations, and
the impact of AI on society.

\textbf{Section 7: Conclusions and Recommendations} -- Summarizes key
insights, provides strategic recommendations for stakeholders, and
outlines areas for future research and development.

We aim to bridge the gap between technological advancements and
practical implementation, providing valuable insights to advance the
field of cloud-based generative AI development. As we navigate the
opportunities and challenges of this rapidly evolving domain, this
review serves as a crucial guide for stakeholders poised to shape the
future of AI innovation.

% Figure 1
\begin{figure}[H]
    \centering
    \includegraphics[width=0.8\textwidth,height=0.8\textheight,keepaspectratio]{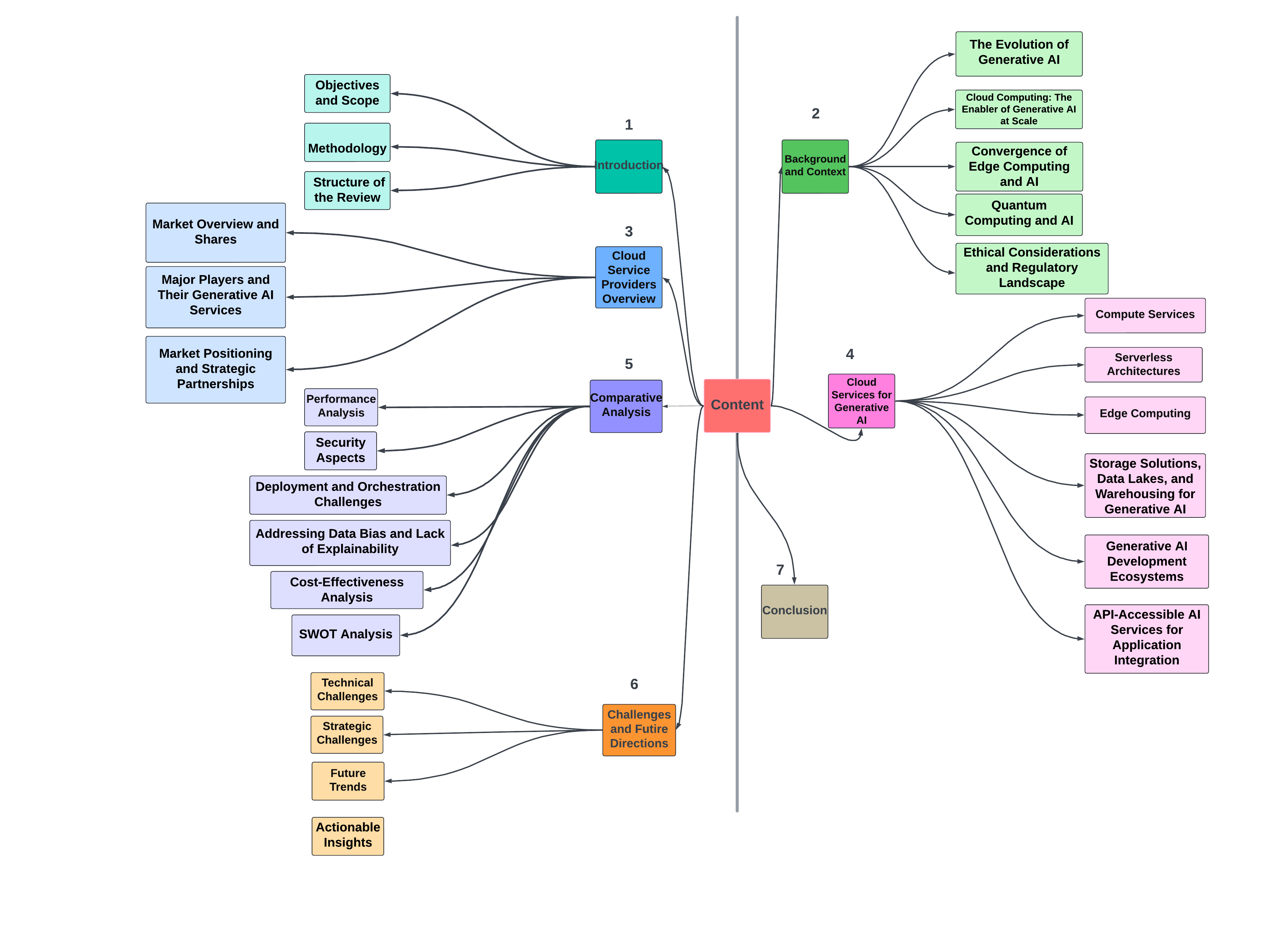}
    \caption{Structured guide through cloud-based generative AI development landscape}
    \label{fig:landscape_guide_Figure_1}  
\end{figure}

\section{Background and Context}

\subsection{The Evolution of Generative AI}

Generative AI signifies a paradigm transitioning from traditional
predictive models to systems capable of creating novel content across
various modalities. This section traces the evolution of generative AI,
highlighting key milestones and recent advancements that have propelled
the technology to the forefront of innovation.

\subsubsection{Historical Perspective}

The concept of generative AI dates to early experiments in computational
creativity{[}9{]}, but its exponential growth in recent years is
attributed to advancements in deep learning architectures and increased
computational power {[}10, 11{]}. \textbf{Figure 2} visually represents
the key developments in the field of AI and Generative AI over the
decades. Notable milestones include the advent of foundational
technologies, the transformative introduction of transformer models, and
the recent advancements in multimodal and generative AI that have
revolutionized interactions between humans and machines.

\textbf{See Supplementary Material Section S1.1 for a detailed timeline
of key developments in AI and generative AI from 1950 to 2024.}

% Figure 2
\begin{figure}[H]
    \centering
    \includegraphics[width=0.7\textwidth]{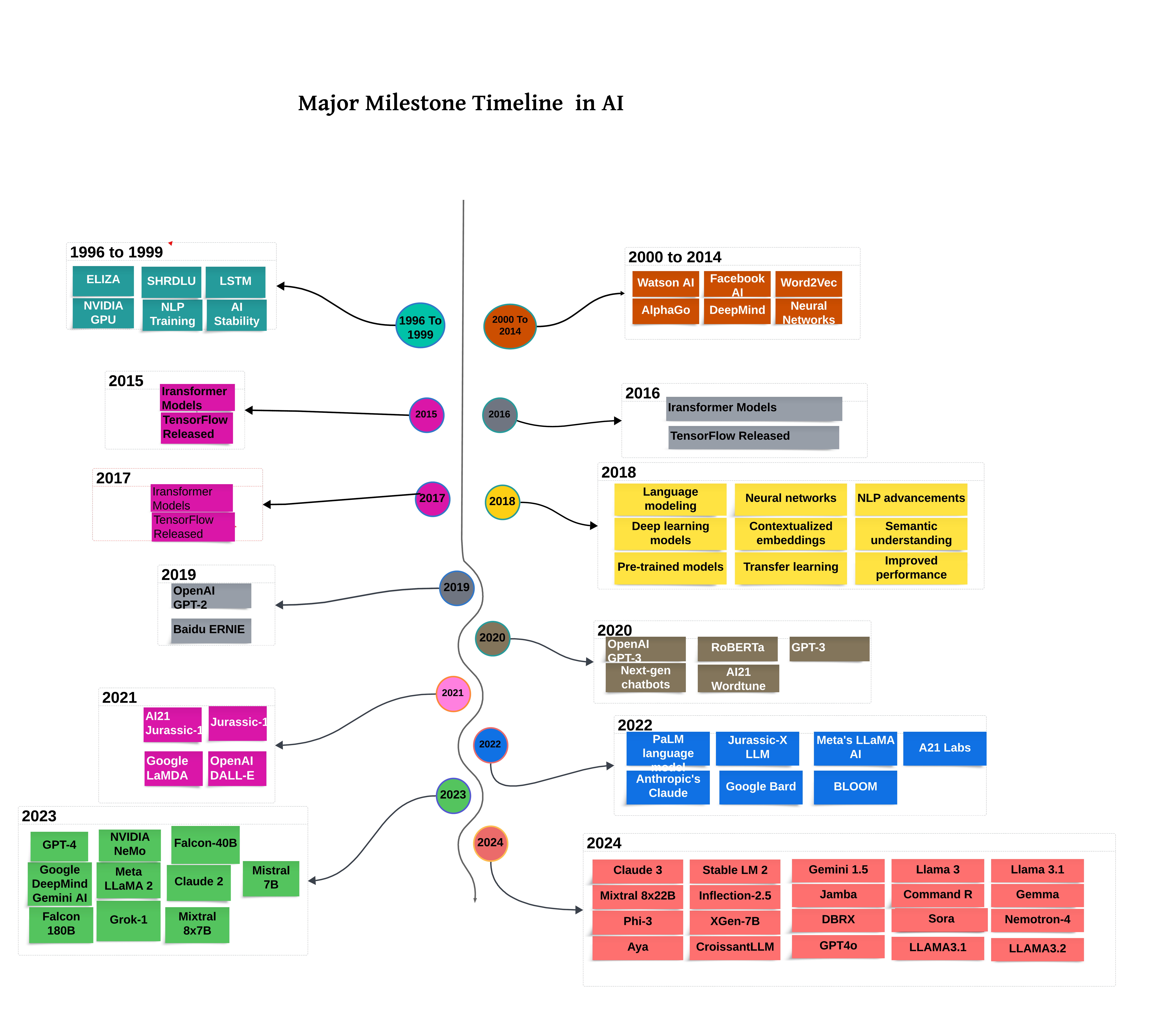}
    \caption{Major Milestone Timeline in AI Development}
    \label{fig:ai_timeline_Figure_2}  
\end{figure}

\subsubsection{Recent Advancements}

In recent years, generative AI has made remarkable strides,
characterized by several key trends:

\begin{enumerate}
\def\labelenumi{\arabic{enumi}.}
\item
  \textbf{Scaling of Model Size and Computational Resources}
\end{enumerate}

Models such as GPT-3 {[}2{]},GPT-4 {[}3{]},GPT-4o {[}24{]},Claude(
Anthropic){[}25{]} , PaLM {[}4{]}, Gemini{[}26{]}, Meta (LLAMA Series
){[}27, 28{]} and Gopher have scaled to hundreds of billions of
parameters , leading to emergent capabilities where models exhibit
skills not explicitly programmed{[}29{]}. While these larger models
demonstrate improved performance, they also pose challenges regarding
computational costs and energy consumption.

\begin{enumerate}
\def\labelenumi{\arabic{enumi}.}
\setcounter{enumi}{1}
\item
  \textbf{Advancements in Multimodal AI}
\end{enumerate}

\begin{quote}
Systems capable of processing and generating content across multiple
modalities (text, images, audio, video) are increasingly prevalent.
Examples include GPT-4 {[}30{]}, DALL·E 3 {[}7{]}, Meta's LLAMA 3.2
Series {[}22, 28{]}and Google\textquotesingle s \emph{Gemini} {[}4{]}.
However, multimodal models require sophisticated training techniques to
align different data types.
\end{quote}

\begin{enumerate}
\def\labelenumi{\arabic{enumi}.}
\setcounter{enumi}{2}
\item
  \textbf{Improved Model Efficiency}
\end{enumerate}

\begin{quote}
Techniques like parameter-efficient fine-tuning (e.g., Low-Rank
Adaptation {[}LoRA{]}{[}31{]}, Adapter Tuning{[}32{]}) and model
distillation are making large models more accessible and
deployable{[}33{]}, significantly reducing resource requirements and
enabling broader adoption in various settings.
\end{quote}

\begin{enumerate}
\def\labelenumi{\arabic{enumi}.}
\setcounter{enumi}{3}
\item
  \textbf{Domain-Specific Models}
\end{enumerate}

\begin{quote}
Development of models tailored for specific domains enhances performance
in specialized tasks. Examples include Med-PaLM for healthcare
{[}34{]}and AlphaFold for protein structure prediction{[}35{]}.This
Domain-specific models address unique challenges and can accelerate
advancements in critical fields like medicine and science.
\end{quote}

\begin{enumerate}
\def\labelenumi{\arabic{enumi}.}
\setcounter{enumi}{4}
\item
  \textbf{Rise of Open-Source Initiatives}
\end{enumerate}

\begin{quote}
Open-source models such as BLOOM {[}36{]} and LLaMA {[}21, 22{]} are
democratizing access to advanced AI capabilities {[}12{]}. These
initiatives foster collaboration, transparency, and innovation but also
raise concerns about misuse and ethical considerations.
\end{quote}

\subsubsection{Advancements in Model Architectures}

Advancements in generative AI are deeply rooted in innovations in model
architectures. Early approaches like Recurrent Neural Networks (RNNs),
including Long Short-Term Memory (LSTM) networks {[}37{]}, were
developed for sequential data processing but faced limitations in
capturing long-range dependencies and challenges in parallelization
{[}38{]}. The introduction of Transformer Architecture by Vaswani et al.
{[}16{]} marked a revolutionary shift, leveraging self-attention
mechanisms for efficient parallel processing and capturing complex
dependencies, forming the backbone of modern LLMs and multimodal models.
Large Language Models (LLMs), such as the GPT series, built upon
transformers, have demonstrated remarkable advancements in language
understanding and generation {[}2{]}, yet bring challenges like high
computational demands and risks of generating misleading or biased
content. In image generation, diffusion models like DALL·E 2 and Stable
Diffusion {[}19{]} have emerged as powerful tools, enabling
high-quality, controllable text-to-image synthesis. Generative
Adversarial Networks (GANs) {[}37{]}, while less prominent today, played
a foundational role in advancing image generation, although they still
face challenges like training instability and mode collapse. These
architectural advancements collectively illustrate the dynamic evolution
and growing capabilities of generative AI.

Enhancing these advancements, Retrieval-Augmented Generation (RAG)
integrates external knowledge sources to improve factual accuracy and
reduce hallucinations in language generation tasks. By retrieving
relevant information from large datasets, RAG models, such as OpenAI's
WebGPT, enhance applications in question answering, dialogue systems,
and content generation {[}39-42{]}. Building on this, Graph RAG
incorporates graph-based data structures, leveraging Graph Neural
Networks (GNNs) to process structured knowledge, making it particularly
effective in knowledge-intensive domains like reasoning over structured
data, entity-centric tasks, and recommendation systems {[}43-44{]}.
However, challenges persist in computational complexity and data
quality, which are critical for reliable outputs.

Advancing further, multimodal AI represents the frontier of generative
models, enabling systems to process and generate content across
modalities such as text, images, audio, and video {[}45{]}. Models like
CLIP {[}46{]} and GPT-4V {[}47{]} showcase the power of integrating
modalities to create more versatile and human-like AI systems. While
aligning different modalities and managing computational demands pose
challenges, advancements in cross-modal transfer learning and
applications in augmented and virtual reality are propelling this field
forward. The implications of multimodal AI are transformative, promising
more immersive human-computer interactions that mimic natural
communication and perception {[}48{]}. Collectively, these advancements
underscore the dynamic evolution of generative AI, highlighting its
expanding capabilities and the challenges that must be addressed for
continued innovation.

\subsection{ Cloud Computing: The Enabler of Generative AI at Scale}

Cloud computing has become instrumental in the development and
deployment of generative AI models, offering the necessary
infrastructure and services to support these computationally intensive
tasks.

\subsubsection{Evolution of Cloud Services for AI}

The relationship between cloud computing and artificial intelligence has
progressed significantly, with various service models emerging to meet
the demands of AI applications. \textbf{Figure 3} provides a visual
summary of these evolutionary stages.

The relationship between cloud computing and AI has evolved
significantly over time. Initially, cloud providers offered
\textbf{Infrastructure as a Service (IaaS)}, enabling organizations to
deploy AI workloads on virtual machines with scalable storage and
networking tailored for high-performance computing {[}49{]}. This
evolved into \textbf{Platform as a Service (PaaS)}, where managed
machine learning services, such as Amazon SageMaker {[}50{]}, Azure
Machine Learning {[}51{]}, and Google Cloud AI Platform {[}52{]},
simplified the training and deployment processes for AI applications.
The emergence of \textbf{AI as a Service (AIaaS)} further broadened
access by providing pre-trained models and APIs, exemplified by services
like AWS Comprehend for NLP {[}53{]}, Azure Cognitive Services {[}54{]},
and Google Cloud Vision AI {[}55{]}. More recently, cloud providers have
introduced \textbf{Specialized AI Infrastructure}, featuring custom
hardware such as Tensor Processing Units (TPUs) {[}56{]} and
AI-optimized instances tailored for deep learning. For instance, AWS
offers EC2 instances with NVIDIA A100 GPUs, while Google Cloud provides
TPU v4 pods for intensive AI workloads {[}57{]}.

% Figure 3
\begin{figure}[H]
    \centering
   \includegraphics[width=0.8\textwidth]{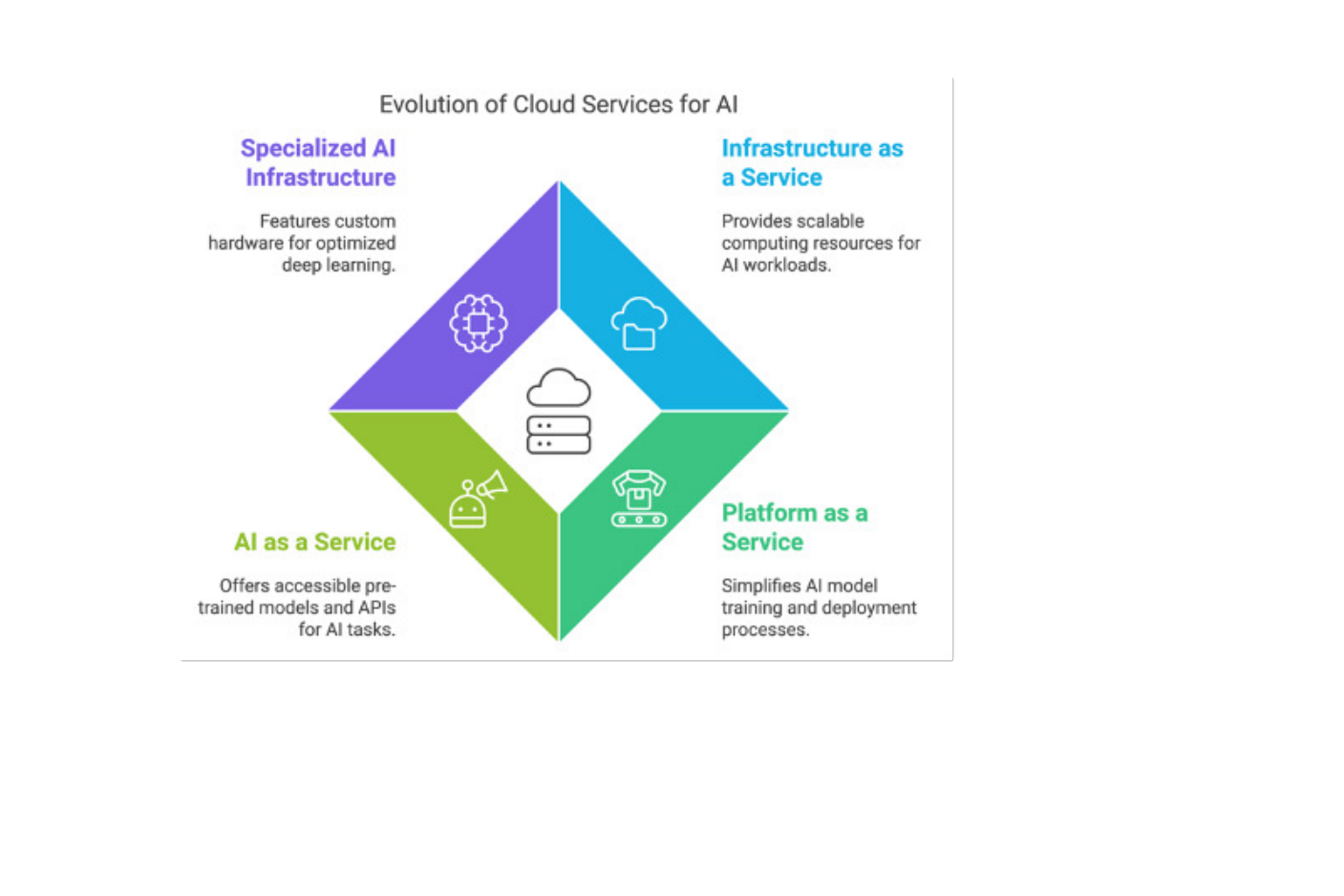}
    \caption{Evolution of Cloud Service Models for AI Development}
    \label{fig:cloud_evolution_Figure_3}  
\end{figure}

\subsubsection{Key Cloud Capabilities for Generative AI}

Several cloud capabilities are particularly crucial for generative AI
development:
\begin{enumerate}
\def\labelenumi{\arabic{enumi}.}
\item
  \textbf{Scalable Compute Resources:} Elastic scaling processes help
  with dynamic allocation of resources based on workload demands,
  ensuring efficiency and cost-effectiveness \cite{ref58}. Access to
  high-performance GPUs and TPUs is crucial for training large models
  \cite{ref59}. Examples include AWS EC2 P4d instances \cite{ref60}, Azure NDv4 Series, and Google Cloud TPU v4 \cite{ref57}.
\item
  \textbf{Managed AI Services:} Automated Machine Learning (AutoML)
  provides pre-built tools and APIs to simplify the development process.
  These tools include capabilities for model selection, hyperparameter
  tuning, and deployment, such as AWS SageMaker Autopilot \cite{ref61}, Azure
  Automated Machine Learning \cite{ref62}, and Google Cloud AutoML \cite{ref52}.
\item
  \textbf{Data Storage and Management:} Scalable storage solutions and
  data management tools are essential for handling large datasets. Data
  lakes and warehousing services provide efficient data processing and
  analysis. Examples include AWS S3 and Redshift \cite{ref63}, Azure Data
  Lake Storage \cite{ref64}, and Google BigQuery \cite{ref65}.
\item
  \textbf{Global Accessibility:} Cloud infrastructure enables
  collaboration and model serving on a global scale. Content Delivery
  Networks (CDNs) provide low-latency model inference worldwide \cite{ref66, ref67}. Examples include AWS Global Accelerator \cite{ref68, ref69}, Azure Front Door \cite{ref70}, and Google Cloud CDN \cite{ref66}.
\item
  \textbf{Cost Optimization:} Pay-as-you-go models and spot instances
  allow for efficient resource utilization \cite{ref71}, which can help lower
  infrastructure costs. Cost management tools provide
  recommendations for optimizing AI workloads. Examples include AWS Cost
  Explorer \cite{ref72}, Azure Cost Management \cite{ref73}, and Google Cloud Cost
  Management \cite{ref74}.
\end{enumerate}

\subsubsection{Cloud-Native AI Development}

Cloud providers have significantly expanded their offerings of
cloud-native tools for AI development, introducing comprehensive
solutions that facilitate efficient development and deployment of AI
applications. At the foundation of these offerings lies containerization
and orchestration technology \textbf{{[}75{]}}, where platforms like
Docker and Kubernetes have become instrumental in creating consistent
environments for AI model development and deployment across diverse
platforms. Docker\textquotesingle s containerization ensures
reproducibility of development environments, while Kubernetes provides
sophisticated automation for scaling and managing containerized AI
workloads, particularly beneficial for large-scale applications spanning
multiple cloud or hybrid environments. In parallel, serverless AI
architectures \textbf{{[}76{]}} have emerged as a transformative
approach, with platforms such as AWS Lambda, Azure Functions, and Google
Cloud Functions enabling event-driven AI execution without the
complexity of server management. This serverless paradigm facilitates
real-time AI model triggering for various applications, from customer
interaction processing to sensor data analysis, while maintaining
automatic scaling capabilities and cost efficiency. The integration of
MLOps \textbf{{[}77, 78{]}} represents another significant advancement,
incorporating DevOps principles into the AI lifecycle. Leading cloud
services including AWS SageMaker, Google Vertex AI, and Azure ML have
implemented comprehensive MLOps tooling for version control, continuous
integration, and model retraining, thereby ensuring the sustained
accuracy and compliance of AI models in production environments.

\subsubsection{Hybrid and Multi-Cloud Strategies}

Organizations are increasingly adopting sophisticated hybrid and
multi-cloud strategies to optimize their AI development capabilities and
meet complex operational requirements. The hybrid cloud approach
{[}79{]} has gained particular traction, combining on-premises
infrastructure with cloud services to address sensitive data handling
needs and meet stringent compliance requirements. This hybrid model
enables organizations to maintain critical data and processes locally
while leveraging the scalability and advanced capabilities of cloud
services. Complementing this, multi-cloud strategies {[}80{]} have
emerged as a crucial approach for organizations seeking to minimize
vendor lock-in and optimize their AI capabilities by selectively
leveraging the unique strengths of different cloud providers. The
integration of edge computing with cloud services {[}81{]} represents
the latest evolution in this space, extending AI capabilities to edge
devices while maintaining seamless cloud connectivity for model updates
and data aggregation. This edge-cloud integration facilitates real-time
processing and decision-making at the network edge while ensuring
consistent model performance through regular updates and centralized
data management.

\subsubsection{Security and Compliance in Cloud AI}

Cloud providers offer a range of security features {[}82{]} to ensure
that AI workloads are both secure and compliant with industry
regulations. Data encryption ensures the confidentiality and integrity
of information by offering end-to-end encryption for data at rest and in
transit. To manage access to resources, Identity and Access Management
(IAM) provides fine-grained control, allowing enforcement of the
principle of least privilege. Moreover, compliance certifications such
as HIPAA, GDPR, and SOC 2 help organizations meet regulatory
requirements. Cloud platforms also address AI model security by offering
tools that protect models from adversarial attacks and unauthorized
access, ensuring the secure hosting and inference of AI models.

\subsection{Convergence of Edge Computing and AI}

The integration of edge computing with cloud-based AI is an emerging
trend, addressing latency issues and enabling real-time AI applications
{[}83{]}. This convergence is particularly relevant for IoT devices,
autonomous systems, and privacy-sensitive applications. where timely and
secure processing is critical. Key developments in this area include
model compression techniques to optimize AI for edge deployment,
federated learning for distributed model training{[}84{]}, and edge-cloud collaborative AI architectures that allow seamless cooperation between cloud and edge resources Edge-cloud collaborative AI architectures {[}85{]}

\subsection{Quantum Computing and AI}

Quantum computing holds promise for accelerating certain AI algorithms
and potentially enabling new AI capabilities. While still in its early
stages, quantum machine learning is an active area of research with
potential implications for generative AI {[}86{]}.Major cloud providers,
including IBM(IBM Quantum{[}87{]}), Microsoft (Azure Quantum{[}88{]}),
Amazon (Amazon Braket {[}89{]}) and Google (Google Quantum AI {[}90{]}),
are beginning to offer quantum computing services, paving the way for
future quantum-enhanced AI applications.

\subsection{Environmental Sustainability}

The environmental impact of training and deploying large AI models has
become a significant concern. Cloud providers are addressing this
challenge by implementing energy-efficient data centers, carbon-neutral
cloud operations, and tools for monitoring and optimizing the carbon
footprint of AI workloads {[}91, 92{]}.This emphasis on sustainability
lays the groundwork for our detailed exploration of cloud platforms
designed for generative AI, highlighting the intricate balance between
technological advancements, infrastructure needs, ethical
considerations, and regulatory demands. A comprehensive understanding of
these foundational elements is essential for researchers, practitioners,
and decision-makers navigating the rapidly evolving generative AI
landscape.

\section{ Cloud Service Providers Overview}

The development and deployment of generative AI solutions are
significantly influenced by the capabilities and offerings of cloud
service providers. This section provides a comprehensive overview of the
major cloud providers, their market positions, and specific services
tailored for generative AI development.

\subsection{ Market Overview and Shares}

The cloud computing market has experienced significant growth,
particularly in AI and machine learning services. \textbf{Table 1} has
showing global cloud AI market is projected to reach \$67.56 billion in
2024 and grow at a CAGR of 32.37\% to reach \$274.54 billion by
2029{[}93{]}. The overall cloud infrastructure market is dominated by
three major players: Amazon Web Services (AWS), Microsoft Azure, and
Google Cloud Platform (GCP). Together, these three providers account for
66\% of total cloud infrastructure spending in the first quarter of
2024{[}94{]}.

% Adjust preamble for better flexibility
\setlength{\emergencystretch}{10pt}
\setlength{\tabcolsep}{4pt}

\begin{longtable}{|p{4.5cm}|p{4.7cm}|p{4.7cm}|}
\hline
\textbf{Provider} & \textbf{Overall Cloud Market Share} & \textbf{AI/ML Services Market Share} \\
\hline
\endfirsthead
\hline
\textbf{Provider} & \textbf{Overall Cloud Market Share} & \textbf{AI/ML Services Market Share} \\
\hline
\endhead
\hline
\multicolumn{3}{r}{\textit{Continued on next page}} \\
\hline
\endfoot
\hline
\endlastfoot
Amazon Web Services (AWS) & 31\% & 34\% \\
\hline
Microsoft Azure & 25\% & 39\% \\
\hline
Google Cloud Platform (GCP) & 10\% & 11\% \\
\hline
Alibaba Cloud & 5\% & 6\% \\
\hline
Others & 29\% & 10\% \\
\hline
\end{longtable}

\textbf{Table 1: Estimated Cloud Market Shares (Overall and AI/ML
Services)}

*\textbf{Note}: It\textquotesingle s important to note that the AI/ML
services market share is more dynamic and rapidly evolving compared to
the overall cloud market share{[}93, 95{]}.

\textbf{AI Usability on Cloud Platforms:}

\begin{enumerate}
\def\labelenumi{\arabic{enumi}.}
\item
  \textbf{Amazon Web Services (AWS)}: AWS offers a wide range of AI and
  ML services, including Amazon SageMaker for building, training, and
  deploying machine learning models. AWS maintains a strong position in
  the AI market, leveraging its extensive cloud infrastructure {[}95{]}.
\item
  \textbf{Microsoft Azure}: Azure AI Services, which includes Azure
  OpenAI, has taken the lead among managed AI services offered by major
  Cloud Service Providers (CSPs). 39\% of organizations are using this
  service, and the number of deployed Azure OpenAI instances rose by
  228\% over a 4-month period in 2023.
\item
  \textbf{Google Cloud Platform (GCP)}: GCP has carved a niche for
  itself through innovations in AI, ML, and data analytics.
  It\textquotesingle s particularly strong in offering AI and ML tools
  for developers and data scientists {[}95{]}.
\item
  \textbf{IBM Cloud}: While not as dominant in the overall cloud market,
  IBM has a strong presence in enterprise AI solutions, particularly
  with its Watson AI platform {[}95{]}.
\item
  \textbf{Alibaba Cloud}: As the largest cloud provider in Asia, Alibaba
  Cloud is making significant investments in AI and ML services,
  particularly targeting the Asian market {[}95{]}.
\end{enumerate}

The AI market within cloud platforms is rapidly evolving, with over 70\%
of organizations now using managed AI service. The adoption of AI
technology in cloud environments now rivals the popularity of managed
Kubernetes services. It\textquotesingle s worth noting that while AWS
remains the overall cloud market leader, Microsoft Azure has made
significant gains in the AI space, particularly with its Azure OpenAI
service. Google Cloud is also showing strong growth in AI and ML
offerings.

.

\subsection{ Major Players and Their Generative AI Services}

\subsubsection{ Amazon Web Services (AWS)}

AWS is a leading cloud provider offering a robust Generative AI Stack
that supports end-to-end AI development. Key services include Amazon
SageMaker, which simplifies model development, training, and deployment,
and Amazon Bedrock, enabling customization of foundation models like
Amazon Titan.

High-performance infrastructure, such as EC2 P4d instances with NVIDIA
GPUs, supports large-scale model training. AWS also enhances developer
productivity through tools like Amazon Code Whisperer, providing
real-time AI-powered coding suggestions.

Recent advancements, such as the introduction of Amazon Bedrock, have
positioned AWS as a critical enabler of generative AI applications,
spanning text generation to image synthesis.

\textbf{See Supplementary Material Figure S2 and Section S2.1 for
detailed descriptions of AWS services and developments.}

\subsubsection{ Microsoft Azure}

Microsoft Azure is the second-largest cloud provider, renowned for its
enterprise-friendly environment and deep Microsoft Azure is the
second-largest cloud provider, renowned for its enterprise-friendly
environment and integration with Microsoft's software ecosystem. The
Azure Generative AI Stack offers comprehensive tools and services to
support AI-driven innovation.

Azure's OpenAI Service provides developers access to advanced language
models, including GPT-3, enabling sophisticated generative AI
applications. Azure Machine Learning serves as a robust platform for
building, training, and deploying machine learning models at scale.

With its exclusive partnership with OpenAI, Azure delivers cutting-edge
generative AI capabilities, further enhanced by seamless integration
with Microsoft 365, Dynamics 365, and Power Platform. Azure also
emphasizes Responsible AI, offering tools for fairness, compliance, and
interpretability.

\textbf{See Supplementary Material Figure S3 and Section S2.2 for
detailed descriptions of Azure services and developments.}

\subsubsection{ Google Cloud Platform (GCP)}

Google Cloud Platform (GCP) is renowned for its advanced AI and machine
learning capabilities, underpinned by Google's leadership in AI
research. The GCP Generative AI Stack includes tools like Vertex AI,
which offers pre-trained models and a Generative AI Studio for
customized solutions. Cloud TPUs provide high-performance infrastructure
optimized for generative AI workloads.

GCP's strengths include natural language processing and computer vision,
supported by a commitment to open-source innovation through projects
like TensorFlow and JAX. Recent advancements, such as the Multitask
Unified Model (MUM) for multimodal applications, have expanded its
generative AI capabilities across diverse use cases.

\textbf{See Supplementary Material Figure S4 and Section S2.3 for
detailed descriptions of GCP's services and developments.}

\subsubsection{ IBM Cloud}

IBM Cloud excels in enterprise AI solutions, particularly in regulated
industries like healthcare and finance. Its Generative AI Stack includes
Watson Studio for building and training models, and Watson Natural
Language Generation (NLG) for creating human-like text. IBM's emphasis
on Explainable AI ensures transparency and compliance, critical for
enterprise trust.

IBM's hybrid cloud capabilities enable flexible deployments, while the
recent launch of watsonx.ai enhances its generative AI workflows for
large-scale applications. These features make IBM Cloud a trusted choice
for enterprise-grade AI projects.

\textbf{See Supplementary Material Figure S5 and Section S2.4 for
detailed descriptions of IBM's services and developments.}

\subsubsection{ Oracle Cloud}

Oracle Cloud is a robust platform for data-intensive workloads,
leveraging its expertise in database management and enterprise
applications. Key offerings include the Oracle Digital Assistant for
conversational interfaces and Oracle Cloud Infrastructure (OCI) Data
Science for collaborative machine learning workflows. Oracle's
integration with its enterprise applications enhances data accessibility
and workflow efficiency while prioritizing data security and compliance.

Recent advancements, such as partnerships with NVIDIA, have strengthened
Oracle's AI capabilities, focusing on streamlining enterprise operations
through generative AI.

\textbf{See Supplementary Material Section S2.5 for detailed descriptions
of Oracle's services and developments.}

\subsubsection{ Alibaba Cloud}

Alibaba Cloud is a leading provider in the Asia-Pacific region, offering
generative AI tools tailored to businesses in Asian markets. Its Machine
Learning Platform for AI (PAI) supports deep learning frameworks, while
NLP services provide APIs for language understanding, translation, and
text generation.

The launch of Tongyi Qianwen 2.0, a large language model, has enhanced
Alibaba's generative AI capabilities, particularly for personalized
e-commerce experiences.

\textbf{See Supplementary Material Section S2.6 for detailed descriptions
of Alibaba Cloud's services and developments.}

\subsubsection{ Databricks}

Databricks excels in unified analytics, integrating data engineering,
analytics, and machine learning into a single platform. Key offerings
include the Databricks Lakehouse Platform, MLflow for the machine
learning lifecycle, and Delta Lake for reliable data storage.Recent
innovations, such as the AI Summit and Databricks Machine Learning
platform, highlight the company's focus on collaborative data science
workflows and scalable machine learning.

\textbf{See Supplementary Material Section S2.7 for detailed descriptions
of Databricks services and developments.}

\subsubsection{ Snowflake}

Snowflake is expanding its leadership in cloud data warehousing into AI
and machine learning services. Key tools include Snowpark for data
processing and Snowflake Data Marketplace for enriched machine learning
models through diverse datasets.

Recent developments, such as the introduction of Snowflake Cortex,
enhance data processing and analytics with AI-powered capabilities.

\textbf{See Supplementary Material Section S2.8 for detailed descriptions
of Snowflake services and developments.}

\subsubsection{ Other Big Providers}

In addition to the major cloud providers---Amazon Web Services (AWS),
Microsoft Azure, Google Cloud Platform (GCP), IBM Cloud, Oracle Cloud,
Alibaba Cloud, Databricks, and Snowflake---there is a growing ecosystem
of platforms contributing to the development and scaling of generative
AI services. The Modern AI Stack: The Emerging Building Blocks for GENAI
provides a comprehensive view of various tools and platforms supporting
the AI lifecycle, spanning from foundational infrastructure to
deployment and observability.

As shown in \textbf{Figure 4}, the \textbf{Modern AI Stack} is divided
into multiple layers, each addressing a critical component of the
generative AI development process:

\textbf{Layer 1:} Compute + Foundation: This foundational layer includes
GPU providers such as Azure, AWS, Google Cloud, NVIDIA, and Lambda,
which offer the necessary infrastructure for training and deploying
large-scale AI models. A significant player in this layer is Hugging
Face, which serves as a repository for open-source models contributed by
various providers. Hugging Face offers models across a wide range of
applications---text generation, image recognition, natural language
understanding---allowing developers to access and fine-tune models
provided by different platform providers. The repository acts as a
collaborative ecosystem, enabling the growth of the generative AI
community and facilitating access to state-of-the-art models.

\textbf{Layer 2:} Data: This layer encompasses tools for managing data,
including ETL (Extract, Transform, Load) pipelines from platforms like
gable, datology ai, and clean lab, and database solutions such as
Pinecone, Weaviate, Neo4j, and Databricks for storing vector embeddings,
metadata, and other structured data crucial for AI applications.

\textbf{Layer 3:} Deployment: Key platforms such as vellum and LangSmith
support prompt management for optimizing model interactions, while
orchestration tools like Martian and Orkes ensure seamless AI workflow
integration. Agent tools such as Relevance AI and LangChain empower
developers to build and automate AI-driven tasks.

\textbf{Layer 4:} Observability: To ensure transparency, security, and
performance, platforms like AgentOps.ai, heliocone, and Patronus AI
provide observability, evaluation, and security solutions that allow
enterprises to monitor AI systems in production environments
effectively.

% Figure 4
\begin{figure}[H]
    \centering
    \includegraphics[width=\textwidth]{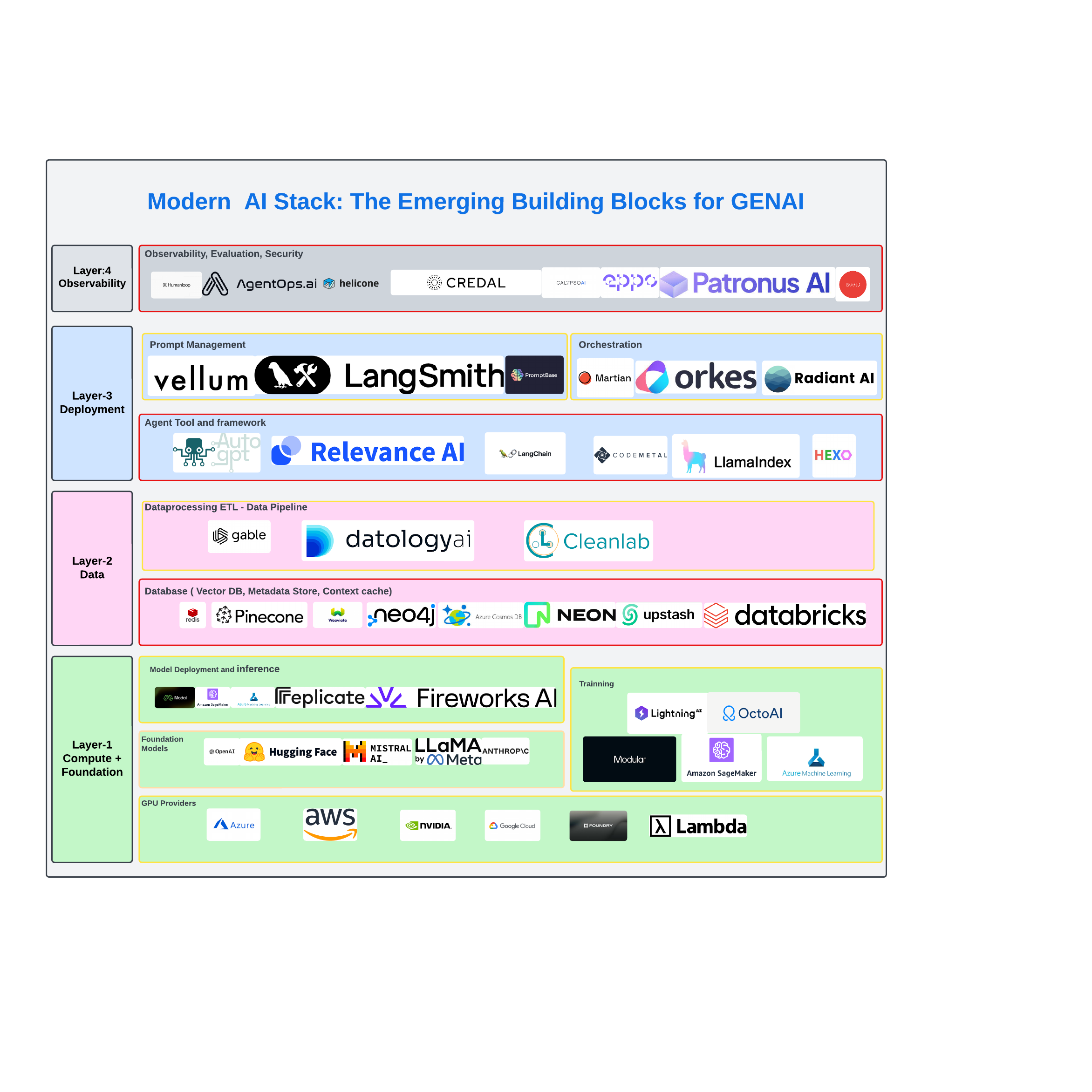}
    \caption{Modern AI Stack: The Emerging Building Blocks for GENAI}
    \label{fig:stack_figure4}
\end{figure}

These platform providers continue to grow, they play a crucial role in
democratizing access to AI and generative models. By providing a
centralized repository for pre-trained models, they support
collaboration and innovation across various sectors. Similarly, the
other layers in the Modern AI Stack showcase how emerging platform
providers are complementing the offerings from the big cloud players,
enabling businesses to deploy and scale cutting-edge generative AI
applications with greater efficiency and flexibility.

\subsection{Market Positioning and Strategic Partnerships}

The cloud AI market is highly competitive, with providers employing
various strategies and forming partnerships to strengthen their
positions.

\subsubsection{Key Market Positioning Strategies}

In the cloud AI market, providers employ a range of strategies and form
alliances to enhance their positions and expand their influence. One
common strategy among leading providers such as AWS, Azure, and GCP is
the development of end-to-end AI solutions. These platforms cover the
entire machine learning lifecycle, from data processing to model
deployment, enabling a streamlined approach for building generative AI
applications. Additionally, some companies have chosen to specialize in
data and AI integration; for instance, Databricks and Snowflake focus on
unifying data warehousing with machine learning, facilitating an
environment where data analytics and AI workflows converge seamlessly.

The emphasis on ethical and responsible AI has also become a key
positioning strategy. Many cloud providers now offer tools designed for
bias detection, explainability, and regulatory compliance, aiming to
foster trust and accountability in AI systems. Moreover, active
participation in open-source communities has proven beneficial for both
innovation and engagement. Major players such as Google and Facebook
contribute significantly to projects like TensorFlow and PyTorch,
respectively, leveraging open-source contributions to drive community
collaboration and accelerate advancements in AI technology.

\subsubsection{ Strategic Partnerships}

Strategic partnerships have become instrumental in shaping the
competitive dynamics of the cloud AI landscape. Microsoft Azure's
exclusive access to OpenAI's GPT models, enabled through a significant
partnership, has greatly enhanced Azure's generative AI offerings,
giving it a distinct advantage in the space {[}96{]}. Similarly, Google
Cloud's collaboration with Anthropic has bolstered its capabilities in
responsible AI, positioning it strongly in an increasingly regulated AI
environment {[}97{]}. AWS\textquotesingle s long-standing alliance with
NVIDIA has been pivotal for GPU-accelerated computing, making AWS a
preferred choice for intensive AI workloads {[}98{]}. Meanwhile,
IBM\textquotesingle s partnership with NASA has focused on leveraging AI
for climate science, contributing to foundational advancements in
AI-driven environmental research {[}99{]}. Oracle's collaboration with
NVIDIA has brought powerful AI tools to Oracle Cloud Infrastructure,
expanding its appeal among enterprises seeking robust AI solutions
{[}100{]}. Alibaba Cloud, through its collaboration with Intel, is
making strides in edge AI solutions, particularly in IoT
applications{[}101{]}.

Together, these alliances create a complex network of partnerships that
not only drive rapid innovation but also pose challenges and
opportunities for organizations aiming to leverage these technologies.
As cloud providers continue to compete, further collaborations and
strategic shifts are anticipated, as each provider seeks to
differentiate and capture market share in the competitive cloud-based AI
services arena.

\section{ Cloud Services for Generative AI}

The development and deployment of generative AI applications require
robust, scalable, and secure infrastructure. Cloud service providers,
such as AWS, Azure, GCP, IBM, Oracle, and Alibaba Cloud, offer a suite
of tools designed to meet these demands. Key services include
High-Performance Computing (HPC), serverless architectures, and AI/ML
platforms, enabling efficient scaling and deployment of generative AI
workloads. Generative AI applications rely on various cloud
infrastructure components, including edge computing, data lakes, and AI
model management. Each provider brings unique strengths in these areas,
offering specialized tools to enhance scalability, efficiency, and
security. Table 2 provides a comprehensive comparison of these offerings
across major cloud providers. \textbf{(See Supplementary Material Table
S1 for detailed comparisons of cloud service offerings across major
providers.)}

Among the critical service categories, High-Performance Computing (HPC)
stands out for enabling large-scale training, while serverless
architectures streamline model deployment. Additionally, edge computing
solutions extend AI capabilities to real-time environments, and robust
data lakes and warehousing solutions facilitate efficient data handling
for AI development.

Cloud providers are continuously evolving their offerings to address
challenges such as sustainability, cost-efficiency, and compliance,
making it essential for organizations to choose platforms that align
with their specific requirements and use cases.

\subsection{Compute Services}

Compute services form the foundation of generative AI development,
providing the necessary processing power for training and deploying
large-scale models. The increasing complexity of generative AI models,
such as GPT-4 {[}3{]}, PaLM 2 {[}102{]}, and Stable Diffusion XL
{[}19{]}, has driven cloud providers to offer more sophisticated and
specialized compute options.

\subsubsection{ High-Performance Computing (HPC) and GPUs}

High-Performance Computing (HPC){[}103{]} and Graphics Processing Units
(GPUs){[}104{]} are crucial for handling the computationally intensive
tasks associated with generative AI model training and
inference{[}105{]}. Cloud providers have developed comprehensive
offerings of GPU-optimized instances and HPC clusters to meet these
demanding computational requirements, as detailed in Table 2 and Table
3.

\begin{longtable}[]{|p{0.18\linewidth}|p{0.18\linewidth}|p{0.2\linewidth}|p{0.18\linewidth}|p{0.18\linewidth}|}
\hline
\textbf{Cloud Provider} & \textbf{HPC Service} & \textbf{GPU Service} & \textbf{Max GPU per Instance} & \textbf{Specialized AI Chips} \\
\hline
\endfirsthead
\hline
\textbf{Cloud Provider} & \textbf{HPC Service} & \textbf{GPU Service} & \textbf{Max GPU per Instance} & \textbf{Specialized AI Chips} \\
\hline
\endhead
AWS & EC2 P4d, P3, and G5 instances & NVIDIA A100, V100, T4 & 8 (P4d) & AWS Trainium, Inferentia \\
\hline
Azure & NC, ND, and NV series & NVIDIA A100, V100, T4 & 8 (NDv4) & Azure NPU (Preview) \\
\hline
Google Cloud & Cloud TPU v4, A2 instances & NVIDIA A100, T4 & 16 (A2) & Cloud TPU \\
\hline
IBM Cloud & HPC instances & NVIDIA T4, V100 & 4 (AC2) & - \\
\hline
Oracle Cloud & HPC instances & NVIDIA A100, V100 & 8 (BM.GPU4.8) & - \\
\hline

\end{longtable}

\textbf{Table 2:} Comparison of High-Performance Computing (HPC) and GPU Services Across Major Cloud Providers

\begin{longtable}[]{|p{0.18\linewidth}|p{0.15\linewidth}|p{0.13\linewidth}|p{0.18\linewidth}|p{0.18\linewidth}|p{0.11\linewidth}|}
\hline
\textbf{Cloud Provider} & \textbf{Service} & \textbf{Launch} & \textbf{GPU Configuration} & \textbf{Key Features} & \textbf{References} \\
\hline
\endfirsthead
\hline
\textbf{Cloud Provider} & \textbf{Service} & \textbf{Launch} & \textbf{GPU Configuration} & \textbf{Key Features} & \textbf{References} \\
\hline
\endhead
AWS & EC2 P5 Instances & Late 2023 & 16 NVIDIA H100 GPUs & 3.35 TB/s memory bandwidth per GPU & {[}106{]} \\
\hline
Azure & ND H100 v5 Series & Early 2024 & 8 NVIDIA H100 80GB GPUs & NVSwitch for high-speed GPU communication & {[}107{]} \\
\hline
Google Cloud & A3 Instances & Mid-2023 & 16 NVIDIA H100 GPUs & Optimized for generative AI workloads & {[}108{]} \\
\hline
Multiple & Specialized AI Chips & 2023-2024 & Custom silicon & AWS Trainium 2, Inferentia 2, Google TPU v5e & {[}109, 110{]} \\
\hline

\end{longtable}

\textbf{Table 3:} Recent Cloud Provider Hardware Innovations for Generative AI (2023-2024)

The most recent developments in cloud-based computing services
demonstrate significant advancement in processing capabilities,
particularly with the introduction of NVIDIA H100-based instances across
major providers. AWS and Google Cloud lead in raw GPU capacity with
their latest offerings supporting up to 16 NVIDIA H100 GPUs per
instance, while Azure emphasizes high-speed inter-GPU communication
through NV Switch technology. Additionally, cloud providers are
increasingly developing specialized AI chips optimized for specific
workloads, such as AWS\textquotesingle s Trainium 2 for training and
Inferentia 2 for inference {[}109{]}, and Google\textquotesingle s
advanced Tensor Processing Units (TPUs) v5e {[}110{]}. These innovations
reflect the industry\textquotesingle s commitment to addressing the
growing computational demands of generative AI while optimizing for both
performance and efficiency.

\subsubsection{ Distributed Training and Model Parallelism}

As generative AI models increase in size and complexity, distributed
training and model parallelism have become crucial to enable scalable
and efficient processing across hardware resources. AWS Sage Maker, for
example, employs model parallelism by automatically partitioning large
models across multiple GPUs, facilitating smoother scaling. Azure
Machine Learning supports distributed training through frameworks such
as DeepSpeed and Megatron-LM, enhancing training efficiency on extensive
datasets. Meanwhile, Google Cloud\textquotesingle s Distribution
Strategy API streamlines the distribution of training across multiple
TPUs or GPUs, with recent updates introducing advanced parallelism
techniques to optimize performance for large-scale models {[}11{]}.

\begin{figure}[H]
    \centering
    \includegraphics[width=\textwidth]{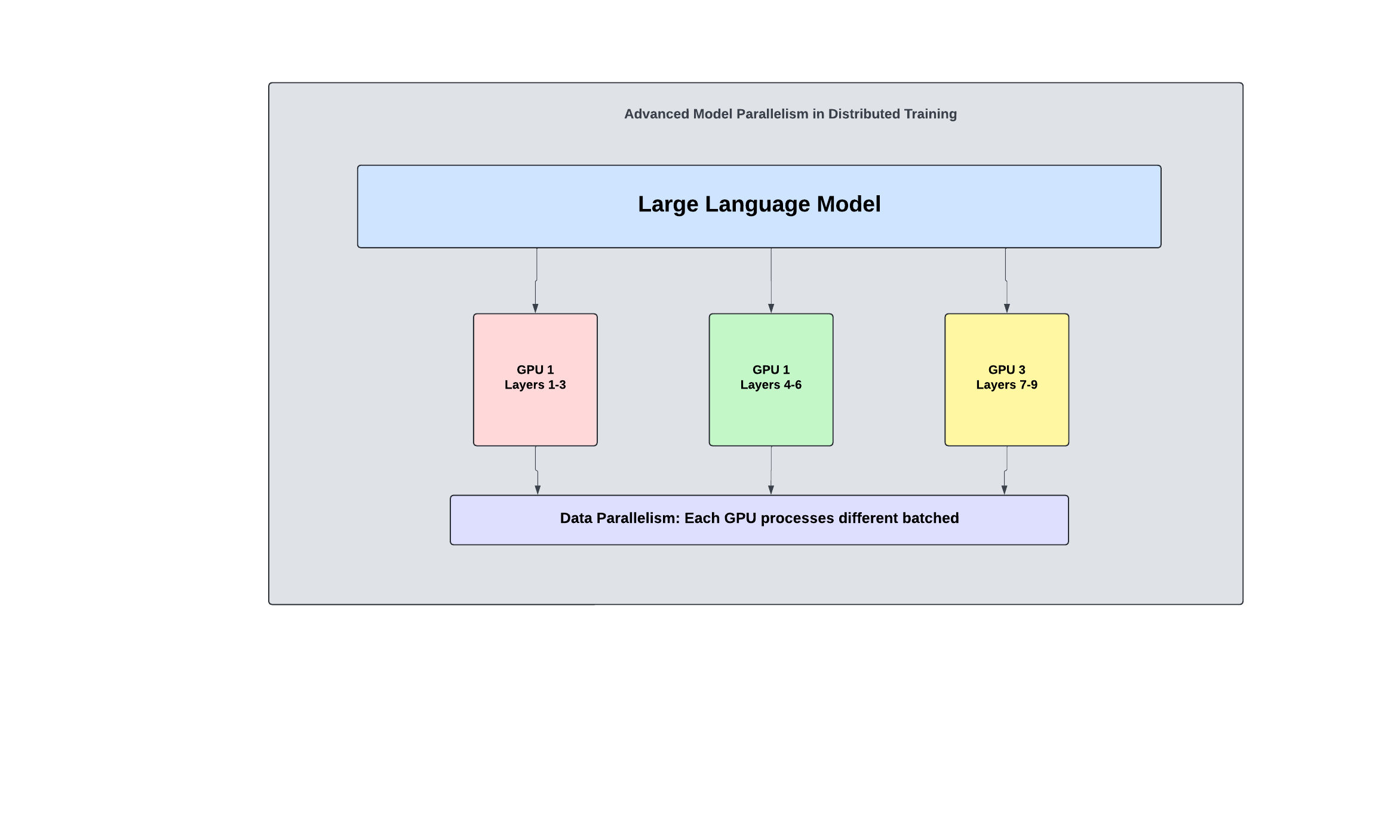}
    \caption{Architecture of distributed training for generative AI system}
    \label{fig:arch1_Figure5}
\end{figure}

As shown in \textbf{Figure 5}, distributed training for generative AI
systems involves splitting model layers across different GPUs, where
each GPU handles a portion of the model\textquotesingle s layers. This
architecture allows for data parallelism, where each GPU processes
different batches of data simultaneously, significantly improving
training performance and scalability.

\subsubsection{ Energy-Efficient AI Computing}

As the computational demands of AI continue to escalate, energy
efficiency has become an essential focus. Cloud providers are
introducing innovative solutions to address this need. For instance,
Microsoft Azure's carbon-aware GPU scheduling dynamically adjusts
computing resources based on carbon intensity, reportedly reducing
emissions by up to 30\% for specific AI workloads. Complementing these
efforts, Azure has also introduced Efficient Net-X, a model architecture
optimized for high performance with minimal energy consumption, which
delivers state-of-the-art results while reducing computational load
{[}111{]}. Additionally, AWS reached a significant milestone in 2023 by
powering all operations with renewable energy, further advancing its
commitment to sustainability by implementing water conservation
techniques to cool AI-centric data centers {[}112{]}. These developments
reflect a growing emphasis on balancing AI innovation with environmental
responsibility.

\subsection{ Serverless Architectures}

Serverless architectures have emerged as a transformative paradigm in
cloud computing, offering unique advantages for deploying and scaling
generative AI models{[}113{]}. This section explores the latest
advancements in serverless technologies and their implications for
generative AI development and deployment{[}76{]}.

\subsubsection{ Evolution of Serverless Computing for AI}

Serverless computing has undergone significant transformation since its
inception, with recent developments specifically addressing the unique
demands of AI workloads. A fundamental advancement in this evolution has
been the enhancement of traditional Function-as-a-Service (FaaS)
platforms to accommodate AI-specific requirements, including support for
extended execution times and expanded memory allocations necessary for
complex AI operations {[}114{]} . This progress has been complemented by
the strategic integration of containerization technologies with
serverless platforms, enabling the deployment of sophisticated AI
workloads in a serverless manner while maintaining the benefits of
container-based isolation and portability {[}115, 116{]}. Perhaps most
significantly, cloud providers have introduced AI-optimized serverless
platforms specifically engineered for AI inference, featuring native
support for popular AI frameworks and model formats {[}115{]}.These
specialized platforms represent a crucial step forward in serverless
architecture, offering purpose-built environments that streamline the
deployment and execution of AI workloads while maintaining the
cost-effectiveness and scalability benefits inherent to serverless
computing. This evolution marks a significant maturation in serverless
technology, making it increasingly viable for production-scale AI
applications.

\begin{longtable}[]{|p{0.10\linewidth}|p{0.2\linewidth}|p{0.17\linewidth}|p{0.1\linewidth}|p{0.15\linewidth}|p{0.18\linewidth}|}
\hline
\textbf{Cloud Provider} & \textbf{Serverless AI Service} & \textbf{Key Features} & \textbf{Max Model Size} & \textbf{Supported Frameworks} & \textbf{Latest Enhancements} \\
\hline
\endfirsthead
\hline
\textbf{Cloud Provider} & \textbf{Serverless AI Service} & \textbf{Key Features} & \textbf{Max Model Size} & \textbf{Supported Frameworks} & \textbf{Latest Enhancements} \\
\hline
\endhead
\hline
AWS & SageMaker Serverless Inference & Auto-scaling, pay-per-use & Up to 20 GB & TensorFlow, PyTorch, MXNet & GPU support, A/B testing \\
\hline
Azure & Azure Functions for AI & Integration with Azure ML, low latency & 14 GB & ONNX, custom containers & Durable Functions for long-running AI tasks \\
\hline
Google Cloud & Cloud Run for AI & Containerized deployments, auto-scaling & 32 GB & TensorFlow, PyTorch, JAX & Support for NVIDIA T4 GPUs \\
\hline
IBM Cloud & Code Engine for AI & Hybrid deployments, scalable inference & 16 GB & TensorFlow, PyTorch, ONNX & Integration with Watson AI services \\
\hline
Oracle Cloud & OCI Functions for AI & Integration with OCI Data Science & 6 GB & Any framework in containers & GPU-accelerated functions (GA) \\
\hline

\end{longtable}

\textbf{Table-4:} A comparison of serverless AI services across major cloud providers.

In \textbf{Table-4} provides a comparison of serverless AI services
across major cloud providers, focusing on key features, supported
frameworks, model size limits, and the latest enhancements.

\subsubsection{ Serverless Inference for Generative AI}

Serverless inference has emerged as a critical focus area for deploying
generative AI models, offering compelling advantages in terms of
cost-efficiency, scalability, and reduced operational overhead {[}115{]}
Recent advances in this domain have introduced several significant
technological improvements that address key challenges in serverless AI
deployment. Advanced serverless platforms now incorporate dynamic
batching capabilities for inference requests, enabling sophisticated
optimization of resource utilization while simultaneously reducing
latency for generative AI models {[}117{]}. This innovation is
complemented by breakthrough developments in serverless architectures
that facilitate efficient caching of large language models, resulting in
substantial reductions in cold start times, a historical bottleneck in
serverless deployments {[}118, 119{]}. Perhaps most significantly,
modern AI-specific serverless platforms have evolved beyond simple
request-volume-based scaling to incorporate more nuanced scaling
mechanisms that consider both model complexity and input size. This
adaptive scaling approach ensures optimal performance for generative AI
tasks by dynamically adjusting resources based on the specific
computational demands of each model and request {[}119{]}.. Together,
these advancements represent a significant maturation of serverless
infrastructure for generative AI deployment, offering organizations more
sophisticated and efficient options for managing their AI workloads.

\subsubsection{ Serverless Workflows for AI Pipelines}

Serverless technologies have become increasingly central to
orchestrating complex AI workflows, encompassing data preprocessing,
model training, and post-processing steps {[}120, 121{]}. As illustrated
in Figure 6, the modern serverless AI workflow implements a streamlined
pipeline that begins with data ingestion, progresses through
preprocessing and model training phases, and culminates in inference and
monitoring stages, creating a feedback loop that ensures continuous
optimization of model performance. Within this framework, cloud
providers have developed sophisticated event-driven architectures that
dynamically trigger various stages of AI workflows in response to data
updates or changes in model performance metrics {[}120{]}. Complementing
these event-driven capabilities, advanced serverless data processing
services have emerged to address the challenges of large-scale data
preparation essential for training generative AI models {[}120{]}. These
services provide robust Extract, Transform, Load (ETL) functionalities
specifically optimized for AI workloads. A particularly significant
advancement in this domain is the integration of federated learning
support, with several serverless platforms now offering native
capabilities for privacy-preserving distributed training of generative
models{[}122, 123{]}. This innovation enables organizations to implement
collaborative learning strategies while maintaining data privacy and
regulatory compliance, representing a crucial step forward in the
evolution of serverless AI architectures.

\begin{figure}[H]
    \centering
    \includegraphics[width=\textwidth]{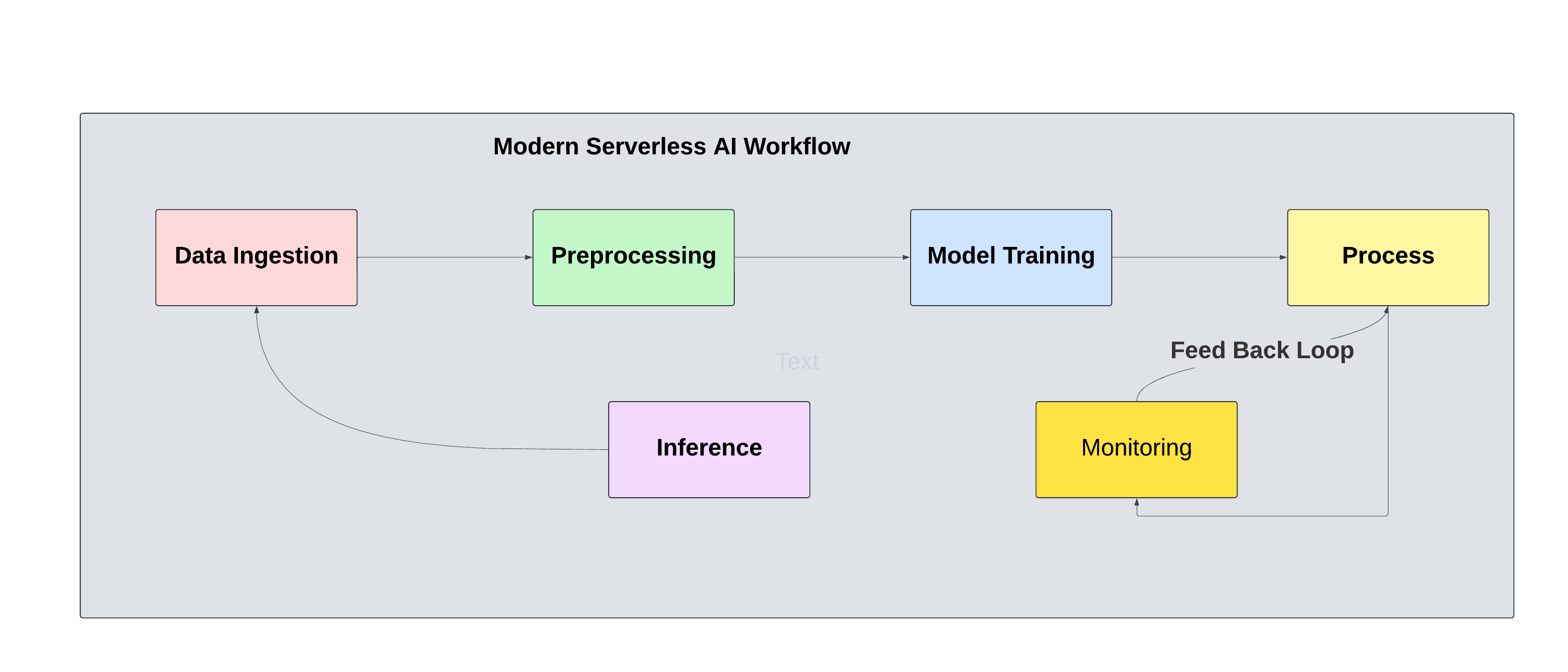}
    \caption{Architecture of serverless pipeline AI Pipeline}
    \label{fig:arch2_Figure6}
\end{figure}

\subsubsection{ Future Trends in Serverless AI}

Future trends in serverless AI include the rise of AI-specific
serverless runtimes, optimized for popular AI frameworks and enhanced
performance, and edge-cloud serverless collaboration, enabling seamless
computation between edge devices and cloud resources for responsive
applications. The emergence of serverless AutoML platforms will automate
the entire machine learning lifecycle, catering to generative AI needs,
while advancements in quantum computing may introduce quantum algorithms
into serverless AI workflows. These innovations will enhance
scalability, cost-efficiency, and accessibility, solidifying serverless
architectures as a cornerstone for generative AI development and
deployment.

\subsection{ Edge Computing}

Edge computing has emerged as a critical paradigm in the deployment of
generative AI models, offering reduced latency, enhanced privacy, and
improved efficiency{[}81{]}. This section explores the latest
advancements in edge computing technologies and their implications for
generative AI development and deployment.

\subsubsection{ Evolution of Edge Computing for Generative AI}

Edge computing has undergone a remarkable transformation to meet the
escalating demands of generative AI, with innovations spanning hardware,
frameworks, and distributed learning approaches. A pivotal advancement
in this evolution has been the development of AI-optimized edge
hardware, featuring specialized devices equipped with AI accelerators
that enable substantially more efficient execution of generative models
at the network edge {[}124{]}. This hardware optimization has been
complemented by the emergence of edge-native AI frameworks, which
represent a new generation of lightweight solutions specifically
engineered for edge deployment. These frameworks excel at optimizing
performance on resource-constrained devices, making sophisticated AI
capabilities accessible in edge environments {[}125{]}. Perhaps most
significantly, the integration of advanced federated learning techniques
has revolutionized edge computing\textquotesingle s capabilities,
enabling collaborative training of generative models across distributed
edge devices while maintaining robust data privacy protections
{[}126{]}.This distributed approach not only preserves data
confidentiality but also reduces network bandwidth requirements and
enables more efficient model training across geographically dispersed
edge nodes. Together, these developments mark a significant maturation
in edge computing\textquotesingle s ability to support increasingly
sophisticated generative AI applications, pushing the boundaries of
what\textquotesingle s possible at the network edge.

\begin{longtable}{|p{0.10\linewidth}|p{0.17\linewidth}|p{0.15\linewidth}|p{0.12\linewidth}|p{0.15\linewidth}|p{0.17\linewidth}|}
\hline
\textbf{Cloud Provider} & \textbf{Edge AI Platform} & \textbf{Key Features} & \textbf{Supported AI Models} & \textbf{Hardware Acceleration} & \textbf{Latest Enhancements} \\
\hline
\endfirsthead
\hline
\textbf{Cloud Provider} & \textbf{Edge AI Platform} & \textbf{Key Features} & \textbf{Supported AI Models} & \textbf{Hardware Acceleration} & \textbf{Latest Enhancements} \\
\hline
\endhead
\hline
AWS & AWS IoT Greengrass & Local inference, ML model management & TensorFlow Lite, MXNet & NVIDIA Jetson, Intel Open VINO & SageMaker Edge Manager integration \\
\hline
Azure & Azure IoT Edge & Custom vision models, speech recognition & ONNX, Custom Vision & Intel Movidius, NVIDIA GPU & Azure Percept integration \\
\hline
Google Cloud & Cloud IoT Edge & TensorFlow Lite support, AutoML Edge & TensorFlow Lite, AutoML Vision Edge & Edge TPU, GPU & Coral AI integration \\
\hline
IBM & IBM Edge Application Manager & Autonomous management, workload orchestration & TensorFlow, PyTorch & NVIDIA Jetson, Intel NCS & Red Hat OpenShift integration \\
\hline
NVIDIA & NVIDIA EGX & GPU-accelerated AI inference & NVIDIA TensorRT, TensorFlow & NVIDIA GPUs, Jetson AGX Orin & Fleet Command for edge AI management \\
\hline

\end{longtable}

\textbf{Table-5}: Comparison of Edge AI Platforms and Key Features
Across Major Cloud Providers

As shown in \textbf{Table 5}, major cloud providers are offering robust
edge AI platforms with support for various AI models and hardware
acceleration options. The latest enhancements focus on integrating edge
capabilities with cloud-based AI services for seamless deployment and
management.

\subsubsection{ Edge Inference for Generative AI}

Edge inference has emerged as a critical capability for deploying
generative AI models in scenarios that demand low latency, offline
operation, or enhanced data privacy protections. Recent advances in
model compression techniques have revolutionized edge deployment
capabilities, with sophisticated quantization and pruning methods
enabling the implementation of large generative models on edge devices
while maintaining minimal performance degradation{[}127-129{]} . This
breakthrough is further enhanced by the development of adaptive edge
inference algorithms, which dynamically optimize model complexity based
on both device capabilities and input complexity, resulting in
significant improvements in both performance and energy efficiency
{[}130{]}. A particularly noteworthy advancement in this domain is the
emergence of distributed inference techniques, which enable the
strategic partitioning of generative models across multiple edge
devices. This distributed approach allows for the execution of
substantially more complex models than what would be possible on any
single device, effectively overcoming the traditional computational
limitations of edge computing {[}125, 131, 132{]}. Together, these
advancements represent a significant leap forward in edge inference
capabilities, making it possible to deploy increasingly sophisticated
generative AI applications at the network edge while maintaining robust
performance and efficiency.

\begin{figure}[H]
    \centering
    \includegraphics[width=\textwidth]{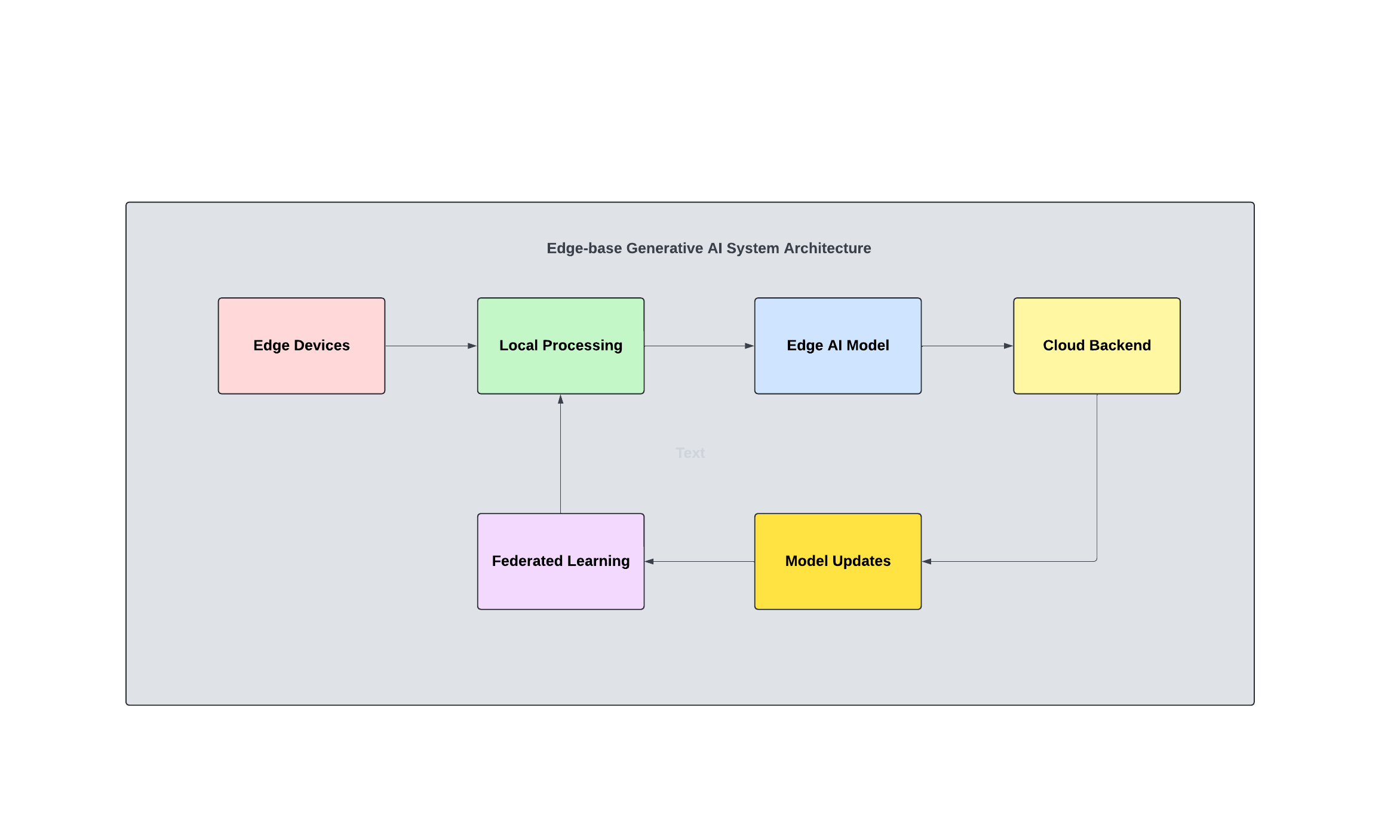}
    \caption{Architecture of a typical edge-based generative AI system}
    \label{fig:arch3_Figure7}
\end{figure}

As depicted in \textbf{Figure 7}, edge-based generative AI systems
involve a complex interplay between edge devices, local processing, edge
AI models, and cloud backends. The architecture emphasizes local
inference while leveraging cloud resources for model updates and
federated learning.

\subsubsection{ Edge-Cloud Collaboration for Generative AI}

The evolving synergy between edge computing and cloud resources has
spawned innovative approaches to generative AI deployment, fundamentally
transforming how these systems operate across distributed architectures.
At the forefront of this evolution is hybrid inference technology, which
employs sophisticated decision-making mechanisms to dynamically
determine whether to execute inference tasks at the edge or in the
cloud, based on real-time analysis of input complexity and network
conditions {[}125{]}. This adaptive approach is complemented by advances
in incremental learning at the edge, where systems continuously update
edge-deployed generative models with new data while maintaining
synchronization with their cloud-based counterparts through periodic
updates {[}85{]}. The collaboration between edge and cloud resources is
further enhanced by edge-triggered cloud generation capabilities, where
edge devices leverage local inference results to intelligently initiate
more complex generative tasks in the cloud {[}133, 134{]}.This
orchestrated interaction between edge and cloud resources represents a
significant advancement in distributed AI architectures, enabling
organizations to optimize performance, resource utilization, and
response times while maintaining the flexibility to handle varying
computational demands and network conditions.

\subsubsection{ Future Trends in Edge Computing for Generative AI}

Future trends in edge computing for generative AI include neuromorphic
computing, which mimics neural structures to enhance processing
efficiency and reduce power consumption, vital for real-time
applications{[}129, 135{]}. The expansion of 5G networks will support
responsive functionalities like augmented reality and autonomous systems
by providing higher bandwidth and lower latency{[}136{]}. Autonomous
edge AI systems with self-optimizing capabilities are expected to adapt
dynamically to real-world conditions without human intervention,
benefiting diverse deployments {[}137{]}. Edge-native generative models,
designed for resource-constrained environments, will optimize
performance, surpassing compressed cloud-based models{[}138{]}. As edge
computing evolves, it will continue to revolutionize generative AI
through innovations in latency, privacy, and offline capabilities.

\subsection{ Storage Solutions, Data Lakes, and Warehousing for Generative AI}

The exponential growth of generative AI models and their associated
datasets has placed unprecedented demands on storage infrastructure and
data management systems. Cloud providers have responded with innovative
solutions that cater to the unique requirements of generative AI
workloads. This section explores the latest advancements in cloud
storage technologies, data lakes, and warehousing solutions, and their
implications for generative AI development and deployment.

\subsubsection{ Evolution of Cloud Storage and Data Management for Generative AI}

Cloud storage and data management solutions have undergone substantial
evolution to address the unique demands of generative AI workloads,
introducing multiple innovative capabilities across the storage stack.
At the foundation of these advances lies next-generation
high-performance object storage systems, specifically optimized for AI
workloads, delivering unprecedented throughput and low-latency access to
large-scale datasets {[}139{]} . Complementing these developments,
AI-optimized file systems have emerged, featuring distributed
architectures specifically engineered for AI workloads, offering
exceptional Input/Output Operations Per Second (IOPS) performance and
robust support for parallel data access patterns {[}140{]}. The
integration of intelligent data tiering represents another significant
advancement, where sophisticated mechanisms automatically orchestrate
data movement between storage tiers based on access patterns and
specific AI workload requirements, optimizing both performance and cost
efficiency. Cloud providers have also revolutionized data lake
architectures, introducing serverless solutions that seamlessly scale to
accommodate varying workloads while providing native integration with AI
and machine learning capabilities {[}141{]}. Further enhancing this
ecosystem, traditional data warehouses have been transformed through AI
augmentation, incorporating advanced capabilities that significantly
improve query optimization and data processing efficiency {[}142{]}.This
comprehensive evolution in cloud storage and data management
technologies has created a robust foundation for supporting the
intensive data requirements of modern generative AI applications.

\begin{longtable}{|p{0.13\linewidth}|p{0.13\linewidth}|p{0.1\linewidth}|p{0.10\linewidth}|p{0.15\linewidth}|p{0.15\linewidth}|p{0.15\linewidth}|}
\hline
\textbf{Cloud Provider} & \textbf{Storage Service} & \textbf{Type} & \textbf{Max Throughput} & \textbf{AI-Specific Features} & \textbf{Data Lake Solution} & \textbf{Data Warehouse Solution} \\
\hline
\endfirsthead
\hline
\textbf{Cloud Provider} & \textbf{Storage Service} & \textbf{Type} & \textbf{Max Throughput} & \textbf{AI-Specific Features} & \textbf{Data Lake Solution} & \textbf{Data Warehouse Solution} \\
\hline
\endhead
\hline
AWS & Amazon S3 & Object Storage & 100 Gbps & Intelligent-Tiering for AI & AWS Lake Formation & Amazon Redshift ML \\
\hline
Azure & Azure Blob Storage & Object Storage & 50 Gbps & Data Lake Storage Gen2 & Azure Synapse Analytics & Azure Synapse Analytics \\
\hline
Google Cloud & Cloud Storage & Object Storage & 240 Gbps* & Dual region buckets for AI & Big Lake & Big Query ML \\
\hline
IBM Cloud & Cloud Object Storage & Object Storage & 120 Gbps & Smart Tier for AI workloads & Cloud Pak for Data & Db2 Warehouse \\
\hline
Oracle Cloud & Object Storage & Object Storage & 160 Gbps & Auto-tiering for AI datasets & Oracle Autonomous Database & Oracle Autonomous Data Warehouse \\
\hline
Snowflake & Snowflake Storage & Object Storage & Varies & Optimized for ML workloads & Snowflake Data Lake & Snowflake Data Warehouse \\
\hline

\end{longtable}

\textbf{Table 6:} landscape of cloud storage and data management
solutions for AI.

As shown in \textbf{Table 6}, major cloud providers offer a range of
storage, data lake, and warehousing solutions optimized for AI
workloads. These solutions are crucial for efficiently handling the
massive datasets required for training and deploying generative AI
models, as well as for managing the entire data lifecycle in AI
projects.

\subsubsection{ Object Storage and File Systems for Generative AI Datasets}

Object storage has emerged as the cornerstone for storing and managing
the massive datasets required for generative AI, while specialized file
systems have evolved to address the demanding requirements of
high-performance training workloads. Cloud providers have made
significant strides in developing object storage solutions capable of
handling exabyte-scale datasets, a critical capability for training
large language models and other computationally intensive generative AI
applications {[}143{]}. These systems are enhanced by sophisticated
content-aware compression techniques that intelligently analyze and
compress AI datasets based on their structural characteristics,
achieving substantial storage cost reductions while maintaining seamless
data accessibility {[}144{]}. The evolution of cloud infrastructure has
also brought about advanced implementations of POSIX-compliant parallel
file systems, including Lustre and GPFS, which provide essential
compatibility for seamless integration with existing AI frameworks
{[}140{]}. Perhaps most significantly, the introduction of GPU-Direct
Storage has revolutionized data access patterns in AI training pipelines
by enabling direct data transfer between storage systems and GPU memory,
effectively eliminating traditional data transfer bottlenecks and
substantially improving training efficiency {[}145{]}. This
comprehensive approach to storage optimization has established a robust
foundation for managing the unprecedented scale of data required for
modern generative AI applications while ensuring high performance and
cost efficiency.

\subsubsection{ Data Lakes for Generative AI Development}

Data lakes have emerged as essential infrastructure components for
managing the diverse and voluminous data requirements inherent in
generative AI development, introducing several innovative capabilities
that address the unique challenges of AI workloads. Cloud providers have
developed sophisticated serverless data lake solutions that
automatically scale to accommodate both growing datasets and fluctuating
query loads, providing the flexibility and dynamism essential for
generative AI projects{[}141, 146{]}. A particularly significant
advancement in this domain is the implementation of schema-on-read
flexibility, where data lake architectures support more adaptable
approaches to data ingestion and experimentation, enabling AI
researchers and developers to work with diverse data formats and
structures more efficiently {[}147{]}. This flexibility is further
enhanced by the integration of AI-powered data cataloging and discovery
tools, which significantly improve the accessibility and usability of
large-scale datasets for generative AI training {[}148{]} .These
intelligent cataloging systems not only facilitate better data
organization but also enable more effective discovery and utilization of
relevant training data, ultimately accelerating the development and
refinement of generative AI models while maintaining data quality and
accessibility at scale.

\begin{figure}[H]
    \centering
    \includegraphics[width=\textwidth]{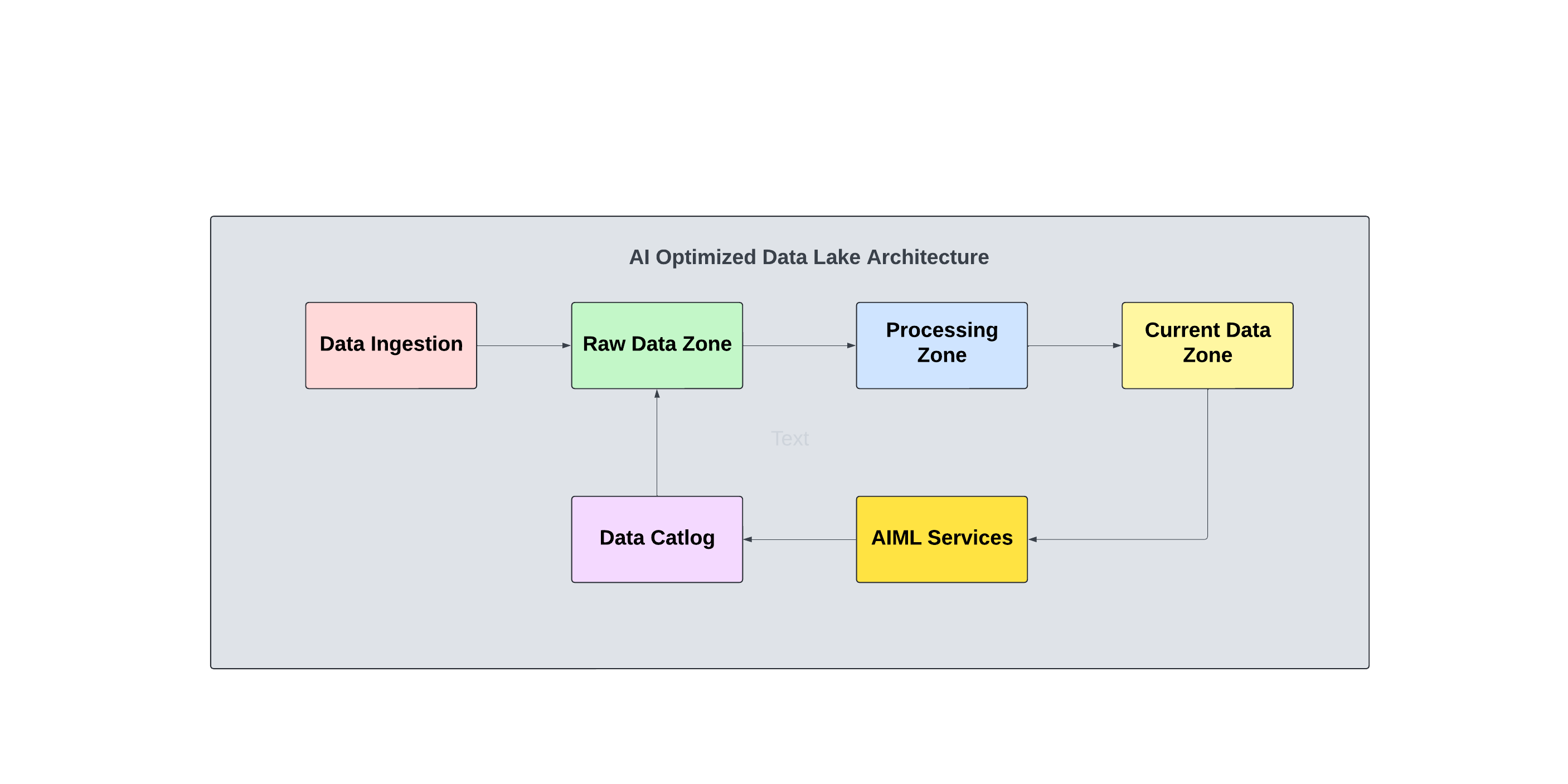}
    \caption{Architecture of a modern data lake optimized for generative AI workloads}
    \label{fig:arch4_Figure8}
\end{figure}

As depicted in \textbf{Figure 8}, AI-optimized data lake architectures
involve multiple zones for data processing and storage, integrated with
data catalogs and AI/ML services to facilitate efficient generative AI
development.

\subsubsection{ Data Warehousing for Generative AI Analytics}

While data lakes excel in providing flexibility for diverse dataset
storage and processing, data warehouses have emerged as crucial
components for structured analytics in generative AI projects, offering
several advanced capabilities. Modern data warehouses now leverage
sophisticated AI techniques for query optimization, significantly
enhancing performance for complex analytics on large-scale generative AI
datasets {[}149{]}. These systems have evolved to support in-database
machine learning capabilities, enabling seamless integration of AI
models with structured data and streamlining the analytical workflow
{[}150{]}. Furthermore, contemporary data warehouses provide robust
real-time analytics capabilities for streaming data, which proves
essential for continuous monitoring and optimization of generative AI
models in production environments {[}151{]}. This combination of
advanced features enables organizations to maintain high-performance
analytical capabilities while effectively managing the structured
aspects of their generative AI initiatives.

\subsubsection{ Challenges and Considerations}

Despite the numerous benefits offered by cloud storage, data lake, and
warehousing solutions for generative AI, organizations face several
significant challenges in implementing and managing these systems
effectively. Data gravity has emerged as a critical concern, as the
immense size of datasets used in generative AI creates substantial
inertia, making it increasingly difficult to move or replicate data
across different regions or cloud providers {[}152, 153{]}. Cost
management presents another significant challenge, as organizations must
carefully balance the requirements for high-performance storage and
analytics capabilities against budget constraints, particularly in
large-scale generative AI projects {[}6{]}. The complexity of data
privacy and compliance requirements adds another layer of challenge,
requiring organizations to ensure adherence to data protection
regulations while maintaining the accessibility and usability of
datasets for AI training and analytics{[}154{]}. Moreover, maintaining
data quality and implementing effective governance practices poses
ongoing challenges, particularly given the diverse and rapidly evolving
nature of datasets used in generative AI development{[}155, 156{]}.
These challenges underscore the need for careful planning and strategic
approaches to data management in generative AI implementations.

\subsubsection{ Future Trends in Cloud Data Management for Generative AI}

Future trends in cloud data management for generative AI include
AI-driven storage and query optimization, which leverage artificial
intelligence to predict access patterns, optimize data placement, and
improve query performance, crucial for handling large dataset {[}157,
158{]}. Additionally, quantum data processing holds the potential to
revolutionize data handling within generative AI workflows. By
integrating quantum computing technologies, cloud platforms may soon be
able to tackle complex data processing tasks that are beyond the
capacity of classical computing, thereby significantly accelerating
generative AI model development and deployment {[}159{]}.

Furthermore, federated data lakes represent a promising architectural
shift, enabling collaborative AI development across distributed
environments while maintaining data sovereignty and privacy. This model
supports the increasingly globalized nature of AI development, where
data-sharing across regions must adhere to diverse regulatory standards,
facilitating compliance and security in cross-border data collaborations
{[}160{]}. Alongside these technological advancements, there is a
growing emphasis on ethical AI data management, which involves
developing practices and tools specifically geared toward ethical AI.
This includes bias detection and mitigation within training datasets, a
necessary step to ensure that generative AI models produce fair and
unbiased outputs {[}161{]}. Together, these innovations in cloud-based
data management will drive the next generation of generative AI
advancements, addressing the growing scale and complexity of AI
workloads.

\subsection{ Generative AI Development Ecosystems}

The rapid evolution of generative AI has led to the emergence of
sophisticated development ecosystems within cloud platforms. These
ecosystems integrate tools, services, and frameworks to streamline the
lifecycle of generative AI models, encompassing data preparation,
training, deployment, and monitoring.

\textbf{Figure S1} provides a comprehensive visualization of the AI and
machine learning ecosystems offered by major cloud platforms, including
Microsoft Azure, AWS, Google Cloud, IBM Cloud, and Oracle Cloud. These
platforms offer distinct strengths, from seamless AI infrastructure to
responsible AI practices.

\textbf{See Supplementary Material Figure S1 and Table S2 for a detailed
comparison of integrated generative AI platforms across major cloud
providers.}

\subsubsection{ Integrated Generative AI Platforms}

Major cloud providers have created end-to-end platforms tailored for
generative AI development. These platforms feature pre-trained models,
prompt engineering tools, and advanced deployment options, empowering
developers to build, fine-tune, and deploy generative AI models
seamlessly.

\textbf{See Supplementary Material Table S2 for a detailed comparison of
integrated generative AI platforms across major cloud providers, and it}
is highlighting their key features, pre-trained models, and development
tools. These platforms offer comprehensive environments for developing,
training, and deploying generative AI models.

\subsubsection{ Advanced Development Tools for Generative AI}

Cloud providers have introduced a sophisticated array of specialized
tools designed to address the unique challenges inherent in generative
AI development. Among these innovations, prompt engineering interfaces
represent a significant advancement, offering intuitive visual
environments for crafting and refining prompts, which play a crucial
role in optimizing generative AI model outputs {[}162-165{]}.
Complementing these interfaces are streamlined fine-tuning pipelines
that enable developers to efficiently adapt pre-trained models to
specific domains or tasks, significantly reducing the time and
complexity involved in model customization {[}166{]}. The development
ecosystem is further enhanced by comprehensive model evaluation
frameworks, which provide robust tools for assessing the quality,
safety, and performance of generative AI models throughout their
development lifecycle {[}3{]}. Together, these advanced development
tools create an integrated environment that significantly streamlines
the generative AI development process, enabling organizations to more
effectively create, customize, and validate their AI models while
maintaining high standards of quality and safety.

\begin{longtable}{|p{0.15\linewidth}|p{0.16\linewidth}|p{0.16\linewidth}|p{0.15\linewidth}|p{0.15\linewidth}|p{0.15\linewidth}|}
\hline
\textbf{Tool Category} & \textbf{Azure} & \textbf{Google Cloud} & \textbf{AWS} & \textbf{IBM} & \textbf{Oracle} \\
\hline
\endfirsthead
\hline
\textbf{Tool Category} & \textbf{Azure} & \textbf{Google Cloud} & \textbf{AWS} & \textbf{IBM} & \textbf{Oracle} \\
\hline
\endhead
\hline
Prompt Engineering Interfaces & Azure AI Studio's Prompt Flow & Vertex AI's Prompt Design & Amazon SageMaker Canvas & watsonx.ai Prompt Lab & OCI Language AI Playground \\
\hline
Fine-tuning Pipelines & Azure Machine Learning's AutoML for NLP & Vertex AI's Model Garden & Amazon SageMaker Automatic Model Tuning & watsonx.ai Tuning Studio & OCI Data Science's AutoML {[}167{]} \\
\hline
Model Evaluation Frameworks & Azure Machine Learning's Model Interpretability & Vertex AI's Model Evaluation & Amazon SageMaker Model Monitor & Watson OpenScale & OCI Data Science's Model Evaluation \\
\hline

\end{longtable}

\textbf{Table 7}: Advanced Development Tools for Generative AI by Major
Cloud Providers.

\subsubsection{ Model Repositories and Versioning}

Efficient management of generative AI models throughout their lifecycle
has become paramount for maintaining reproducibility and enabling
effective collaboration in modern AI development environments. Central
to this management approach are cloud-based centralized model
registries, which provide robust repositories for storing, versioning,
and sharing generative AI models across development teams and
organizations. These registries are enhanced by sophisticated model
lineage tracking capabilities, offering comprehensive tools that
maintain detailed records of a model\textquotesingle s entire
development history, including training data sources, hyperparameter
configurations, and evaluation metrics. The automation of model
versioning represents another crucial advancement in this domain, with
systems intelligently creating new versions based on significant
performance improvements or meaningful changes in model architecture or
behavior. This integrated approach to model management ensures
consistent versioning practices, facilitates collaboration among team
members, and maintains a clear audit trail of model development and
optimization efforts, ultimately supporting the reproducibility and
reliability of generative AI systems at scale.As shown in \textbf{Table
8}, centralized model registries and model versioning systems employed
by each cloud provider, emphasizing their importance for model
management, lineage tracking, and collaboration in AI development.

\begin{longtable}{|p{0.2\linewidth}|p{0.16\linewidth}|p{0.16\linewidth}|p{0.16\linewidth}|p{0.16\linewidth}|p{0.16\linewidth}|}
\hline
\textbf{Feature} & \textbf{Azure} & \textbf{Google Cloud} & \textbf{AWS} & \textbf{IBM} & \textbf{Oracle} \\
\hline
\endfirsthead
\hline
\textbf{Feature} & \textbf{Azure} & \textbf{Google Cloud} & \textbf{AWS} & \textbf{IBM} & \textbf{Oracle} \\
\hline
\endhead
Centralized Model Registries & Azure Machine Learning Model Registry {[}168{]} & Vertex AI Model Registry {[}169{]} & Amazon SageMaker Model Registry {[}170{]} & Watson Machine Learning Model Management & OCI Model Catalog {[}171{]} \\
\hline
Model Lineage Tracking & Azure Machine Learning MLflow Tracking {[}172{]} & Vertex AI Metadata {[}173{]} & Amazon SageMaker Lineage Tracking {[}174{]} & Watson Studio Experiments {[}175{]} & OCI Data Science Model Provenance {[}176{]} \\
\hline
Automated Model Versioning & Azure DevOps for ML {[}172{]} & Vertex AI Pipelines {[}177{]} & Amazon SageMaker Projects {[}178{]} & Watson Machine Learning Versioning & OCI DevOps for ML {[}179{]} \\
\hline
\caption{Comparison of Centralized Model Registries and Versioning Tools Across Major Cloud Providers}
\label{tab:model_registries_comparison_duplicate}
\end{longtable}

\textbf{Table 8:} Comparison of Centralized Model Registries and
Versioning Tools Across Major Cloud Providers

\begin{figure}[H]
    \centering
    \includegraphics[width=\textwidth]{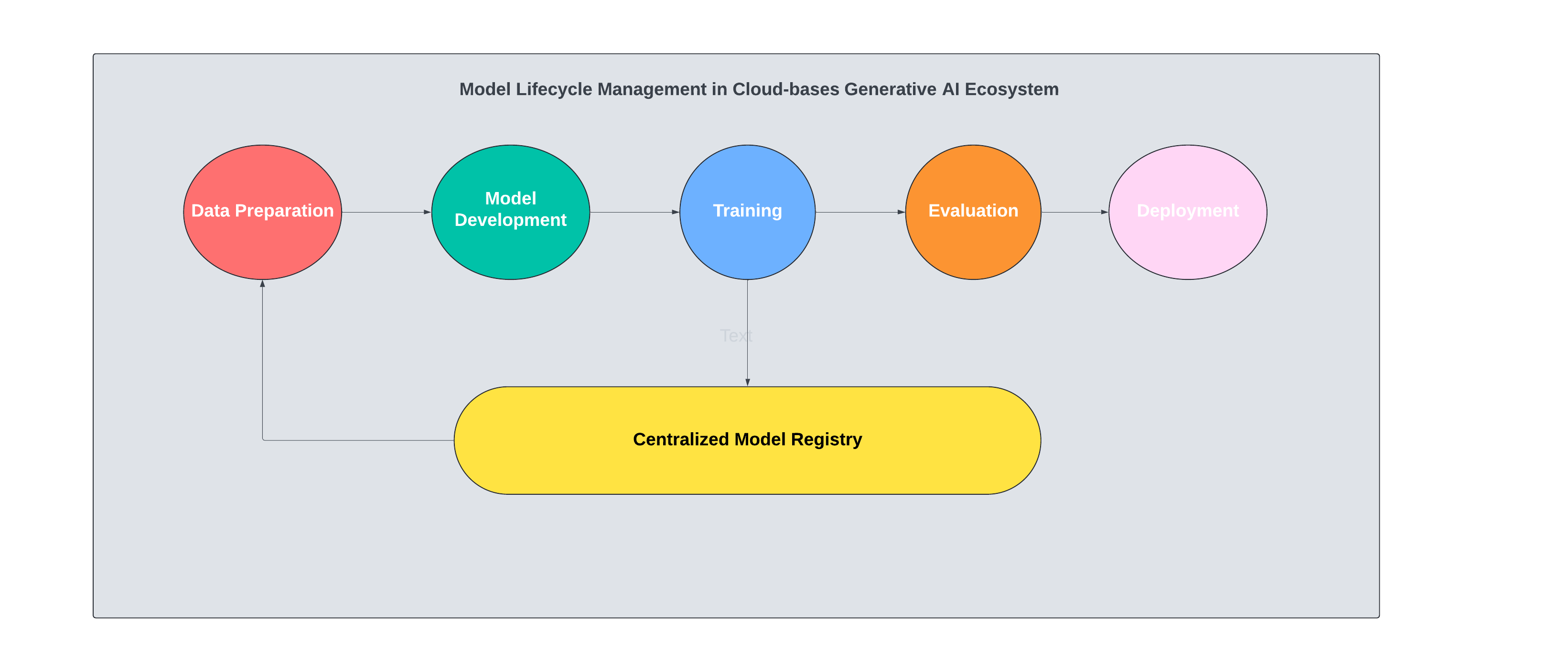}
    \caption{Model lifecycle management process}
    \label{fig:arch5_Figure9}
\end{figure}

\textbf{In Figure 9} shows the model lifecycle management process in a
cloud-based generative AI ecosystem, highlighting the central role of
the model registry in facilitating collaboration and versioning
throughout the development process.

\subsubsection{ Deployment and Inference Optimization}

Cloud providers have developed a comprehensive suite of solutions for
optimizing the deployment and inference of generative AI models,
addressing key challenges in performance, scalability, and efficiency.
At the forefront of these innovations is serverless inference
technology, which offers scalable, pay-per-use inference endpoints
capable of automatically adjusting to fluctuating workloads, enabling
organizations to optimize resource utilization while maintaining
consistent performance {[}7{]}. This flexibility is complemented by
sophisticated model compression techniques, including advanced tools for
quantization and pruning, which effectively reduce model size while
preserving inference accuracy and improving execution speed {[}8{]}.
Further enhancing these capabilities are hardware-specific optimizations
that automatically tune model performance for various AI accelerators,
including GPUs, TPUs, and FPGAs, ensuring optimal utilization of
specialized computing resources {[}9{]}. This multi-faceted approach to
deployment and inference optimization enables organizations to achieve
superior performance and cost-efficiency in their generative AI
implementations while maintaining the flexibility to adapt to evolving
computational demands.

In \textbf{Table 9} focuses on the monitoring and observability
solutions provided by these platforms, detailing the AI-specific
metrics, drift detection, and explainability tools that help ensure the
quality and reliability of generative AI models in production
environments.

\begin{longtable}{|p{0.16\linewidth}|p{0.16\linewidth}|p{0.16\linewidth}|p{0.16\linewidth}|p{0.16\linewidth}|p{0.16\linewidth}|}
\hline
\textbf{Feature} & \textbf{Azure} & \textbf{Google Cloud} & \textbf{AWS} & \textbf{IBM} & \textbf{Oracle} \\
\hline
\endfirsthead
\hline
\textbf{Feature} & \textbf{Azure} & \textbf{Google Cloud} & \textbf{AWS} & \textbf{IBM} & \textbf{Oracle} \\
\hline
\endhead
Centralized Model Registries & Azure Machine Learning Model Registry {[}168{]} & Vertex AI Model Registry {[}169{]} & Amazon SageMaker Model Registry {[}170{]} & Watson Machine Learning Model Management & OCI Model Catalog {[}171{]} \\
\hline
Model Lineage Tracking & Azure Machine Learning MLflow Tracking {[}172{]} & Vertex AI Metadata {[}173{]} & Amazon SageMaker Lineage Tracking {[}174{]} & Watson Studio Experiments {[}175{]} & OCI Data Science Model Provenance {[}176{]} \\
\hline
Automated Model Versioning & Azure DevOps for ML {[}172{]} & Vertex AI Pipelines {[}177{]} & Amazon SageMaker Projects {[}178{]} & Watson Machine Learning Versioning & OCI DevOps for ML {[}179{]} \\
\hline
\caption{Comparison of Centralized Model Registries and Versioning Tools Across Major Cloud Providers}
\label{tab:model_registries_comparison}
\end{longtable}

\textbf{Table 9:} Comparison of Monitoring and Observability Solutions
for Generative AI Across Major Cloud Providers

\subsubsection{ Deployment and Inference Optimization}

Effective monitoring and observability are essential for maintaining the
performance and reliability of generative AI models in production
environments. As shown in \textbf{Table 10}, cloud platforms offer
comprehensive monitoring solutions that include AI-specific metrics,
drift detection, and explainability tools, each designed to ensure model
quality and reliability. AI-specific metrics provide a focused approach
to tracking the quality and performance of generative AI outputs,
enabling teams to monitor model effectiveness. Drift detection tools are
invaluable for identifying data or concept drift in deployed models,
promptly alerting stakeholders to potential issues that may compromise
model accuracy {[}11{]}. Additionally, explainability dashboards
facilitate a deeper understanding of model decisions and outputs by
offering visual interfaces that demystify the underlying processes,
thereby enhancing transparency and trust in AI applications {[}12{]}.

\begin{longtable}{|p{0.18\linewidth}|p{0.18\linewidth}|p{0.18\linewidth}|p{0.18\linewidth}|p{0.18\linewidth}|p{0.1\linewidth}|}
\hline
\textbf{Feature} & \textbf{Azure} & \textbf{Google Cloud} & \textbf{AWS} & \textbf{IBM} & \textbf{Oracle} \\
\hline
\endfirsthead
\hline
\textbf{Feature} & \textbf{Azure} & \textbf{Google Cloud} & \textbf{AWS} & \textbf{IBM} & \textbf{Oracle} \\
\hline
\endhead
AI-Specific Metrics & Azure Monitor for AI {[}191{]} & Cloud Monitoring for AI {[}192{]} & Amazon SageMaker Model Monitor {[}193{]} & Watson OpenScale Monitoring {[}194{]} & OCI Monitoring for AI {[}176{]} \\
\hline
Drift Detection & Azure Machine Learning Data Drift Monitoring {[}191{]} & Vertex AI Continuous Monitoring {[}195{]} & Amazon SageMaker Model Monitor Drift Detection {[}196{]} & Watson OpenScale Drift Detection {[}197{]} & OCI Data Science Drift Detection \\
\hline
Explainability Dashboards & Azure Machine Learning Interpretability Dashboard {[}198{]} & Vertex Explainable AI {[}199{]} & Amazon SageMaker Clarify {[}200{]} & Watson OpenScale Explainability {[}201{]} & OCI Data Science Explainable AI {[}202{]} \\
\hline
\end{longtable}

\textbf{Table 10: Comparison of Monitoring and Observability Solutions
for Generative AI Across Major Cloud Providers}

\subsubsection{ Monitoring and Observability}

In exploring the challenges and future directions of cloud-based
generative AI ecosystems, several key areas emerge. One primary
challenge is interoperability, which involves ensuring that models and
workflows can seamlessly integrate and remain portable across various
cloud platforms. This capability is essential for maintaining
flexibility and avoiding vendor lock-in in increasingly multi-cloud
environments. Another area of focus is ethical AI development, where
there is a growing need to incorporate tools and frameworks that
facilitate bias detection, fairness assessments, and overall responsible
AI practices. These efforts are crucial in addressing the ethical
concerns surrounding AI, especially as it becomes more embedded in
diverse applications and sectors.

Additionally, the integration of edge AI into cloud-based ecosystems
presents both a challenge and an opportunity. Extending these generative
AI development platforms to support edge deployment and federated
learning scenarios could enable real-time processing closer to data
sources, making AI more responsive and enhancing data privacy. The
convergence of quantum computing and AI, often termed Quantum-AI, is
also anticipated to play a transformative role in future generative AI
workflows. Preparing for this integration will require advancements in
quantum computing capabilities and adaptations in AI frameworks to
leverage quantum technologies effectively.

In summary, cloud-based generative AI development ecosystems are
advancing rapidly to address the sophisticated demands of AI
practitioners. These platforms now offer a holistic approach to the
generative AI lifecycle, encompassing data preparation, model
deployment, and continuous monitoring. As the field of generative AI
progresses, these ecosystems are expected to become instrumental in
driving innovation and democratizing access to powerful AI tools.

\subsubsection{ Challenges and Future Directions}

The integration of generative AI capabilities into existing applications
and workflows has become a cornerstone of enterprise AI adoption. Cloud
providers have responded by offering a wide range of API-accessible AI
services that enable developers to seamlessly incorporate generative AI
functionalities into their applications. This section explores the
state-of-the-art in API-accessible AI services, their importance in the
generative AI ecosystem, and the offerings from major cloud providers.

\subsection{ API-Accessible AI Services for Application Integration}

API-accessible AI services have emerged as a cornerstone in
democratizing access to advanced generative AI capabilities, offering a
comprehensive suite of advantages that facilitate broader adoption and
implementation. These services fundamentally transform the development
landscape by enabling rapid prototyping capabilities, allowing
developers to swiftly integrate sophisticated AI functionalities without
requiring extensive machine learning expertise, thereby accelerating
innovation cycles across industries. The inherent scalability of
cloud-based APIs provides a crucial technical advantage, seamlessly
accommodating workloads that range from small-scale experimental
implementations to full production-level deployments, ensuring
organizations can grow their AI capabilities without infrastructure
constraints. From a financial perspective, the cost-effectiveness of
pay-as-you-go models has dramatically lowered the barrier to entry,
enabling organizations to access cutting-edge AI technologies without
substantial upfront investments, effectively democratizing AI adoption
across different organizational scales. The continuous update mechanism
of API services ensures that organizations consistently have access to
state-of-the-art capabilities and models without the burden of manual
upgrades, keeping them at the forefront of technological advancement.
Furthermore, the cross-platform compatibility inherent in API
architectures enables AI integration across diverse platforms and
programming languages, fostering widespread adoption and providing
organizations with the flexibility to implement AI solutions within
their existing technical ecosystems.

\subsubsection{ Importance of API-Accessible AI Services}

Cloud providers offer a diverse range of API-accessible generative AI
services, catering to various use cases. \textbf{Table 11} provides a
comprehensive comparison of these services across major cloud providers.

% Corrected table structure with proper width specifications
\begin{longtable}{|p{0.15\linewidth}|p{0.15\linewidth}|p{0.18\linewidth}|p{0.19\linewidth}|p{0.15\linewidth}|p{0.18\linewidth}|}
\hline
\textbf{Service Type} & \textbf{Azure} & \textbf{Google Cloud} & \textbf{AWS} & \textbf{IBM} & \textbf{Oracle} \\
\hline
\endfirsthead
\hline
\textbf{Service Type} & \textbf{Azure} & \textbf{Google Cloud} & \textbf{AWS} & \textbf{IBM} & \textbf{Oracle} \\
\hline
\endhead
Text Generation & Azure OpenAI Service (GPT-4) {[}96{]} & Vertex AI (PaLM 2) {[}203{]} & Amazon Bedrock (Claude 2, Jurassic-2) {[}204{]} & Watson NLP {[}205{]} & OCI Language AI {[}206{]} \\
\hline
Image Generation & Azure OpenAI Service (DALL-E 3) {[}96{]} & Imagen API {[}207{]} & Amazon Bedrock (Stable Diffusion) {[}204{]} & Watson Media Generator {[}208{]} & OCI Vision AI {[}209{]} \\
\hline
Code Generation & GitHub Copilot {[}210{]} & Vertex AI Code Generation {[}211{]} & Amazon CodeWhisperer {[}212{]} & Watson Code Assistant {[}213{]} & OCI DevOps AI {[}179{]} \\
\hline
Speech Synthesis & Azure Neural TTS {[}214{]} & Cloud Text-to-Speech {[}215{]} & Amazon Polly {[}216{]} & Watson Text to Speech {[}217{]} & OCI Speech AI {[}218{]} \\
\hline
Multimodal AI & Azure Cognitive Services {[}62{]} & Vertex AI Multimodal {[}219{]} & Amazon Bedrock (Titan) {[}204{]} & Watson Multimodal AI & OCI Multimodal AI {[}167{]} \\
\hline
Conversational AI & Azure Bot Service {[}220{]} & Dialogflow CX {[}221{]} & Amazon Lex {[}222{]} & Watson Assistant {[}223{]} & OCI Digital Assistant {[}224{]} \\
\hline
Language Translation & Azure Translator {[}225{]} & Cloud Translation AI {[}226{]} & Amazon Translate {[}227{]} & Watson Language Translator & OCI Translation AI {[}228{]} \\
\hline
Sentiment Analysis & Azure Text Analytics {[}62{]} & Cloud Natural Language API {[}229{]} & Amazon Comprehend {[}53{]} & Watson NLU & OCI Language AI {[}206{]} \\
\hline
\end{longtable}

\textbf{Table 11}: Comparison of Generative AI Services Across Major
Cloud Providers

As shown in \textbf{Table 11}, the landscape of API-accessible
generative AI services is rich and diverse, with each cloud provider
offering a comprehensive suite of tools. The recent advancements in
large language models (LLMs) have significantly enhanced the
capabilities of these services, particularly in text generation and
multimodal AI {[}6{]}.

\subsubsection{ Types of API-Accessible Generative AI Services}

Integrating API-accessible AI services effectively into existing systems
necessitates a strategic approach to architecture and design patterns. A
common best practice is the adoption of microservices architecture,
where AI services operate as independent microservices, thereby
enhancing scalability and ease of maintenance {[}230, 231{]}. Utilizing
API gateways plays a crucial role in managing access, as they streamline
authentication, enforce rate limits, and monitor service interactions,
ensuring robust control over API usage {[}8{]}. Additionally, caching
strategies are essential to optimize performance, as they reduce the
frequency of API calls and enhance response times for commonly requested
generations. To further ensure resilience, robust error handling and
fallback mechanisms should be established, enabling the system to
recover gracefully from disruptions. Finally, asynchronous processing is
recommended for handling long-running AI tasks, improving the user
experience by maintaining system responsiveness and minimizing latency.

\subsubsection{ Integration Patterns and Best Practices}

When integrating API-accessible AI services, addressing performance and
scalability considerations is essential for optimal functionality. A
foundational approach involves conducting comprehensive load testing,
which helps identify and understand the performance characteristics of
AI services under varying conditions. To efficiently handle fluctuating
workloads, the implementation of auto-scaling mechanisms is recommended,
allowing for dynamic adjustments based on demand {[}13{]}. Additionally,
batching multiple requests, where feasible, minimizes network overhead
and enhances overall throughput, making the integration more efficient.
For frequently accessed AI-generated content, edge caching proves
beneficial, as it reduces latency and the volume of API calls, thus
improving response times and reducing server load.

\textbf{Figure 10} illustrates these performance considerations and best
practices, demonstrating a typical architecture for integrating
API-accessible AI services. The diagram highlights how client
applications interact with AI services via API gateways and load
balancers, with optimizations such as caching, rate limiting, and
request batching to improve scalability and efficiency.

\begin{figure}[H]
    \centering
    \includegraphics[width=\textwidth]{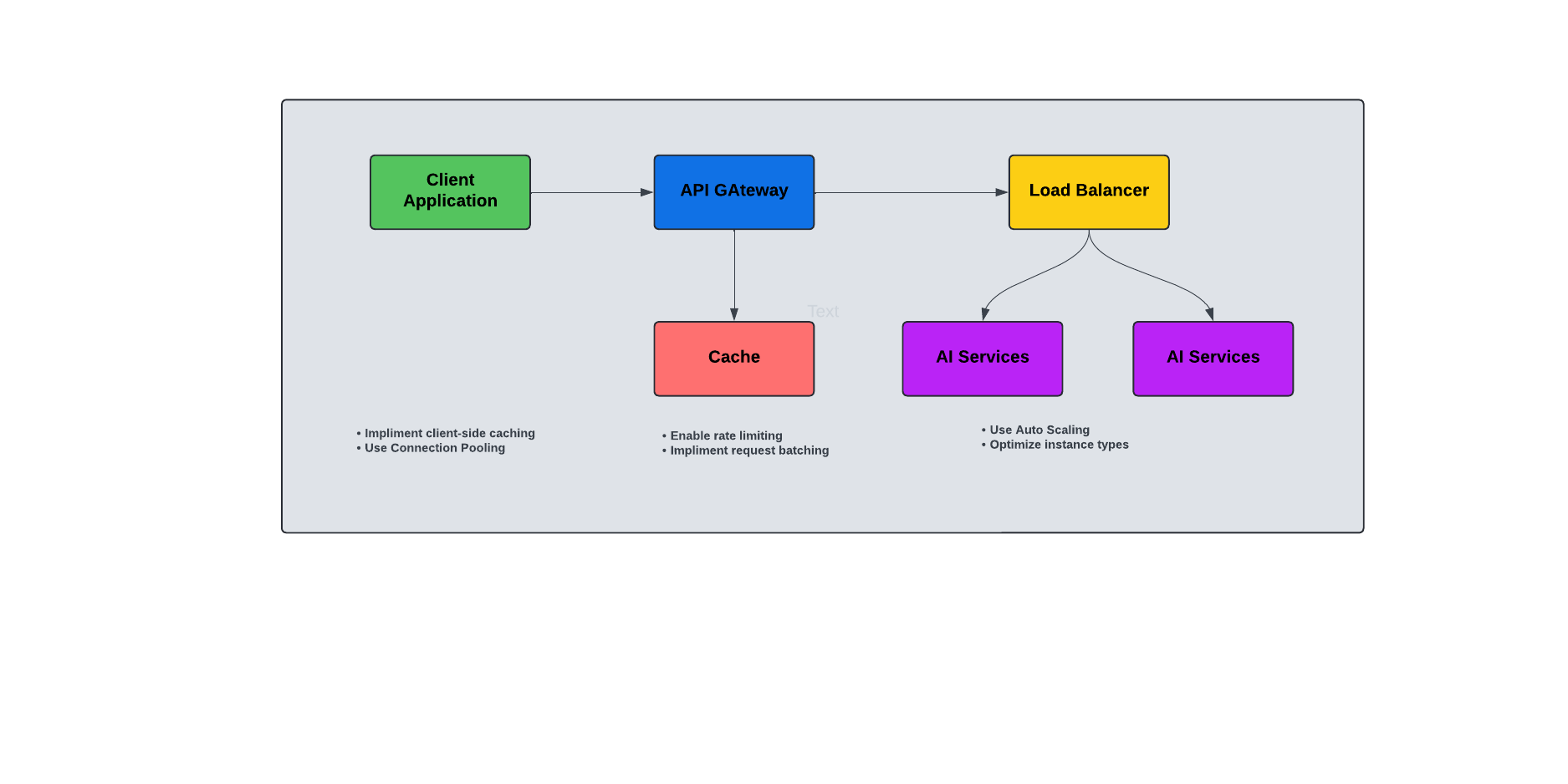}
    \caption{Performance Optimization and Scalability Best Practices for Integrating API-Accessible AI Services}
    \label{fig:arch6_Figure10}
\end{figure}

\subsubsection{ Performance and Scalability Considerations}

Ensuring security and compliance is paramount when integrating
API-accessible AI services. Data encryption must be maintained both in
transit and at rest to protect sensitive information throughout its
interaction with AI APIs {[}16{]}. Access control practices, including
fine-grained control and adherence to the principle of least privilege,
are essential to prevent unauthorized access and minimize security
risks. Comprehensive audit logging of API usage is critical for meeting
security and compliance standards, offering traceability and
accountability. Additionally, understanding data residency requirements
is crucial for choosing API endpoints that align with regional
regulatory demands. Finally, implementing model governance practices
supports responsible and ethical AI usage, reinforcing trust in AI
integrations and facilitating compliance with emerging regulations.

\subsubsection{ Security and Compliance Considerations}

Ensuring security and compliance is paramount when integrating API-accessible AI services. Data encryption must be maintained both in transit and at rest to protect sensitive information throughout its interaction with AI APIs [16]. Access control practices, including fine-grained control and adherence to the principle of least privilege, are essential to prevent unauthorized access and minimize security risks. Comprehensive audit logging of API usage is critical for meeting security and compliance standards, offering traceability and accountability. Additionally, understanding data residency requirements is crucial for choosing API endpoints that align with regional regulatory demands. Finally, implementing model governance practices supports responsible and ethical AI usage, reinforcing trust in AI integrations and facilitating compliance with emerging regulations.

\subsubsection{ Future Trends in API-Accessible AI Services}

Future trends in API-accessible AI services include fine-tuning APIs,
enabling users to customize pre-trained models for domain-specific
applications, enhancing relevance and efficiency. Federated learning
APIs are advancing decentralized model training across multiple data
sources while preserving data privacy, vital for sensitive sectors. Edge
AI APIs are emerging, combining edge computing and cloud-based AI for
low-latency, real-time inferencing. AI-powered API generation is also
anticipated, where AI automates and optimizes API integrations from
natural language inputs, simplifying development. Additionally,
quantum-enhanced AI APIs are expected to integrate quantum computing,
offering performance gains for complex AI algorithms. These innovations
will democratize generative AI capabilities, driving industry-wide
advancements and accessibility.

\section{ Comparative Analysis}

Performance is a critical factor in generative AI development. Recent
benchmarks conducted across various generative AI tasks show that
performance can vary significantly depending on the specific use case
and model size. For instance, Google Cloud\textquotesingle s TPU v4 has
shown superior performance for large-scale language model training,
while AWS\textquotesingle s EC2 P5 instances excel in computer vision
tasks {[}57{]}.

It\textquotesingle s important to note that performance can be heavily
influenced by factors such as model architecture, dataset
characteristics, and optimization techniques. Therefore, organizations
should conduct their own benchmarks tailored to their specific use cases
when evaluating cloud platforms {[}232, 233{]}.

\subsection{ Performance Analysis}

Security is a critical concern in generative AI development,
particularly when handling sensitive data or model outputs. Cloud
providers have made significant advancements in this area, offering a
range of security features to ensure data protection and model
integrity.

As shown in \textbf{Table 12}, all major cloud providers---Azure, AWS,
Google Cloud, IBM Cloud, and Oracle Cloud---offer comprehensive security
measures, including data encryption at rest and data encryption in
transit. These features are essential to maintain confidentiality and
prevent unauthorized access during the AI development lifecycle.

Moreover, each provider offers specialized AI model security tools, such
as Azure AI Shield, SageMaker Model Monitor, and Watson Open Scale,
which provide advanced monitoring and defense mechanisms against
adversarial attacks. Additionally, compliance certifications like GDPR,
HIPAA, and ISO 27001 ensure that these platforms adhere to strict
regulatory standards, safeguarding AI models and data

\begin{longtable}{|p{0.2\linewidth}|p{0.15\linewidth}|p{0.15\linewidth}|p{0.15\linewidth}|p{0.15\linewidth}|p{0.2\linewidth}|}
\hline
\textbf{Security Feature} & \textbf{Azure} & \textbf{AWS} & \textbf{Google Cloud} & \textbf{IBM Cloud} & \textbf{Oracle Cloud} \\
\hline
\endfirsthead
\hline
\textbf{Security Feature} & \textbf{Azure} & \textbf{AWS} & \textbf{Google Cloud} & \textbf{IBM Cloud} & \textbf{Oracle Cloud} \\
\hline
\endhead
Data Encryption at Rest & \checkmark & \checkmark & \checkmark & \checkmark & \checkmark \\
\hline
Data Encryption in Transit & \checkmark & \checkmark & \checkmark & \checkmark & \checkmark \\
\hline
AI Model Security & Azure AI Shield {[}234{]} & SageMaker Model Monitor {[}193{]} & Vertex AI Model Monitoring {[}195{]} & Watson OpenScale {[}194{]} & OCI AI Model Security {[}235{]} \\
\hline
Privacy-Preserving ML & Confidential Computing {[}236{]} & AWS Nitro Enclaves {[}237{]} & Confidential Computing {[}238{]} & IBM Fully Homomorphic Encryption {[}239{]} & OCI Confidential Computing {[}235{]} \\
\hline
Compliance Certifications & GDPR, HIPAA, ISO 27001 & GDPR, HIPAA, ISO 27001 & GDPR, HIPAA, ISO 27001 & GDPR, HIPAA, ISO 27001 & GDPR, HIPAA, ISO 27001 \\
\hline

\end{longtable}

\textbf{Table 12:} Comparison of Security Features for Generative AI
(2024)

In the realm of security for AI and cloud services, major providers
present robust yet distinct offerings. Microsoft Azure's AI Shield, for
instance, delivers sophisticated defenses against model theft and
adversarial attacks, providing a fortified layer of security for AI
deployments {[}234{]}. Amazon Web Services approaches security through
its Nitro Enclaves, which offer an innovative method for secure enclave
computing, effectively isolating sensitive workloads {[}237{]}.
Meanwhile, Google Cloud and IBM have made significant strides in
privacy-preserving machine learning, a field essential for protecting
data in AI processes, placing them at the forefront of privacy-conscious
AI solutions {[}238, 239{]}. These specialized security measures
underscore the varying strategies each provider employs to safeguard AI
models and data, addressing diverse aspects of security in cloud-based
AI ecosystems.

\subsection{ Security Aspects}

Deploying and orchestrating generative AI models at scale brings forth a
series of challenges that cloud providers address with a variety of
solutions. Automation plays a pivotal role, with AWS's Generative AI
Application Builder and Azure's Machine Learning platform designed to
streamline and automate routine deployment tasks {[}240,
241{]}.Additionally, flexibility in deployment options is essential, and
all major providers support managed cloud services, custom virtual
machines, and hybrid configurations. Google Cloud's Anthos particularly
excels in supporting hybrid and multi-cloud deployments, enhancing
cross-platform flexibility {[}242{]}. Orchestration is typically managed
through Kubernetes-based solutions, with Amazon EKS, Azure Kubernetes
Service, and Google Kubernetes Engine as popular choices for
coordinating complex generative AI workflows at scale {[}243, 244{]}.

\subsection{ Deployment and Orchestration Challenges}
Deploying and orchestrating generative AI models at scale brings forth a series of challenges that cloud providers address with a variety of solutions. Automation plays a pivotal role, with AWS’s Generative AI Application Builder and Azure’s Machine Learning platform designed to streamline and automate routine deployment tasks [220, 221].Additionally, flexibility in deployment options is essential, and all major providers support managed cloud services, custom virtual machines, and hybrid configurations. Google Cloud’s Anthos particularly excels in supporting hybrid and multi-cloud deployments, enhancing cross-platform flexibility [222]. Orchestration is typically managed through Kubernetes-based solutions, with Amazon EKS, Azure Kubernetes Service, and Google Kubernetes Engine as popular choices for coordinating complex generative AI workflows at scale [223, 224].

\subsection{ Addressing Data Bias and Lack of Explainability}

As generative AI models become more prevalent, addressing issues of bias
and explainability has become crucial. Cloud providers are developing
tools and strategies to tackle these challenges. As outlined in
\textbf{Table 13}, each major cloud provider offers dedicated tools for
bias detection and model explainability.

\begin{longtable}{|p{0.2\linewidth}|p{0.15\linewidth}|p{0.15\linewidth}|p{0.2\linewidth}|p{0.15\linewidth}|p{0.15\linewidth}|}
\hline
\textbf{Feature} & \textbf{Azure} & \textbf{AWS} & \textbf{Google Cloud} & \textbf{IBM Cloud} & \textbf{Oracle Cloud} \\
\hline
\endfirsthead
\hline
\textbf{Feature} & \textbf{Azure} & \textbf{AWS} & \textbf{Google Cloud} & \textbf{IBM Cloud} & \textbf{Oracle Cloud} \\
\hline
\endhead
Bias Detection & Azure Fairness {[}245{]} & Amazon SageMaker Clarify {[}61, 200{]} & Vertex AI Explainable AI {[}199{]} & AI Fairness 360 {[}246{]} & OCI AI Fairness {[}247{]} \\
\hline
Model Explainability & Interpretable Machine Learning {[}248, 249{]} & Amazon SageMaker Debugger {[}250{]} & Vertex Explainable AI {[}199{]} & AI Explainability 360 {[}201{]} & OCI Model Explanation \\
\hline
Responsible AI Dashboards & Azure Responsible AI Dashboard {[}198{]} & AWS AI Governance {[}251, 252{]} & Responsible AI Toolkit {[}253{]} & Watson OpenScale {[}254{]} & OCI Responsible AI {[}255{]} \\
\hline

\end{longtable}

\textbf{Table 13:} Tools for Addressing Bias and Explainability in
Generative AI

The latest trends in bias mitigation and explainability focus on
integrating detection mechanisms directly into the model development
pipelines, with major cloud providers now offering tools designed to
identify and address bias throughout the AI lifecycle. Additionally,
providers are implementing advanced explainability techniques,
incorporating methods like SHAP (Shapley Additive Explanations) and LIME
(Local Interpretable Model-agnostic Explanations) to provide deeper
insights into model behavior {[}256{]} .Furthermore, there is an
increased emphasis on developing responsible AI frameworks, which
holistically address bias detection, model transparency, and ethical
considerations to ensure trustworthiness and accountability in AI
systems.

\subsection{ Cost-Effectiveness Analysis}
Cost is a critical factor when selecting a cloud platform for generative
AI development. Although pricing models are complex and vary by specific
use case, certain general trends have emerged. AWS often provides
cost-effective solutions for large-scale deployments, especially when
utilizing reserved instances {[}257{]}. Google Cloud offers preemptible
VMs, which can lead to significant cost savings for fault-tolerant
workloads {[}258{]}. Meanwhile, Azure's spot instances and hybrid
benefits for Windows workloads can present cost advantages in specific
scenarios {[}259{]}. To make informed decisions, organizations should
conduct a thorough total cost of ownership (TCO) analysis based on their
unique usage patterns and requirements.

\section{ Challenges and Future Directions}

The increasing adoption of cloud services for generative AI application development offers numerous benefits, including scalability, flexibility, and cost-effectiveness. However, this reliance on cloud platforms also presents several technical and strategic challenges that enterprises must navigate to fully leverage the potential of generative AI.

\subsection{ Technical Challenges}
\subsubsection{ Integration Complexity}

Integrating cloud AI services with existing enterprise systems is
challenging due to interoperability issues across different systems,
applications, and data formats. This complexity is heightened by the
need for specialized skills to manage disparate systems. To address
these challenges, enterprises can adopt standardized APIs and data
formats (e.g., JSON, XML) to facilitate smoother integration, employ
middleware solutions like enterprise service buses (ESBs) to bridge
system gaps, and invest in skill development for managing integrations
effectively.

\subsubsection{ Data Management}

Managing large data volumes in cloud environments introduces issues
related to data latency and hybrid cloud complexities, particularly
impacting real-time AI applications. Data distributed across multiple
providers complicates governance and consistency. Mitigation strategies
include employing data caching, compression, and replication techniques
to reduce latency, using unified data platforms like Azure Arc and AWS
Outposts to manage hybrid environments, and establishing data governance
policies to ensure data quality, consistency, and compliance.

\subsubsection{ Security Concerns}

Security is a critical issue in cloud-based AI deployments, involving
protection of sensitive data, regulatory compliance, and intellectual
property security. Effective measures include encryption for data at
rest and in transit, multi-factor authentication, and regular security
audits to ensure compliance with standards such as ISO 27001, GDPR, and
HIPAA. Additionally, adopting DevSecOps methodologies can integrate
security into every stage of the AI development lifecycle.

\subsubsection{ Latency and Bandwidth Limitations}

Latency and bandwidth constraints can impact the performance of
cloud-based AI applications, especially those requiring real-time data
processing, like autonomous vehicles or financial trading systems.
Solutions to these challenges include deploying AI models closer to data
sources through edge computing, optimizing networks with content
delivery networks (CDNs) and high-speed solutions, and designing
adaptive algorithms that perform efficiently under varying network
conditions.

\subsection{ Strategic Challenges}

The deployment of AI and cloud solutions presents numerous strategic
challenges, encompassing industry-specific compliance, vendor
dependence, sustainability concerns, regulatory shifts, talent
acquisition, data integrity, model transparency, and security.

\subsubsection{ Compliance and Regulatory Challenges}

Different industries face unique compliance requirements that demand
tailored strategies. For example, healthcare organizations must comply
with HIPAA regulations, employing cloud services that offer data
encryption, access controls, and audit logging. Financial institutions
must meet PCI-DSS and SOX standards to protect sensitive financial data,
while retail and government entities adhere to PCI-DSS and FedRAMP,
respectively, to ensure secure handling of sensitive information.

Beyond industry-specific compliance, the evolving regulatory landscape
for AI introduces new compliance complexities. Organizations need to
stay informed on updates like GDPR, CCPA, and emerging AI-specific
regulations. Implementing a comprehensive compliance framework and
consulting with technology law experts help to address both current and
forthcoming regulatory demands.

\subsubsection{ Operational and Resource Challenges}

Vendor lock-in is a significant risk when organizations become overly
dependent on a single cloud provider, complicating transitions to
alternative services. Multi-cloud and hybrid strategies, open standards
like containerization, and robust contract negotiations with data
portability provisions can mitigate this risk.

Sustainability is another pressing concern, as the computational demands
of AI models can lead to environmental impacts, including greenhouse gas
emissions. Organizations can opt for cloud providers committed to
renewable energy, adopt model efficiency techniques (e.g., model
pruning, quantization), and incorporate sustainability goals into their
corporate policies, reporting on their environmental efforts.

Cybersecurity also presents operational challenges, with AI systems
vulnerable to cyber threats, including adversarial attacks. Ensuring
robust security protocols, training models to recognize adversarial
inputs, and maintaining secure cloud environments are essential steps in
safeguarding AI systems against potential risks.

\subsubsection{ Human and Data-Centric Challenges}

Acquiring and retaining skilled professionals in AI and cloud
technologies is increasingly difficult due to high demand. Organizations
can address this by investing in training and certification programs,
developing talent pipelines through educational partnerships, and
fostering an inclusive, innovative workplace.

Data quality and integrity are fundamental for reliable AI outcomes.
Poor data quality can lead to inaccurate predictions, eroding trust in
AI systems. Implementing data governance policies, utilizing AI-powered
data cleaning tools, and continuously monitoring data quality are
essential steps in maintaining data integrity.

\subsubsection{ Model Transparency and Security}

The complexity of AI models often raises concerns about transparency,
fairness, and bias, particularly as models grow in sophistication. To
address these concerns, organizations should adopt explainable AI
techniques, such as SHAP values or LIME, which provide insights into
model behavior. Documenting model development processes and adhering to
ethical guidelines from organizations like IEEE or the EU also support
transparency and accountability.

\subsection{ Future Trends}

The future of AI is marked by significant advancements in model
efficiency, cloud-native innovations, hardware development, and
real-world applications, shaping its transformative potential across
industries. Efforts are directed toward optimizing models through
techniques like quantization, pruning, and knowledge distillation,
enabling real-time processing on edge devices with reduced computational
and environmental costs. Cloud-native AI, enhanced by quantum computing,
AI as a Service (AIaaS), IoT integration, and blockchain, is
revolutionizing deployment capabilities and data security.
Simultaneously, specialized AI chips and 5G connectivity are driving
high-performance and responsive applications, particularly in IoT
ecosystems. With increasing adoption, governments and organizations are
developing ethical and regulatory frameworks, such as the EU's proposed
AI Act, to ensure responsible AI usage. From supply chain optimization
to personalized healthcare, companies are leveraging AI to solve
complex challenges, reduce costs, and improve efficiency, paving the way
for a more integrated, efficient, and ethical AI-driven future.

\subsection{ Actionable Insights}

To navigate these challenges and capitalize on future trends, enterprises should:The path to successful AI adoption requires organizations to balance multiple critical elements. A flexible cloud strategy serves as the foundation, embracing multi-cloud and hybrid approaches to avoid vendor lock-in while building resilient systems. Simultaneously, investment in cutting-edge technologies like quantum computing, AI-as-a-Service platforms, and edge computing innovations remains vital for maintaining competitive advantage in a rapidly evolving technological landscape.
Data quality and talent development form the core pillars of effective AI implementation. Organizations must prioritize robust data practices, including comprehensive security measures and clear governance frameworks, as these directly impact AI system reliability. Building strong talent pipelines through internal development programs and strategic partnerships helps address skill shortages, making human capital investment as crucial as technological advancement.
The final piece of the puzzle involves ethical considerations and regulatory compliance. Organizations need to develop clear frameworks for responsible AI development and deployment, establishing transparent policies that ensure fairness and accountability. As the legal landscape around AI continues to evolve, staying informed and adaptable helps avoid potential penalties while building stakeholder trust. Success ultimately depends on thoughtfully balancing these elements while maintaining focus on long-term strategic goals.

\section{ Conclusion}

The analysis of cloud platforms and their services for generative AI has
provided essential insights into the evolving landscape of cloud-based
AI development. This review emphasizes that selecting the right cloud
platform is critical for organizations aiming to develop generative AI
applications. Each cloud provider offers unique strengths, but the
optimal platform should align with an enterprise\textquotesingle s
specific needs, including scalability, cost efficiency, and AI service
capabilities.

Our comprehensive review reveals the extensive ecosystem of generative
AI services offered by major cloud providers including AWS, Azure, GCP,
IBM, and Oracle. These providers have developed sophisticated service
portfolios encompassing essential capabilities such as high-performance
computing infrastructure, serverless architectures, and comprehensive
AI/ML platforms. While these platforms deliver the computational power
and flexibility crucial for AI development, and their Generative AI
tools and API architectures enable efficient model deployment,
organizations must carefully navigate challenges including
cost-performance trade-offs and potential vendor lock-in risks.

The relationship between cloud computing and AI has evolved into a
powerful symbiosis, driving unprecedented advancement in both domains.
As generative AI models grow in sophistication and computational
requirements, cloud platforms are rapidly evolving to meet these demands
through innovations in hardware, software, and service offerings. The
transformative potential of this technology extends across numerous
sectors, from healthcare and finance to creative arts and education,
with many applications still waiting to be discovered. Cloud platforms
are positioned to play a crucial role in democratizing access to AI
technologies, enabling organizations of all sizes to harness the power
of generative AI.

This technological convergence presents both opportunities and
challenges for organizations. To maximize the benefits while minimizing
risks, we recommend that organizations conduct thorough assessments of
their AI needs and align them with the strengths of different cloud
providers. They should prioritize scalability and flexibility in their
cloud strategy to accommodate the rapid pace of AI advancements, while
investing in building internal AI expertise alongside leveraging managed
cloud services. Implementation of robust data governance and security
measures is essential to address the unique challenges posed by AI
applications. Additionally, organizations should consider adopting a
multi-cloud or hybrid cloud approach to mitigate vendor lock-in risks
and optimize performance across different use cases. These strategic
considerations will be crucial as the field continues to evolve and
mature.

For enterprises considering cloud-based generative AI solutions, we
offer the following strategic recommendations:

\begin{enumerate}
\def\labelenumi{\arabic{enumi}.}
\item
  Conduct a thorough assessment of your organization\textquotesingle s
  AI needs and align them with the strengths of different cloud
  providers.
\item
  Prioritize scalability and flexibility in your cloud strategy to
  accommodate the rapid pace of AI advancements.
\item
  Invest in building internal AI expertise while leveraging the managed
  services offered by cloud providers.
\item
  Implement robust data governance and security measures to address the
  unique challenges posed by AI applications.
\item
  Consider a multi-cloud or hybrid cloud approach to mitigate vendor
  lock-in and optimize performance across different use cases.
\end{enumerate}

\textbf{Long-Term Outlook}

The future landscape of cloud AI services is poised to be shaped by
several transformative technological trends and societal imperatives.
The integration of 5G technology with cloud platforms represents a
particularly promising development, promising to unlock new
possibilities for edge AI applications, especially in the domains of
Internet of Things (IoT) and smart city initiatives. This evolution will
be further accelerated by anticipated breakthroughs in AI chip
architecture and quantum computing technologies, which are expected to
deliver substantial performance improvements in cloud-based AI services,
potentially revolutionizing both model training and inference
capabilities. As the computational demands of AI systems continue to
grow, sustainability considerations are becoming increasingly critical,
driving the development of more energy-efficient AI models and cloud
infrastructure. This emphasis on environmental responsibility is
complemented by a growing focus on AI transparency and ethics, with
explainable AI and ethical AI frameworks likely to play an increasingly
influential role in shaping the development of cloud AI services and
tools. These parallel developments in technology and governance
frameworks suggest a future where cloud AI services will not only be
more powerful and efficient but also more responsible and sustainable.

In conclusion, the synergy between cloud computing and generative AI is
poised to drive innovation across industries, offering unprecedented
opportunities for organizations to create value and solve complex
problems. As this field continues to evolve, staying informed and
adaptable will be key to leveraging the full potential of cloud-based
generative AI solutions.

% Before the references section, add these formatting commands
\newpage
\section*{References}
\addcontentsline{toc}{section}{References}

% Add these settings for better reference formatting
\begingroup
\setlength{\parindent}{0pt}
\setlength{\hangindent}{2em}
\makeatletter
\renewcommand\@biblabel[1]{[#1]}
\makeatother

{[}1{]} M. U. Hadi \emph{et al.}, "Large language models: a
comprehensive survey of its applications, challenges, limitations, and
future prospects", \emph{Authorea Preprints,} 2023.

{[}2{]} T. B. Brown \emph{et al.}, "Language Models are Few-Shot
Learners," p. arXiv:2005.14165doi: 10.48550/arXiv.2005.14165.

{[}3{]} J. Achiam \emph{et al.}, "Gpt-4 technical report," \emph{arXiv
preprint arXiv:2303.08774,} 2023.

{[}4{]} G. Team \emph{et al.}, "Gemini: A Family of Highly Capable
Multimodal Models", p. arXiv:2312.11805doi: 10.48550/arXiv.2312.11805.

{[}5{]} R. Grand View. "Generative AI Market Size, Share \& Trends
Analysis Report By Component (Service, Software), By Technology (GAN,
Transformer), By End-use, By Regional Outlook, And Segment Forecasts,
2023 - 2030."
\url{https://www.grandviewresearch.com/industry-analysis/generative-ai-market-report}
(accessed.

{[}6{]} McKinsey and Company. "The Economic Potential of Generative AI:
The Next Productivity Frontier."
\url{https://www.mckinsey.com/capabilities/mckinsey-digital/our-insights/the-economic-potential-of-generative-ai-the-next-productivity-frontier}
(accessed.

{[}7{]} Z. Shi, X. Zhou, X. Qiu, and X. Zhu, "Improving Image Captioning
with Better Use of Captions," p. arXiv:2006.11807doi:
10.48550/arXiv.2006.11807.

{[}8{]} H. Arksey and L. O\textquotesingle Malley, "Scoping studies:
towards a methodological framework," \emph{International Journal of
Social Research Methodology,} vol. 8, no. 1, pp. 19-32, 2005/02/01 2005,
doi: 10.1080/1364557032000119616.

{[}9{]} M. A. Boden, "Creativity and artificial intelligence,"
\emph{Artificial intelligence,} vol. 103, no. 1-2, pp. 347-356, 1998.

{[}10{]} Y. LeCun, Y. Bengio, and G. Hinton, "Deep learning,"
\emph{nature,} vol. 521, no. 7553, pp. 436-444, 2015.

{[}11{]} C. Akkus \emph{et al.}, "Multimodal Deep Learning," p.
arXiv:2301.04856doi: 10.48550/arXiv.2301.04856.

{[}12{]} A. Newell, J. C. Shaw, and H. A. Simon, "Report on a general
problem solving program," in \emph{IFIP congress}, 1959, vol. 256:
Pittsburgh, PA, p. 64.

{[}13{]} D. E. Rumelhart, G. E. Hinton, and R. J. Williams, "Learning
representations by back-propagating errors," \emph{nature,} vol. 323,
no. 6088, pp. 533-536, 1986.

{[}14{]} J. H. Holland, "Genetic algorithms," \emph{Scientific
american,} vol. 267, no. 1, pp. 66-73, 1992.

{[}15{]} I. Goodfellow \emph{et al.}, "Generative adversarial nets,"
\emph{Advances in neural information processing systems,} vol. 27, 2014.

{[}16{]} A. Vaswani \emph{et al.}, "Attention Is All You Need," p.
arXiv:1706.03762doi: 10.48550/arXiv.1706.03762.

{[}17{]} J. Devlin, M.-W. Chang, K. Lee, and K. Toutanova, "BERT:
Pre-training of Deep Bidirectional Transformers for Language
Understanding," p. arXiv:1810.04805doi: 10.48550/arXiv.1810.04805.

{[}18{]} A. Ramesh, P. Dhariwal, A. Nichol, C. Chu, and M. Chen,
"Hierarchical Text-Conditional Image Generation with CLIP Latents," p.
arXiv:2204.06125doi: 10.48550/arXiv.2204.06125.

{[}19{]} R. Rombach, A. Blattmann, D. Lorenz, P. Esser, and B. Ommer,
"High-resolution image synthesis with latent diffusion models," in
\emph{Proceedings of the IEEE/CVF conference on computer vision and
pattern recognition}, 2022, pp. 10684-10695.

{[}20{]} OpenAi. "Hello GPT-4: OpenAI\textquotesingle s Latest Language
Model." \url{https://openai.com/index/hello-gpt-4o/} (accessed.

{[}21{]} A. Dubey \emph{et al.}, "The llama 3 herd of models,"
\emph{arXiv preprint arXiv:2407.21783,} 2024.

{[}22{]} Meta. "Meta LLaMA 3: Advancing Open-Source Language Models."
\url{https://ai.meta.com/blog/meta-llama-3/} (accessed.

{[}23{]} T. Gunter \emph{et al.}, "Apple intelligence foundation
language models," \emph{arXiv preprint arXiv:2407.21075,} 2024.

{[}24{]} 2024, "What is GPT-4 and Why Does it Matter?." {[}Online{]}.
Available: \url{https://www.datacamp.com/blog/what-we-know-gpt4}.

{[}25{]} Anthropic. "Claude 3.5 Sonnet: Advancing Conversational AI with
Ethical AI Principles."
\url{https://www.anthropic.com/news/claude-3-5-sonnet} (accessed.

{[}26{]} G. Team \emph{et al.}, "Gemini 1.5: Unlocking multimodal
understanding across millions of tokens of context," p.
arXiv:2403.05530doi: 10.48550/arXiv.2403.05530.

{[}27{]} H. Touvron \emph{et al.}, "Llama 2: Open Foundation and
Fine-Tuned Chat Models," p. arXiv:2307.09288doi:
10.48550/arXiv.2307.09288.

{[}28{]} A. Dubey \emph{et al.}, "The Llama 3 Herd of Models," p.
arXiv:2407.21783doi: 10.48550/arXiv.2407.21783.

{[}29{]} O. Sharir, B. Peleg, and Y. Shoham, "The cost of training nlp
models: A concise overview," \emph{arXiv preprint arXiv:2004.08900,}
2020.

{[}30{]} OpenAI \emph{et al.}, "GPT-4 Technical Report," p.
arXiv:2303.08774doi: 10.48550/arXiv.2303.08774.

{[}31{]} E. J. Hu \emph{et al.}, "Lora: Low-rank adaptation of large
language models," \emph{arXiv preprint arXiv:2106.09685,} 2021.

{[}32{]} N. Houlsby \emph{et al.}, "Parameter-efficient transfer
learning for NLP," in \emph{International conference on machine
learning}, 2019: PMLR, pp. 2790-2799.

{[}33{]} G. Hinton, "Distilling the Knowledge in a Neural Network,"
\emph{arXiv preprint arXiv:1503.02531,} 2015.

{[}34{]} K. Singhal \emph{et al.}, "Large language models encode
clinical knowledge," \emph{arXiv preprint arXiv:2212.13138,} 2022.

{[}35{]} J. Jumper \emph{et al.}, "Highly accurate protein structure
prediction with AlphaFold," \emph{nature,} vol. 596, no. 7873, pp.
583-589, 2021.

{[}36{]} T. Le Scao \emph{et al.}, "Bloom: A 176b-parameter open-access
multilingual language model," 2023.

{[}37{]} S. Hochreiter and J. Schmidhuber, "Long Short-Term Memory,"
\emph{Neural Computation,} vol. 9, no. 8, pp. 1735-1780, 1997, doi:
10.1162/neco.1997.9.8.1735.

{[}38{]} Y. Bengio, P. Simard, and P. Frasconi, "Learning long-term
dependencies with gradient descent is difficult," \emph{IEEE
Transactions on Neural Networks,} vol. 5, no. 2, pp. 157-166, 1994, doi:
10.1109/72.279181.

{[}39{]} C. Mission, "RAG \& Generative AI: Enhancing AI Solutions with
Retrieval-Augmented Generation," 2024. {[}Online{]}. Available:
\url{https://www.missioncloud.com/blog/rag-generative-ai}.

{[}40{]} G. Izacard and E. Grave, "Leveraging passage retrieval with
generative models for open domain question answering," \emph{arXiv
preprint arXiv:2007.01282,} 2020.

{[}41{]} R. Nakano \emph{et al.}, "WebGPT: Browser-assisted
question-answering with human feedback," p. arXiv:2112.09332doi:
10.48550/arXiv.2112.09332.

{[}42{]} OpenAi. "SearchGPT Prototype: Exploring the Future of AI-Driven
Search." \url{https://openai.com/index/searchgpt-prototype/} (accessed.

{[}43{]} B. Peng \emph{et al.}, "Graph Retrieval-Augmented Generation: A
Survey," p. arXiv:2408.08921doi: 10.48550/arXiv.2408.08921.

{[}44{]} B. Sarmah, B. Hall, R. Rao, S. Patel, S. Pasquali, and D.
Mehta, "HybridRAG: Integrating Knowledge Graphs and Vector Retrieval
Augmented Generation for Efficient Information Extraction," p.
arXiv:2408.04948doi: 10.48550/arXiv.2408.04948.

{[}45{]} T. Baltrušaitis, C. Ahuja, and L.-P. Morency, "Multimodal
machine learning: A survey and taxonomy," \emph{IEEE transactions on
pattern analysis and machine intelligence,} vol. 41, no. 2, pp. 423-443,
2018.

{[}46{]} A. Radford \emph{et al.}, "Learning transferable visual models
from natural language supervision," presented at the International
Conference on Machine Learning, 2021. {[}Online{]}. Available:
\url{https://openai.com/blog/clip/}.

{[}47{]} "GPT-4V(ision) System Card," 2023.

{[}48{]} J. Andreas, D. Klein, and S. Levine, "Learning with latent
language," \emph{arXiv preprint arXiv:1711.00482,} 2017.

{[}49{]} P. Mell, "The NIST Definition of Cloud Computing," \emph{NIST
Special Publication,} pp. 800-145, 2011.

{[}50{]} S. Amazon Web. "Amazon SageMaker."
\url{https://aws.amazon.com/sagemaker/} (accessed.

{[}51{]} A. Microsoft. "Azure Machine Learning."
\url{https://azure.microsoft.com/en-us/products/machine-learning}
(accessed.

{[}52{]} C. Google. "AI \& Machine Learning Products."
\url{https://cloud.google.com/products/ai} (accessed.

{[}53{]} S. Amazon Web. "AWS Comprehend: Natural Language Processing and
Text Analytics Service." \url{https://aws.amazon.com/comprehend/}
(accessed.

{[}54{]} M. Azure. "Azure AI."
\url{https://cloud.google.com/ai/generative-ai} (accessed.

{[}55{]} C. Google. "Google Cloud Vision API: AI-Powered Image and Video
Recognition." \url{https://cloud.google.com/vision?hl=en} (accessed.

{[}56{]} N. P. Jouppi \emph{et al.}, "In-Datacenter Performance Analysis
of a Tensor Processing Unit," p. arXiv:1704.04760doi:
10.48550/arXiv.1704.04760.

{[}57{]} C. Google. "Google Showcases Cloud TPU v4 Pods for Large Model
Training."
\url{https://cloud.google.com/blog/topics/tpus/google-showcases-cloud-tpu-v4-pods-for-large-model-training}
(accessed.

{[}58{]} E. F. Coutinho, F. R. de Carvalho Sousa, P. A. L. Rego, D. G.
Gomes, and J. N. de Souza, "Elasticity in cloud computing: a survey,"
\emph{annals of telecommunications - annales des télécommunications,}
vol. 70, no. 7, pp. 289-309, 2015/08/01 2015, doi:
10.1007/s12243-014-0450-7.

{[}59{]} A. Aghajanyan \emph{et al.}, "Scaling Laws for Generative
Mixed-Modal Language Models," p. arXiv:2301.03728doi:
10.48550/arXiv.2301.03728.

{[}60{]} S. Amazon Web. "EC2 Instance Types: P4."
\url{https://aws.amazon.com/ec2/instance-types/p4/} (accessed.

{[}61{]} S. Amazon Web. "AWS SageMaker Autopilot: Automate Machine
Learning Model Building and Deployment."
\url{https://aws.amazon.com/sagemaker/autopilot/} (accessed.

{[}62{]} A. Microsoft. "Azure AI Services."
\url{https://azure.microsoft.com/en-us/products/ai-services} (accessed.

{[}63{]} S. Amazon Web. "Amazon Redshift: Cloud Data Warehousing for
Analytics." \url{https://aws.amazon.com/redshift/} (accessed.

{[}64{]} A. Microsoft. "Azure Data Lake Analytics: On-Demand Analytics
Job Service."
\url{https://azure.microsoft.com/en-us/products/data-lake-analytics/}
(accessed.

{[}65{]} C. Google. "Introducing BigQuery Omni."
\url{https://cloud.google.com/blog/products/data-analytics/introducing-bigquery-omni}
(accessed.

{[}66{]} C. Google. "Google Cloud CDN: High-Performance Content Delivery
Network." \url{https://cloud.google.com/cdn?hl=en} (accessed.

{[}67{]} A. Microsoft. "Azure Front Door Overview: Secure and Scalable
Application Delivery."
\url{https://learn.microsoft.com/en-us/azure/frontdoor/front-door-overview}
(accessed.

{[}68{]} S. Amazon Web. "AWS Global Accelerator: Improve Availability
and Performance of Global Applications."
\url{https://aws.amazon.com/global-accelerator/} (accessed.

{[}69{]} S. Amazon Web. "AWS Global Accelerator: Key Features for
Enhanced Global Application Performance."
\url{https://aws.amazon.com/global-accelerator/features/} (accessed.

{[}70{]} A. Microsoft. "Azure Front Door: Scalable and Secure
Application Delivery Network."
\url{https://azure.microsoft.com/en-us/products/frontdoor/} (accessed.

{[}71{]} DigitalOcean. "Pay-As-You-Go Cloud Computing: Flexible and
Cost-Effective Cloud Solutions."
\url{https://www.digitalocean.com/resources/articles/pay-as-you-go-cloud-computing}
(accessed.

{[}72{]} S. Amazon Web. "AWS Cost Explorer: Visualize and Manage Your
AWS Costs and Usage."
\url{https://aws.amazon.com/aws-cost-management/aws-cost-explorer/}
(accessed.

{[}73{]} A. Microsoft. "Azure Cost Management: Optimize Your Cloud
Spending and Improve Efficiency."
\url{https://azure.microsoft.com/en-us/products/cost-management}
(accessed.

{[}74{]} C. Google. "Google Cloud Cost Management: Tools to Optimize
Your Cloud Spend." \url{https://cloud.google.com/cost-management?hl=en}
(accessed.

{[}75{]} D. Bernstein, "Containers and Cloud: From LXC to Docker to
Kubernetes," \emph{IEEE Cloud Computing,} vol. 1, no. 3, pp. 81-84,
2014, doi: 10.1109/MCC.2014.51.

{[}76{]} I. Baldini \emph{et al.}, "Serverless Computing: Current Trends
and Open Problems," p. arXiv:1706.03178doi: 10.48550/arXiv.1706.03178.

{[}77{]} P. Singh, A. Kaur, and S. S. Gill, "Machine learning for cloud,
fog, edge and serverless computing environments: Comparisons,
performance evaluation benchmark and future directions,"
\emph{International Journal of Grid and Utility Computing,} vol. 13, no.
4, pp. 447-457, 2022.

{[}78{]} D. Kreuzberger, N. Kühl, and S. Hirschl, "Machine learning
operations (mlops): Overview, definition, and architecture," \emph{IEEE
access,} vol. 11, pp. 31866-31879, 2023.

{[}79{]} R. Buyya, C. S. Yeo, S. Venugopal, J. Broberg, and I. Brandic,
"Cloud computing and emerging IT platforms: Vision, hype, and reality
for delivering computing as the 5th utility," \emph{Future Generation
Computer Systems,} vol. 25, no. 6, pp. 599-616, 2009/06/01/ 2009, doi:
\url{https://doi.org/10.1016/j.future.2008.12.001}.

{[}80{]} J. Hong, T. Dreibholz, J. A. Schenkel, and J. A. Hu, "An
overview of multi-cloud computing," in \emph{Web, Artificial
Intelligence and Network Applications: Proceedings of the Workshops of
the 33rd International Conference on Advanced Information Networking and
Applications (WAINA-2019) 33}, 2019: Springer, pp. 1055-1068.

{[}81{]} W. Shi, J. Cao, Q. Zhang, Y. Li, and L. Xu, "Edge Computing:
Vision and Challenges," \emph{IEEE Internet of Things Journal,} vol. 3,
no. 5, pp. 637-646, 2016, doi: 10.1109/JIOT.2016.2579198.

{[}82{]} L. Alhenaki, A. Alwatban, B. Alahmri, and N. Alarifi, "Security
in cloud computing: a survey," \emph{International Journal of Computer
Science and Information Security (IJCSIS),} vol. 17, no. 4, pp. 67-90,
2019.

{[}83{]} K. Zhang, Y. Mao, S. Leng, Y. He, and Y. Zhang, "Mobile-edge
computing for vehicular networks: A promising network paradigm with
predictive off-loading," \emph{IEEE vehicular technology magazine,} vol.
12, no. 2, pp. 36-44, 2017.

{[}84{]} X. Wang, Y. Han, C. Wang, Q. Zhao, X. Chen, and M. Chen,
"In-Edge AI: Intelligentizing Mobile Edge Computing, Caching and
Communication by Federated Learning," p. arXiv:1809.07857doi:
10.48550/arXiv.1809.07857.

{[}85{]} H. Li, K. Ota, and M. Dong, "Learning IoT in edge: Deep
learning for the Internet of Things with edge computing," \emph{IEEE
network,} vol. 32, no. 1, pp. 96-101, 2018.

{[}86{]} M. Schuld, I. Sinayskiy, and F. Petruccione, "An introduction
to quantum machine learning," \emph{Contemporary Physics,} vol. 56, no.
2, pp. 172-185, 2015.

{[}87{]} Ibm. "IBM Quantum: Pioneering the Future of Quantum Computing."
\url{https://www.ibm.com/quantum} (accessed.

{[}88{]} A. Microsoft. "Azure Quantum Computing: Solutions for
Next-Generation Computing."
\url{https://azure.microsoft.com/en-us/solutions/quantum-computing/}
(accessed.

{[}89{]} S. Amazon Web. "AWS Quantum Computing: Tools and Services for
Quantum Innovation." \url{https://aws.amazon.com/products/quantum/}
(accessed.

{[}90{]} Google. "Google Quantum AI: Advancing Quantum Computing for
Real-World Impact." \url{https://quantumai.google/} (accessed.

{[}91{]} Dell. "Addressing the Environmental Footprint While Optimizing
the Handprint of AI."
\url{https://www.dell.com/en-us/blog/addressing-the-environmental-footprint-while-optimizing-the-handprint-of-ai/}
(accessed.

{[}92{]} F. World Economic. "How to Optimize AI While Minimizing Your
Carbon Footprint."
\url{https://www.weforum.org/agenda/2024/01/how-to-optimize-ai-while-minimizing-your-carbon-footprint/}
(accessed.

{[}93{]} C. R. N. Staff. "Cloud Market Share in Q2: Microsoft Drops,
Google Gains, AWS Remains Leader."
\url{https://www.crn.com/news/cloud/2024/cloud-market-share-in-q2-microsoft-drops-google-gains-aws-remains-leader}
(accessed.

{[}94{]} Statista. "Worldwide Cloud Infrastructure Services Market Share
by Vendor."
\url{https://www.statista.com/statistics/967365/worldwide-cloud-infrastructure-services-market-share-vendor/}
(accessed.

{[}95{]} T. CloudZero. "Cloud Service Providers: A Comprehensive Guide."
\url{https://www.cloudzero.com/blog/cloud-service-providers/} (accessed.

{[}96{]} A. Microsoft. "Azure OpenAI Service."
\url{https://azure.microsoft.com/en-us/products/ai-services/openai-service/}
(accessed.

{[}97{]} Anthropic. "Anthropic Partners with Google Cloud to Scale AI
Research and Development."
\url{https://www.anthropic.com/news/anthropic-partners-with-google-cloud}
(accessed.

{[}98{]} Nvidia. "AWS-NVIDIA Strategic Collaboration for Generative AI."
\url{https://nvidianews.nvidia.com/news/aws-nvidia-strategic-collaboration-for-generative-ai}
(accessed.

{[}99{]} Ibm. "IBM and NASA Release Open-Source AI Model on Hugging Face
for Weather and Climate Applications."
\url{https://newsroom.ibm.com/2024-09-23-ibm-and-nasa-release-open-source-ai-model-on-hugging-face-for-weather-and-climate-applications}
(accessed.

{[}100{]} Nvidia. "Oracle and NVIDIA Announce Sovereign AI Cloud for
Europe."
\url{https://nvidianews.nvidia.com/news/oracle-nvidia-sovereign-ai}
(accessed.

{[}101{]} Intel. "Alibaba Cloud Analytics Zoo: AI-Powered Analytics on
Intel Architecture."
\url{https://www.intel.com/content/www/us/en/customer-spotlight/stories/alibaba-cloud-analytics-zoo-customer-story.html}
(accessed.

{[}102{]} A. Chowdhery \emph{et al.}, "Palm: Scaling language modeling
with pathways," \emph{Journal of Machine Learning Research,} vol. 24,
no. 240, pp. 1-113, 2023.

{[}103{]} T. Sterling, M. Brodowicz, and M. Anderson, \emph{High
performance computing: modern systems and practices}. Morgan Kaufmann,
2017.

{[}104{]} W. J. Dally, S. W. Keckler, and D. B. Kirk, "Evolution of the
graphics processing unit (GPU)," \emph{IEEE Micro,} vol. 41, no. 6, pp.
42-51, 2021.

{[}105{]} A. Aghajanyan \emph{et al.}, "Scaling laws for generative
mixed-modal language models," in \emph{International Conference on
Machine Learning}, 2023: PMLR, pp. 265-279.

{[}106{]} S. Amazon Web. "AWS EC2 P5 Instance Types: High-Performance
Computing for Machine Learning and AI."
\url{https://aws.amazon.com/ec2/instance-types/p5/} (accessed.

{[}107{]} A. Microsoft. "Azure ND H100v5 Series: High-Performance
GPU-Accelerated Virtual Machines for AI and ML."
\url{https://learn.microsoft.com/en-us/azure/virtual-machines/sizes/gpu-accelerated/ndh100v5-series?tabs=sizebasic}
(accessed.

{[}108{]} C. Google. "Google Cloud On-Demand A3 VMs: Flexible and
High-Performance Virtual Machines for AI Workloads."
\url{https://cloud.google.com/skus/sku-groups/on-demand-a3-vms}
(accessed.

{[}109{]} S. Amazon Web. "AWS Trainium: High-Performance Machine
Learning Training Chips."
\url{https://aws.amazon.com/machine-learning/trainium/} (accessed.

{[}110{]} C. Google. "Google Cloud TPU v5e Documentation:
High-Performance AI Accelerators for Machine Learning."
\url{https://cloud.google.com/tpu/docs/v5e} (accessed.

{[}111{]} D. Patterson \emph{et al.}, "Carbon emissions and large neural
network training," \emph{arXiv preprint arXiv:2104.10350,} 2021.

{[}112{]} Amazon. "Amazon Cloud Sustainability: Reducing Environmental
Impact through the Cloud."
\url{https://sustainability.aboutamazon.com/products-services/the-cloud}
(accessed.

{[}113{]} D. Chahal, M. Ramesh, R. Ojha, and R. Singhal, "High
performance serverless architecture for deep learning workflows," in
\emph{2021 IEEE/ACM 21st International Symposium on Cluster, Cloud and
Internet Computing (CCGrid)}, 2021: IEEE, pp. 790-796.

{[}114{]} E. Paraskevoulakou and D. Kyriazis, "ML-FaaS: Toward
Exploiting the Serverless Paradigm to Facilitate Machine Learning
Functions as a Service," \emph{IEEE Transactions on Network and Service
Management,} vol. 20, no. 3, pp. 2110-2123, 2023.

{[}115{]} K. Kojs, "A Survey of Serverless Machine Learning Model
Inference," p. arXiv:2311.13587doi: 10.48550/arXiv.2311.13587.

{[}116{]} V. Ishakian, V. Muthusamy, and A. Slominski, "Serving deep
learning models in a serverless platform," in \emph{2018 IEEE
International conference on cloud engineering (IC2E)}, 2018: IEEE, pp.
257-262.

{[}117{]} A. Gallego, U. Odyurt, Y. Cheng, Y. Wang, and Z. Zhao,
"Machine Learning Inference on Serverless Platforms Using Model
Decomposition," in \emph{Proceedings of the IEEE/ACM 16th International
Conference on Utility and Cloud Computing}, 2023, pp. 1-6.

{[}118{]} M. S. Aslanpour \emph{et al.}, "Serverless edge computing:
vision and challenges," in \emph{Proceedings of the 2021 Australasian
computer science week multiconference}, 2021, pp. 1-10.

{[}119{]} Y. Fu \emph{et al.}, "\{ServerlessLLM\}:\{Low-Latency\}
Serverless Inference for Large Language Models," in \emph{18th USENIX
Symposium on Operating Systems Design and Implementation (OSDI 24)},
2024, pp. 135-153.

{[}120{]} O. Donati, M. Macario, and M. H. Karim, "Event-Driven AI
Workflows in Serverless Computing: Enabling Real-Time Data Processing
and Decision-Making," 2024.

{[}121{]} A. Arjona, P. G. López, J. Sampé, A. Slominski, and L.
Villard, "Triggerflow: Trigger-based orchestration of serverless
workflows," \emph{Future Generation Computer Systems,} vol. 124, pp.
215-229, 2021/11/01/ 2021, doi:
\url{https://doi.org/10.1016/j.future.2021.06.004}.

{[}122{]} D. Loconte, S. Ieva, A. Pinto, G. Loseto, F. Scioscia, and M.
Ruta, "Expanding the cloud-to-edge continuum to the IoT in serverless
federated learning," \emph{Future Generation Computer Systems,} vol.
155, pp. 447-462, 2024/06/01/ 2024, doi:
\url{https://doi.org/10.1016/j.future.2024.02.024}.

{[}123{]} S. Qi, K. Ramakrishnan, and M. Lee, "LIFL: A Lightweight,
Event-driven Serverless Platform for Federated Learning,"
\emph{Proceedings of Machine Learning and Systems,} vol. 6, pp. 408-425,
2024.

{[}124{]} Z. Zhou, X. Chen, E. Li, L. Zeng, K. Luo, and J. Zhang, "Edge
Intelligence: Paving the Last Mile of Artificial Intelligence With Edge
Computing," \emph{Proceedings of the IEEE,} vol. 107, no. 8, pp.
1738-1762, 2019, doi: 10.1109/JPROC.2019.2918951.

{[}125{]} S. Teerapittayanon, B. McDanel, and H.-T. Kung, "Distributed
deep neural networks over the cloud, the edge and end devices," in
\emph{2017 IEEE 37th international conference on distributed computing
systems (ICDCS)}, 2017: IEEE, pp. 328-339.

{[}126{]} T. Li, A. K. Sahu, A. Talwalkar, and V. Smith, "Federated
learning: Challenges, methods, and future directions," \emph{IEEE signal
processing magazine,} vol. 37, no. 3, pp. 50-60, 2020.

{[}127{]} T. Liang, J. Glossner, L. Wang, S. Shi, and X. Zhang, "Pruning
and quantization for deep neural network acceleration: A survey,"
\emph{Neurocomputing,} vol. 461, pp. 370-403, 2021.

{[}128{]} A. Kuzmin, M. Nagel, M. Van Baalen, A. Behboodi, and T.
Blankevoort, "Pruning vs quantization: which is better?," \emph{Advances
in neural information processing systems,} vol. 36, 2024.

{[}129{]} Y. Mao, C. You, J. Zhang, K. Huang, and K. B. Letaief, "A
survey on mobile edge computing: The communication perspective,"
\emph{IEEE communications surveys \& tutorials,} vol. 19, no. 4, pp.
2322-2358, 2017.

{[}130{]} A. Gholami \emph{et al.}, "Squeezenext: Hardware-aware neural
network design," in \emph{Proceedings of the IEEE conference on computer
vision and pattern recognition workshops}, 2018, pp. 1638-1647.

{[}131{]} Y. Jia \emph{et al.}, "Model Pruning-enabled Federated Split
Learning for Resource-constrained Devices in Artificial Intelligence
Empowered Edge Computing Environment," \emph{ACM Transactions on Sensor
Networks,} 2024.

{[}132{]} P. Patel \emph{et al.}, "Splitwise: Efficient generative llm
inference using phase splitting," in \emph{2024 ACM/IEEE 51st Annual
International Symposium on Computer Architecture (ISCA)}, 2024: IEEE,
pp. 118-132.

{[}133{]} Y. Long, I. Chakraborty, G. Srinivasan, and K. Roy,
"Complexity-aware adaptive training and inference for edge-cloud
distributed AI systems," in \emph{2021 IEEE 41st International
Conference on Distributed Computing Systems (ICDCS)}, 2021: IEEE, pp.
573-583.

{[}134{]} S. Nayak, R. Patgiri, L. Waikhom, and A. Ahmed, "A review on
edge analytics: Issues, challenges, opportunities, promises, future
directions, and applications," \emph{Digital Communications and
Networks,} vol. 10, no. 3, pp. 783-804, 2024/06/01/ 2024, doi:
\url{https://doi.org/10.1016/j.dcan.2022.10.016}.

{[}135{]} E. Forno, A. Spitale, E. Macii, and G. Urgese, "Configuring an
embedded neuromorphic coprocessor using a risc-v chip for enabling edge
computing applications," in \emph{2021 IEEE 14th International Symposium
on Embedded Multicore/Many-core Systems-on-Chip (MCSoC)}, 2021: IEEE,
pp. 328-332.

{[}136{]} P. Madduru, "Artificial Intelligence as a service in
distributed multi access edge computing on 5G extracting data using IoT
and including AR/VR for real-time reporting," \emph{Information
Technology In Industry,} vol. 9, no. 1, pp. 912-931, 2021.

{[}137{]} Y. Shen \emph{et al.}, "Large language models empowered
autonomous edge AI for connected intelligence," \emph{IEEE
Communications Magazine,} 2024.

{[}138{]} Y. Tian \emph{et al.}, "An Edge-Cloud Collaboration Framework
for Generative AI Service Provision with Synergetic Big Cloud Model and
Small Edge Models," \emph{arXiv preprint arXiv:2401.01666,} 2024.

{[}139{]} R. De Prisco, B. Lampson, and N. Lynch, "Revisiting the Paxos
algorithm," \emph{Theoretical Computer Science,} vol. 243, no. 1-2, pp.
35-91, 2000.

{[}140{]} F. Chang \emph{et al.}, "Bigtable: A distributed storage
system for structured data," \emph{ACM Transactions on Computer Systems
(TOCS),} vol. 26, no. 2, pp. 1-26, 2008.

{[}141{]} H. Fang, "Managing data lakes in big data era:
What\textquotesingle s a data lake and why has it became popular in data
management ecosystem," in \emph{2015 IEEE International Conference on
Cyber Technology in Automation, Control, and Intelligent Systems
(CYBER)}, 2015: IEEE, pp. 820-824.

{[}142{]} A. Singhal, \emph{Data warehousing and data mining techniques
for cyber security}. Springer Science \& Business Media, 2007.

{[}143{]} K. Shvachko, H. Kuang, S. Radia, and R. Chansler, "The hadoop
distributed file system," in \emph{2010 IEEE 26th symposium on mass
storage systems and technologies (MSST)}, 2010: Ieee, pp. 1-10.

{[}144{]} B. Chen and J. Zhang, "Content-aware scalable deep compressed
sensing," \emph{IEEE Transactions on Image Processing,} vol. 31, pp.
5412-5426, 2022.

{[}145{]} Nvidia. "NVIDIA GPUDirect Storage: Accelerating Data Movement
for GPU Computing." \url{https://developer.nvidia.com/gpudirect-storage}
(accessed.

{[}146{]} Q. D. Tran, "Toward a serverless data pipeline."

{[}147{]} A. L'heureux, K. Grolinger, H. F. Elyamany, and M. A. Capretz,
"Machine learning with big data: Challenges and approaches," \emph{Ieee
Access,} vol. 5, pp. 7776-7797, 2017.

{[}148{]} O. Olesen-Bagneux, \emph{The Enterprise Data Catalog}. "
O\textquotesingle Reilly Media, Inc.", 2023.

{[}149{]} C. Google, "Data Analytics Innovations to Fuel AI
Initiatives," 2024/10/06 2024. {[}Online{]}. Available:
\url{https://cloud.google.com/blog/products/data-analytics/data-analytics-innovations-to-fuel-ai-initiatives}.

{[}150{]} Oracle, "Machine Learning Platform on Oracle Autonomous Data
Warehouse," 2024/10/06 2024. {[}Online{]}. Available:
\url{https://docs.oracle.com/en/solutions/ml-platform-on-adw/index.html}.

{[}151{]} P. Russom, D. Stodder, and F. Halper, "Real-time data, BI, and
analytics," \emph{Accelerating Business to Leverage Customer Relations,
Competitiveness, and Insights. TDWI best practices report, fourth
quarter,} pp. 5-25, 2014.

{[}152{]} R. Ramakrishnan and J. Gehrke, \emph{Database management
systems}. McGraw-Hill, Inc., 2002.

{[}153{]} T. Dell, "What Does Generative AI Mean for Data Gravity and IT
Infrastructure?," 2024/10/06 2024. {[}Online{]}. Available:
\url{https://www.dell.com/en-us/blog/what-does-generative-ai-mean-for-data-gravity-and-it-infrastructure/}.

{[}154{]} D. Solove, \emph{The Digital Person: Technology and Privacy in
the Information Age}. New York University Press, 2004.

{[}155{]} T. Koltay, "Data governance, data literacy and the management
of data quality," \emph{IFLA journal,} vol. 42, no. 4, pp. 303-312,
2016.

{[}156{]} Dataiku, "Why Data Quality Matters in the Age of Generative
AI," 2024/10/06 2024. {[}Online{]}. Available:
\url{https://blog.dataiku.com/why-data-quality-matters-in-the-age-of-generative-ai}.

{[}157{]} V. Panwar, "AI-Driven Query Optimization: Revolutionizing
Database Performance and Efficiency."

{[}158{]} T. Juopperi, "AI-driven SQL query optimization techniques,"
2024.

{[}159{]} S. Arunachalam and R. De Wolf, "Guest column: A survey of
quantum learning theory," \emph{ACM Sigact News,} vol. 48, no. 2, pp.
41-67, 2017.

{[}160{]} F. Bonchi, B. Malin, and Y. Saygin, "Recent advances in
preserving privacy when mining data," \emph{Data \& Knowledge
Engineering,} vol. 65, no. 1, pp. 1-4, 2008.

{[}161{]} K. Holstein, J. Wortman Vaughan, H. Daumé III, M. Dudik, and
H. Wallach, "Improving fairness in machine learning systems: What do
industry practitioners need?," in \emph{Proceedings of the 2019 CHI
conference on human factors in computing systems}, 2019, pp. 1-16.

{[}162{]} Ibm, "IBM Watsonx AI Prompt Tuner," 2024 2024. {[}Online{]}.
Available:
\url{https://ibm.github.io/watsonx-ai-python-sdk/prompt_tuner.html}.

{[}163{]} C. Google, "Google Cloud Vertex AI Generative Text Prompts,"
2024 2024. {[}Online{]}. Available:
\url{https://cloud.google.com/vertex-ai/generative-ai/docs/text/text-prompts}.

{[}164{]} S. Amazon Web, "AWS Bedrock Prompt Flows," 2024 2024.
{[}Online{]}. Available:
\url{https://aws.amazon.com/bedrock/prompt-flows/}.

{[}165{]} Microsoft, "Microsoft Azure Prompt Flow," 2024 2024.
{[}Online{]}. Available:
\url{https://learn.microsoft.com/en-us/azure/ai-studio/how-to/prompt-flow}.

{[}166{]} S. Amazon Web. "AWS SageMaker: End-to-End Machine Learning
Platform for Building, Training, and Deploying Models."
\url{https://aws.amazon.com/sagemaker/} (accessed.

{[}167{]} Oracle, "Oracle AI Data Science," 2024/10/06. {[}Online{]}.
Available:
\url{https://www.oracle.com/artificial-intelligence/data-science/}.

{[}168{]} Microsoft, "Microsoft Azure Machine Learning Registries and
MLOps," 2024 2024. {[}Online{]}. Available:
\url{https://learn.microsoft.com/en-us/azure/machine-learning/concept-machine-learning-registries-mlops?view=azureml-api-2}.

{[}169{]} C. Google, "Google Cloud Vertex AI Model Registry
Introduction," 2024 2024. {[}Online{]}. Available:
\url{https://cloud.google.com/vertex-ai/docs/model-registry/introduction}.

{[}170{]} S. Amazon Web, "AWS SageMaker Model Registry," 2024 2024.
{[}Online{]}. Available:
\url{https://docs.aws.amazon.com/sagemaker/latest/dg/model-registry-version.html}.

{[}171{]} Oracle, "Oracle Data Science Model Catalog," 2024 2024.
{[}Online{]}. Available:
\url{https://docs.oracle.com/en-us/iaas/data-science/using/models-about.htm}.

{[}172{]} Microsoft, "Microsoft Azure MLOps Setup," 2024 2024.
{[}Online{]}. Available:
\url{https://learn.microsoft.com/en-us/azure/machine-learning/how-to-setup-mlops-azureml?view=azureml-api-2&tabs=azure-shell}.

{[}173{]} C. Google, "Google Cloud Vertex AI ML Metadata," 2024 2024.
{[}Online{]}. Available:
\url{https://cloud.google.com/vertex-ai/docs/ml-metadata/introduction}.

{[}174{]} S. Amazon Web, "AWS SageMaker Lineage Tracking," 2024 2024.
{[}Online{]}. Available:
\url{https://docs.aws.amazon.com/sagemaker/latest/dg/lineage-tracking.html}.

{[}175{]} Ibm. "IBM Watson Studio: Collaborative Environment for AI
Model Building and Deployment."
\url{https://www.ibm.com/products/watson-studio} (accessed.

{[}176{]} Oracle, "Oracle Data Science Models Overview," 2024 2024.
{[}Online{]}. Available:
\url{https://docs.oracle.com/en-us/iaas/data-science/using/models-about.htm}.

{[}177{]} C. Google, "Google Cloud Vertex AI Pipelines Introduction,"
2024 2024. {[}Online{]}. Available:
\url{https://cloud.google.com/vertex-ai/docs/pipelines/introduction}.

{[}178{]} S. Amazon Web, "AWS SageMaker Projects Overview," 2024 2024.
{[}Online{]}. Available:
\url{https://docs.aws.amazon.com/sagemaker/latest/dg/sagemaker-projects-whatis.html}.

{[}179{]} Oracle, "Oracle DevOps Service," 2024/10/06. {[}Online{]}.
Available: \url{https://www.oracle.com/devops/devops-service/}.

{[}180{]} S. Amazon Web, "Amazon SageMaker Serverless Inference: Machine
Learning Inference Without Worrying About Servers," 2024 2024.
{[}Online{]}. Available:
\url{https://aws.amazon.com/blogs/aws/amazon-sagemaker-serverless-inference-machine-learning-inference-without-worrying-about-servers/}.

{[}181{]} S. Amazon Web, "Amazon SageMaker Neo," 2024 2024.
{[}Online{]}. Available: \url{https://aws.amazon.com/sagemaker/neo/}.

{[}182{]} Microsoft, "Microsoft Azure Container Instances," 2024 2024.
{[}Online{]}. Available:
\url{https://azure.microsoft.com/en-us/products/container-instances}.

{[}183{]} Microsoft, "Microsoft Azure ONNX Concept Overview," 2024 2024.
{[}Online{]}. Available:
\url{https://learn.microsoft.com/en-us/azure/machine-learning/concept-onnx?view=azureml-api-2}.

{[}184{]} Microsoft, "Microsoft Support: All About Neural Processing
Units (NPUs)," 2024 2024. {[}Online{]}. Available:
\url{https://support.microsoft.com/en-us/windows/all-about-neural-processing-units-npus-e77a5637-7705-4915-96c8-0c6a975f9db4}.

{[}185{]} C. Google. "Google Cloud Run."
\url{https://cloud.google.com/run?hl=en} (accessed.

{[}186{]} A. I. Google, "Google AI Edge LiteRT," 2024 2024.
{[}Online{]}. Available: \url{https://ai.google.dev/edge/litert}.

{[}187{]} C. Google. "Accelerate AI development with Google Cloud TPUs."
\url{https://cloud.google.com/tpu?hl=en} (accessed.

{[}188{]} I. B. M. Cloud. "IBM Cloud for AI."
\url{https://www.ibm.com/cloud/ai} (accessed.

{[}189{]} Oracle. "Oracle Functions Overview."
\url{https://docs.oracle.com/en-us/iaas/Content/Functions/Concepts/functionsoverview.htm}
(accessed.

{[}190{]} Oracle, "Oracle Cloud GPU Compute," 2024/10/06. {[}Online{]}.
Available: \url{https://www.oracle.com/cloud/compute/gpu/}.

{[}191{]} C. Microsoft. "Concept of Model Monitoring in Azure Machine
Learning."
\url{https://learn.microsoft.com/en-us/azure/machine-learning/concept-model-monitoring?view=azureml-api-2}
(accessed.

{[}192{]} C. Google, "Google Cloud Monitoring," 2024 2024. {[}Online{]}.
Available: \url{https://cloud.google.com/monitoring?hl=en}.

{[}193{]} S. Amazon Web, "Amazon SageMaker Model Monitor: Fully Managed
Automatic Monitoring for Your Machine Learning Models," 2024/10/06 2024.
{[}Online{]}. Available:
\url{https://aws.amazon.com/blogs/aws/amazon-sagemaker-model-monitor-fully-managed-automatic-monitoring-for-your-machine-learning-models/}.

{[}194{]} Ibm, "Watson OpenScale \textbar{} IBM Cloud Paks," 2024/10/06
2024. {[}Online{]}. Available:
\url{https://www.ibm.com/docs/en/cloud-paks/cp-data/5.0.x?topic=new-watson-openscale}.

{[}195{]} C. Google, "Model Monitoring Overview \textbar{} Google Cloud
Vertex AI," 2024/10/06 2024. {[}Online{]}. Available:
\url{https://cloud.google.com/vertex-ai/docs/model-monitoring/overview}.

{[}196{]} S. Amazon Web. "Detecting Data Drift Using Amazon SageMaker."
Amazon Web Services.
\url{https://aws.amazon.com/blogs/architecture/detecting-data-drift-using-amazon-sagemaker/}
(accessed.

{[}197{]} I. B. M. Corporation. "Watson Drift Metrics in IBM Watson
OpenScale."
\url{https://www.ibm.com/docs/en/watsonx/w-and-w/1.1.x?topic=openscale-watson-drift-metrics}
(accessed.

{[}198{]} A. Microsoft. "How to Use the Responsible AI Dashboard in
Azure Machine Learning."
\url{https://learn.microsoft.com/en-us/azure/machine-learning/how-to-responsible-ai-dashboard?view=azureml-api-2}
(accessed.

{[}199{]} C. Google. "Explainable AI in Google Cloud Vertex AI: Overview
of Interpretability Tools."
\url{https://cloud.google.com/vertex-ai/docs/explainable-ai/overview}
(accessed.

{[}200{]} S. Amazon Web, "SageMaker Clarify," 2024/10/06. {[}Online{]}.
Available: \url{https://aws.amazon.com/sagemaker/clarify/}.

{[}201{]} Ibm. "AI Explainability 360: IBM\textquotesingle s Open-Source
Toolkit for Explainable AI." \url{https://aix360.res.ibm.com/}
(accessed.

{[}202{]} Oracle. "Let\textquotesingle s Be Fair: Using AI Fairness
Tools to Ensure Ethical AI."
\url{https://www.ateam-oracle.com/post/lets-be-fair-using-ai} (accessed.

{[}203{]} C. Google. "Vertex AI."
\url{https://cloud.google.com/vertex-ai} (accessed.

{[}204{]} S. Amazon Web. "AWS Bedrock: Build and Scale Generative AI
Applications with Pre-Trained Models."
\url{https://aws.amazon.com/bedrock/} (accessed.

{[}205{]} Ibm, "IBM Natural Language Understanding," 2024/10/06 2024.
{[}Online{]}. Available:
\url{https://www.ibm.com/products/natural-language-understanding}.

{[}206{]} Oracle, "Oracle AI Language Service," 2024/10/06.
{[}Online{]}. Available:
\url{https://www.oracle.com/artificial-intelligence/language/}.

{[}207{]} Google, "Google Imagen Research," 2024/10/06. {[}Online{]}.
Available: \url{https://imagen.research.google/}.

{[}208{]} Ibm, "IBM Video Streaming," 2024/10/06. {[}Online{]}.
Available: \url{https://www.ibm.com/products/video-streaming}.

{[}209{]} Oracle, "Oracle AI Vision," 2024/10/06. {[}Online{]}.
Available: \url{https://www.oracle.com/artificial-intelligence/vision/}.

{[}210{]} GitHub, "GitHub Copilot," 2024/10/06. {[}Online{]}. Available:
\url{https://github.com/features/copilot}.

{[}211{]} Google, "Google Cloud AI Code Generation," 2024/10/06.
{[}Online{]}. Available:
\url{https://cloud.google.com/use-cases/ai-code-generation?hl=en}.

{[}212{]} Aws, "AWS CodeWhisperer," 2024/10/06. {[}Online{]}. Available:
\url{https://docs.aws.amazon.com/codewhisperer/latest/userguide/what-is-cwspr.html}.

{[}213{]} Ibm, "IBM Watsonx Code Assistant," 2024/10/06. {[}Online{]}.
Available: \url{https://www.ibm.com/products/watsonx-code-assistant}.

{[}214{]} Microsoft, "Azure AI Speech Launches New Zero-Shot TTS Models
for Personal Use," 2024/10/06. {[}Online{]}. Available:
\url{https://techcommunity.microsoft.com/t5/ai-azure-ai-services-blog/azure-ai-speech-launches-new-zero-shot-tts-models-for-personal/ba-p/4044521}.

{[}215{]} C. Google. "Google Cloud Speech-to-Text: Speech Recognition
and Transcription Services."
\url{https://cloud.google.com/speech-to-text?hl=en} (accessed.

{[}216{]} S. Amazon Web. "AWS Polly: Text-to-Speech Service for
Generating Realistic Speech." \url{https://aws.amazon.com/polly/}
(accessed.

{[}217{]} Ibm, "IBM Text-to-Speech," 2024/10/06. {[}Online{]}.
Available: \url{https://www.ibm.com/products/text-to-speech}.

{[}218{]} Oracle, "Oracle AI Speech," 2024/10/06. {[}Online{]}.
Available: \url{https://www.oracle.com/artificial-intelligence/speech/}.

{[}219{]} C. Google, "Get Multimodal Embeddings with Vertex AI,"
2024/10/06 2024. {[}Online{]}. Available:
\url{https://cloud.google.com/vertex-ai/generative-ai/docs/embeddings/get-multimodal-embeddings}.

{[}220{]} A. Microsoft. "Microsoft Azure AI Bot Service."
\url{https://azure.microsoft.com/en-us/products/ai-services/ai-bot-service}
(accessed.

{[}221{]} C. Google. "Cloud Google Dialogflow CX Documentation."
\url{https://cloud.google.com/dialogflow/cx/docs} (accessed.

{[}222{]} S. Amazon Web. "Amazon Lex V2 Documentation."
\url{https://docs.aws.amazon.com/lexv2/latest/dg/what-is.html}
(accessed.

{[}223{]} Ibm. "IBM WatsonX Assistant Documentation."
\url{https://www.ibm.com/products/watsonx-assistant} (accessed.

{[}224{]} Oracle. "Oracle Digital Assistant Documentation."
\url{https://www.oracle.com/chatbots/what-is-a-digital-assistant/}
(accessed.

{[}225{]} A. Microsoft. "Microsoft Azure Translator Service Overview."
\url{https://learn.microsoft.com/en-us/azure/ai-services/translator/translator-overview}
(accessed.

{[}226{]} C. Google. "Google Cloud Translate Service Overview."
\url{https://cloud.google.com/translate?hl=en} (accessed.

{[}227{]} S. Amazon Web. "Amazon Translate Service Overview."
\url{https://docs.aws.amazon.com/translate/latest/dg/what-is.html}
(accessed.

{[}228{]} Oracle. "Oracle AI Language Service Overview."
\url{https://www.oracle.com/artificial-intelligence/language/}
(accessed.

{[}229{]} C. Google. "Google Cloud Natural Language API: Text Analysis
and Language Understanding Platform."
\url{https://cloud.google.com/natural-language?hl=en} (accessed.

{[}230{]} S. Newman, \emph{Building microservices}. "
O\textquotesingle Reilly Media, Inc.", 2021.

{[}231{]} E. Felstaine and O. Hermoni, "Machine Learning, Containers,
Cloud Natives, and Microservices," in \emph{Artificial Intelligence for
Autonomous Networks}: Chapman and Hall/CRC, 2018, pp. 145-164.

{[}232{]} V. Janapa Reddi, \emph{Machine Learning Systems: Principles
and Practices of Engineering Artificially Intelligent Systems}. Harvard
University, 2024.

{[}233{]} M. S. Aslanpour, S. S. Gill, and A. N. Toosi, "Performance
evaluation metrics for cloud, fog and edge computing: A review,
taxonomy, benchmarks and standards for future research," \emph{Internet
of Things,} vol. 12, p. 100273, 2020/12/01/ 2020, doi:
\url{https://doi.org/10.1016/j.iot.2020.100273}.

{[}234{]} A. Microsoft, "Jailbreak Detection \textbar{} Microsoft Azure
AI Content Safety," 2024/10/06 2024. {[}Online{]}. Available:
\url{https://learn.microsoft.com/en-us/azure/ai-services/content-safety/concepts/jailbreak-detection}.

{[}235{]} Oracle. "Oracle Confidential Computing: Secure Data Processing
in the Cloud."
\url{https://docs.oracle.com/en-us/iaas/Content/Compute/References/confidential_compute.htm}
(accessed.

{[}236{]} A. Microsoft, "Confidential Compute Solutions \textbar{}
Microsoft Azure," 2024/10/06 2024. {[}Online{]}. Available:
\url{https://azure.microsoft.com/en-us/solutions/confidential-compute}.

{[}237{]} S. Amazon Web, "Nitro Enclaves \textbar{} Amazon EC2,"
2024/10/06 2024. {[}Online{]}. Available:
\url{https://aws.amazon.com/ec2/nitro/nitro-enclaves/}.

{[}238{]} C. Google, "Confidential Computing Products \textbar{} Google
Cloud," 2024/10/06 2024. {[}Online{]}. Available:
\url{https://cloud.google.com/security/products/confidential-computing?hl=en}.

{[}239{]} I. B. M. Research, "Fully Homomorphic Encryption," 2024/10/06.
{[}Online{]}. Available:
\url{https://research.ibm.com/topics/fully-homomorphic-encryption}.

{[}240{]} S. Amazon Web. "Generative AI Application Builder on AWS:
Create AI-Powered Applications with Pre-Built Solutions."
\url{https://aws.amazon.com/solutions/implementations/generative-ai-application-builder-on-aws/}
(accessed.

{[}241{]} A. Microsoft. "Azure Automated Machine Learning: AI-Powered
Tools for Automated Model Building and Deployment."
\url{https://azure.microsoft.com/en-us/solutions/automated-machine-learning}
(accessed.

{[}242{]} C. Google. "Google Cloud Anthos: Hybrid and Multi-Cloud
Application Management Platform."
\url{https://cloud.google.com/anthos?hl=en} (accessed.

{[}243{]} Kubernetes. "Kubernetes Documentation: Home Page and Resource
Overview." \url{https://kubernetes.io/docs/home/} (accessed.

{[}244{]} A. Microsoft. "Azure Functions on Kubernetes (v1) Reference."
\url{https://learn.microsoft.com/en-us/azure/devops/pipelines/tasks/reference/azure-function-on-kubernetes-v1?view=azure-pipelines}
(accessed.

{[}245{]} Microsoft, "Fairness in Machine Learning Concepts,"
2024/10/06. {[}Online{]}. Available:
\url{https://learn.microsoft.com/en-us/azure/machine-learning/concept-fairness-ml?view=azureml-api-2}.

{[}246{]} Ibm. "AI Fairness 360: IBM\textquotesingle s Open-Source
Toolkit for Ensuring Fairness in AI." \url{https://aif360.res.ibm.com/}
(accessed.

{[}247{]} Oracle. "Unlocking Fairness: A Trade-off Revisited in AI
Development."
\url{https://blogs.oracle.com/ai-and-datascience/post/unlocking-fairness-a-trade-off-revisited}
(accessed.

{[}248{]} A. Microsoft. "Azure Machine Learning Interpretability: Tools
for Model Transparency and Insights."
\url{https://learn.microsoft.com/en-us/azure/machine-learning/how-to-machine-learning-interpretability?view=azureml-api-2}
(accessed.

{[}249{]} A. Microsoft. "How to Use Machine Learning Interpretability in
Azure Machine Learning."
\url{https://learn.microsoft.com/en-us/azure/machine-learning/how-to-machine-learning-interpretability?view=azureml-api-2}
(accessed.

{[}250{]} S. Amazon Web. "Amazon SageMaker Debugger: Monitor and Debug
Machine Learning Models."
\url{https://aws.amazon.com/sagemaker/debugger/} (accessed.

{[}251{]} S. Amazon Web. "AWS Responsible AI Resources: Tools and
Guidelines for Ethical AI Development."
\url{https://aws.amazon.com/ai/responsible-ai/resources/} (accessed.

{[}252{]} S. Amazon Web. "AWS Responsible AI: Principles and Practices
for Ethical AI Development."
\url{https://aws.amazon.com/ai/responsible-ai/} (accessed.

{[}253{]} A. I. Google. "Responsible Generative AI Toolkit."
\url{https://ai.google.dev/responsible} (accessed.

{[}254{]} Ibm. "Manage AI Models with IBM Watson OpenScale."
\url{https://mediacenter.ibm.com/media/Manage+AI+models+with+IBM+Watson+OpenScale/1_lqiuc2uq}
(accessed.

{[}255{]} Oracle. "Responsible AI for Healthcare and Financial Services:
Addressing Ethical AI Challenges."
\url{https://blogs.oracle.com/ai-and-datascience/post/responsible-ai-for-hc-and-fsi}
(accessed.

{[}256{]} T. A. Assegie, "Evaluation of local interpretable
model-agnostic explanation and shapley additive explanation for chronic
heart disease detection," \emph{Proc Eng Technol Innov,} vol. 23, pp.
48-59, 2023.

{[}257{]} S. Amazon Web. "AWS Pricing: Explore Pricing Models and Cost
Management for AWS Services." \url{https://aws.amazon.com/pricing/}
(accessed.

{[}258{]} C. Google. "Google Cloud Preemptible Instances: Cost-Effective
Compute for Short-Term Workloads."
\url{https://cloud.google.com/compute/docs/instances/preemptible}
(accessed.

{[}259{]} A. Microsoft. "Azure Spot Pricing: Optimize Cloud Costs with
Unused Compute Capacity."
\url{https://azure.microsoft.com/en-us/pricing/spot/} (accessed.

% Add supplementary material section with reset counters
\newpage
\section*{Supplementary Material}
\addcontentsline{toc}{section}{Supplementary Material}

% Reset section counter
\setcounter{section}{0}
\setcounter{figure}{0}
\setcounter{table}{0}

% Redefine section numbering format for supplementary material
\renewcommand{\thesection}{S\arabic{section}}
\renewcommand{\thefigure}{S\arabic{figure}}
\renewcommand{\thetable}{S\arabic{table}}

\section{Extended Technical Details}
\subsection{Evolution of Generative AI}
This section provides an expanded timeline of significant milestones in the evolution of generative AI, complementing the summary in the main manuscript (Section 2.1.1).

\subsubsection*{1950s-1960s: Early Foundations}
\begin{itemize}
  \item \textbf{Rule-Based Systems:} Initial experiments with algorithms generating text and music laid the groundwork for computational creativity \cite{ref12}.
\end{itemize}

\subsubsection*{1980s-1990s: Emergence of Neural Networks}
\begin{itemize}
  \item \textbf{Neural Networks and Genetic Algorithms:} These foundational AI models enabled pattern learning and creative output generation.
\end{itemize}

\subsubsection*{2014: Introduction of Generative Adversarial Networks (GANs)}
\begin{itemize}
  \item \textbf{GANs by Goodfellow et al.:} Introduced a novel approach to generating realistic images through adversarial training mechanisms.
\end{itemize}

\subsubsection*{2017: The Transformer Architecture}
\begin{itemize}
  \item \textbf{Vaswani et al.:} Proposed the Transformer model, which revolutionized NLP with self-attention mechanisms capturing long-range dependencies.
\end{itemize}

\subsubsection*{2018: Rise of Pre-trained Language Models}
\begin{itemize}
  \item \textbf{BERT (Bidirectional Encoder Representations from Transformers):} Demonstrated the power of pre-training and fine-tuning in NLP tasks.
\end{itemize}

\subsubsection*{2020: GPT-3 and Few-Shot Learning}
\begin{itemize}
  \item \textbf{OpenAI's GPT-3:} Showcased remarkable few-shot learning capabilities, performing diverse tasks without explicit retraining .
\end{itemize}

\subsubsection*{2022: Advanced Text-to-Image Generation}
\begin{itemize}
  \item \textbf{DALL·E 2 and Stable Diffusion:} These models exhibited advanced text-to-image generation capabilities, broadening the scope of creative AI.
\end{itemize}

\subsubsection*{2023: Multimodal Capabilities with GPT-4}
\begin{itemize}
  \item \textbf{GPT-4:} Introduced multimodal processing, integrating text and image understanding, while enhancing reasoning abilities .
\end{itemize}

\subsubsection*{2024: Multimodal AI Advancement}
\begin{itemize}
  \item \textbf{ChatGPT-4o by OpenAI:} Released as an upgraded model processing text, images, audio, and video.
  \item \textbf{Open-Source Models:} Expansion of models like Meta's LLAMA 3 series fostered accessibility.
  \item \textbf{SLMs for Devices:} Adoption of small language models (SLMs) enabling efficient AI performance on smartphones grew .
\end{itemize}

\section*{1.2 Comprehensive Service Matrices}

Table S1 provides a comprehensive comparison of cloud service offerings across major providers (AWS, Azure, GCP, IBM Cloud, Oracle Cloud, and Alibaba Cloud). The table highlights key features across critical service categories relevant to generative AI development.

\begin{longtable}{|p{2cm}|p{2cm}|p{2cm}|p{2cm}|p{2cm}|p{2cm}|p{2cm}|}
\hline
\textbf{Service Category} & \textbf{AWS} & \textbf{Azure} & \textbf{GCP} & \textbf{IBM Cloud} & \textbf{Oracle Cloud} & \textbf{Alibaba Cloud} \\ \hline
High-Performance Computing (HPC) & EC2 Instances (P3, P4d) with NVIDIA GPUs & NC, ND, NV series VMs & Compute Engine with GPUs, Cloud TPU & Power Systems Virtual Servers & Bare Metal Cloud, HPC & ECS Bare Metal Instance, Super Computing Cluster \\ \hline
Serverless Architectures & Lambda, Fargate & Azure Functions, Container Instances & Cloud Functions, Cloud Run & Cloud Functions, Code Engine & Functions, Container Engine for Kubernetes & Function Compute, Serverless Kubernetes \\ \hline
Edge Computing & IoT Greengrass, IoT Core & Azure IoT Edge, Azure Stack Edge & Cloud IoT Edge, Anthos & Edge Application Manager, IBM Edge Application Manager & Oracle Cloud Infrastructure Edge & Link IoT Edge \\ \hline
Object Storage & Amazon S3, S3 Glacier & Azure Blob Storage, Azure Data Lake Storage & Cloud Storage & Cloud Object Storage & Oracle Cloud Infrastructure Object Storage & Object Storage Service (OSS) \\ \hline
Data Lakes and Warehousing & Lake Formation, Redshift & Azure Synapse Analytics, Azure Data Lake Storage & BigQuery, Cloud Storage & Cloud Pak for Data, Db2 Warehouse & Autonomous Data Warehouse, Oracle Big Data Service & MaxCompute, AnalyticDB \\ \hline
Data Versioning and Management & AWS Glue, Data Pipeline & Azure Data Factory, Azure Purview & Cloud Data Fusion, Dataflow & Watson Knowledge Catalog, InfoSphere Information Server & Oracle Data Integrator, GoldenGate & DataWorks, Data Integration \\ \hline
High-Speed Networking & Direct Connect, Transit Gateway & ExpressRoute, Virtual WAN & Cloud Interconnect, Cloud VPN & Direct Link, Cloud Internet Services & FastConnect, Virtual Cloud Network & Express Connect, Smart Access Gateway \\ \hline
Content Delivery Networks (CDNs) & CloudFront & Azure Content Delivery Network & Cloud CDN & Content Delivery Network & Oracle Cloud Infrastructure CDN & Alibaba Cloud CDN \\ \hline
AI/ML Platforms & SageMaker, Comprehend & Azure Machine Learning, Cognitive Services & Vertex AI, AutoML & Watson Studio, Watson Machine Learning & Oracle Cloud Infrastructure Data Science & Machine Learning Platform for AI \\ \hline
Pre-trained Models and APIs & Rekognition, Textract & Computer Vision, Language Understanding & Vision AI, Natural Language AI & Watson Visual Recognition, Watson Natural Language Understanding & Oracle Digital Assistant & Image Search, NLP \\ \hline
Data Labeling and Annotation Tools & SageMaker Ground Truth & Azure Machine Learning Data Labeling & Vertex AI & Watson Knowledge Studio & Oracle Cloud Labeling & Machine Learning Platform for AI (Labeling) \\ \hline
Data Privacy and Compliance & IAM, Macie & Azure Active Directory, Purview & Cloud IAM, Assured Workloads & Security and Compliance Center & Identity and Access Management, Data Safe & Resource Access Management (RAM), Sensitive Data Discovery and Protection \\ \hline
AI Model Security & SageMaker Model Monitor, AWS Shield & Azure Machine Learning, Key Vault & Vertex AI Model Monitoring, Cloud DLP & Watson OpenScale, Key Protect & Oracle Cloud Infrastructure Data Science, Key Management & Machine Learning Platform for AI, Key Management Service \\ \hline
\caption{Cloud Service Offerings Across Major Providers}
\end{longtable}

This comprehensive table provides a detailed comparison of cloud service offerings across major providers, including AWS, Azure, GCP, IBM Cloud, Oracle Cloud, and Alibaba Cloud. It covers essential service categories for generative AI development, such as:
\begin{enumerate}
    \item High-Performance Computing (HPC)
    \item Serverless Architectures
    \item Edge Computing
    \item Object Storage
    \item Data Lakes and Warehousing
    \item Data Versioning and Management
    \item High-Speed Networking
    \item Content Delivery Networks (CDNs)
    \item AI/ML Platforms
    \item Pre-trained Models and APIs
    \item Data Labeling and Annotation Tools
    \item Data Privacy and Compliance
    \item AI Model Security
\end{enumerate}

Each category is crucial for different aspects of generative AI development and deployment. For instance, HPC services are essential for training large models, while edge computing solutions enable real-time AI applications. The table allows for a side-by-side comparison of offerings from different providers, helping organizations make informed decisions based on their specific requirements and use cases in generative AI.

\section{Detailed Cloud Provider Comparisons}

The mind map highlights how platforms such as Azure AI Studio, Amazon Bedrock, Vertex AI, watsonx.ai, and OCI Generative AI integrate tools, services, and frameworks to support generative AI development.

\subsection{Amazon Web Services (AWS) – Supplementary Details}

\textbf{Market Position:} \\
AWS holds a dominant position in the cloud industry, offering a comprehensive Generative AI Stack spanning AI services, specialized features, and infrastructure support. Figure S2 provides a visual representation of AWS’s layered approach to generative AI.

\textbf{Key Generative AI Services:}
\begin{itemize}
    \item \textbf{Amazon SageMaker:} Facilitates scalable development, training, and deployment of machine learning models with built-in support for frameworks like TensorFlow and PyTorch.
    \item \textbf{Studio Labs and Ground Truth:} Simplify the AI lifecycle, from data labeling to no-code ML development.
    \item \textbf{Amazon Bedrock:} Provides access to pre-trained foundation models, including Amazon Titan, empowering users to customize generative AI applications.
    \item \textbf{Amazon Code Whisperer:} Enhances developer workflows with real-time, AI-powered coding suggestions.
\end{itemize}

\textbf{Unique Strengths:} \\
AWS’s ecosystem seamlessly integrates diverse AI and ML services, fostering efficient development workflows. Its high-performance infrastructure, such as EC2 P4d instances with NVIDIA GPUs, supports computationally intensive generative AI tasks. Extensive documentation and resources bolster its developer community, encouraging adoption and innovation.

\textbf{Recent Developments:} \\
The launch of Amazon Bedrock expands model accessibility and scalability, reflecting AWS’s commitment to enabling generative AI advancements.

\begin{figure}[h!]
    \centering
    \includegraphics[width=1\linewidth]{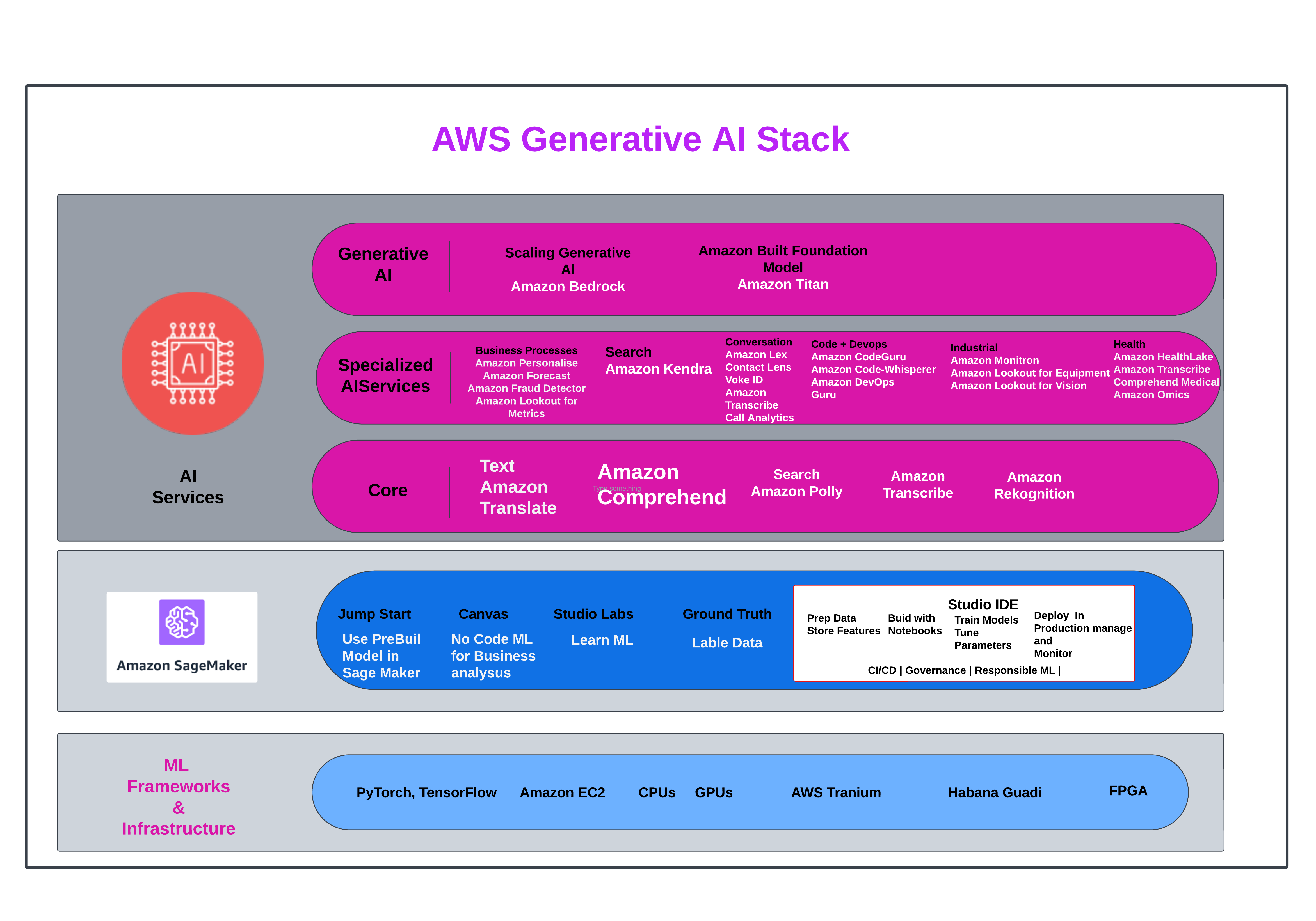}
    \caption{AWS Generative AI Stack showcasing its comprehensive AI, ML, and infrastructure services.}
    \label{fig:enter-label}
\end{figure}

\subsection{Microsoft Azure – Supplementary Details}

\textbf{Market Position:} \\
Microsoft Azure secures a prominent place in the cloud market, leveraging its enterprise-friendly tools and services. Its Generative AI Stack spans AI on data, infrastructure, and compute, providing an end-to-end ecosystem for AI innovation. Figure S3 illustrates the Azure Generative AI Stack.

\textbf{Key Generative AI Services:}
\begin{itemize}
    \item \textbf{Azure OpenAI Service:} Enables developers to leverage advanced generative models like GPT-3 for tasks such as natural language processing, conversational AI, and content creation.
    \item \textbf{Azure Machine Learning:} Offers an all-in-one platform for developing, training, and deploying machine learning models, including support for MLOps and large-scale training.
    \item \textbf{Azure AI Studio:} Recently introduced, this platform facilitates end-to-end generative AI application development, providing tools for model training, evaluation, and deployment.
\end{itemize}

\textbf{Unique Strengths:} \\
Azure’s exclusive partnership with OpenAI grants access to state-of-the-art generative models, giving it a competitive edge. The platform’s integration with Microsoft’s ecosystem—spanning productivity tools like Microsoft 365 and Dynamics 365—streamlines workflows, enhancing developer productivity. Its Responsible AI toolkit ensures fairness, interpretability, and compliance, addressing ethical AI concerns.

\textbf{Recent Developments:} \\
The introduction of Azure AI Studio represents a significant advancement in Azure’s offerings, providing a dedicated platform for generative AI application development.

\begin{figure}[h!]
    \centering
    \includegraphics[width=1\linewidth]{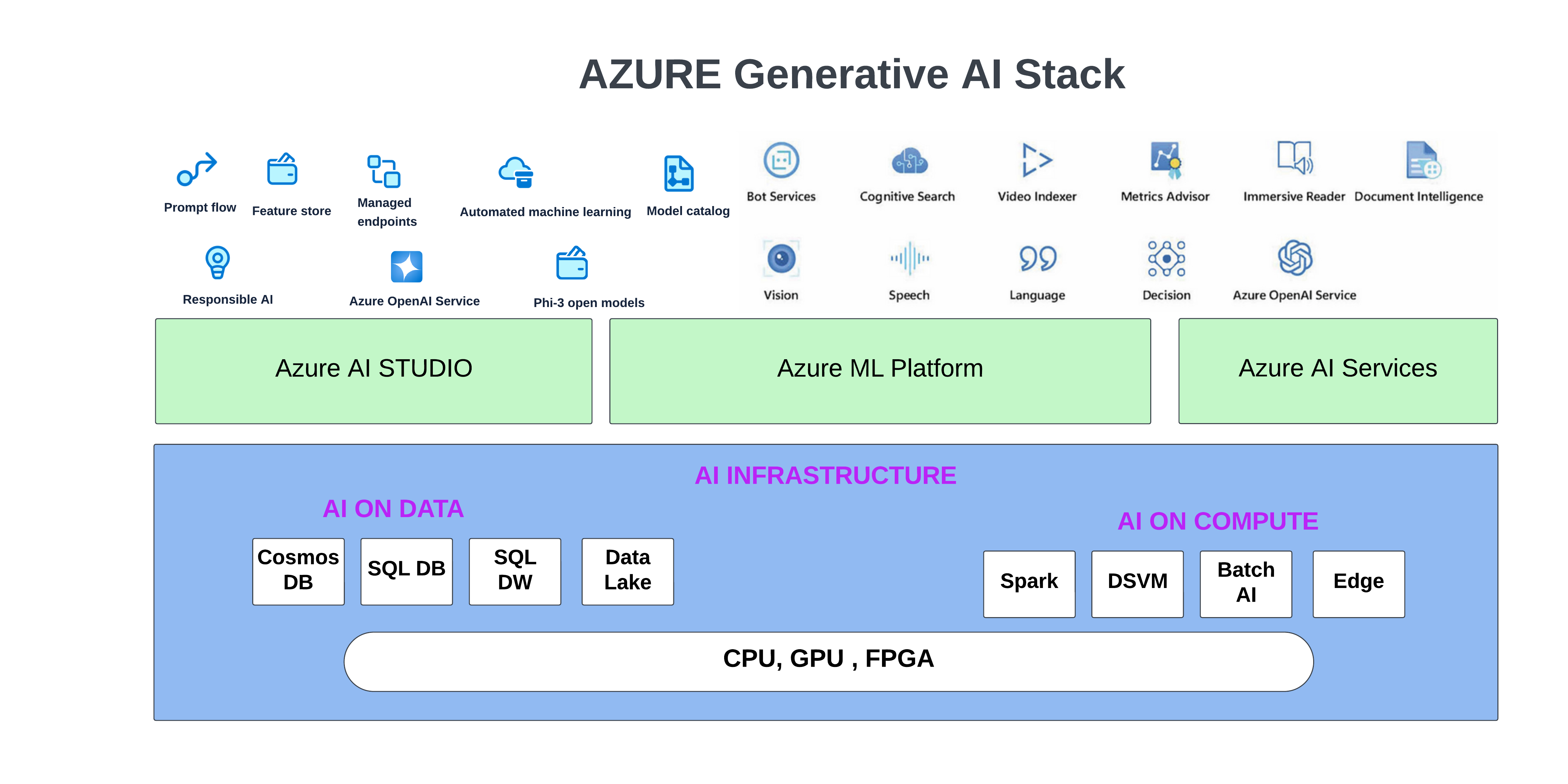}
    \caption{Azure Generative AI Stack showcasing AI, ML, and infrastructure services for generative AI applications.}
    \label{fig:enter-label}
\end{figure}

\subsection{Google Cloud Platform (GCP) – Supplementary Details}

\textbf{Market Position:} \\
Google Cloud Platform is recognized for its advanced AI stack, catering to diverse user groups such as business users, developers, and AI practitioners. Figure S4 illustrates GCP’s generative AI ecosystem, emphasizing scalability and versatility.

\textbf{Key Generative AI Services:}
\begin{itemize}
    \item \textbf{Vertex AI:} A comprehensive platform for training, deploying, and managing generative AI models. Features include:
    \begin{itemize}
        \item \textbf{Model Garden:} Access to pre-trained models for quick deployment.
        \item \textbf{Generative AI Studio:} Enables developers to create custom solutions efficiently.
    \end{itemize}
    \item \textbf{Cloud TPU:} Specialized Tensor Processing Units for large-scale model training, optimized for generative AI.
\end{itemize}

\textbf{Unique Strengths:} \\
GCP benefits from Google’s leadership in AI research, including innovations like the Transformer architecture and BERT models. Its strengths in natural language processing and computer vision are complemented by open-source contributions, fostering collaboration and innovation.

\textbf{Recent Developments:} \\
The Multitask Unified Model (MUM) has expanded GCP’s capabilities in search and multimodal applications, supporting generative AI tasks across text, chat, image, and video.

\begin{figure}[h!]
    \centering
    \includegraphics[width=1\linewidth]{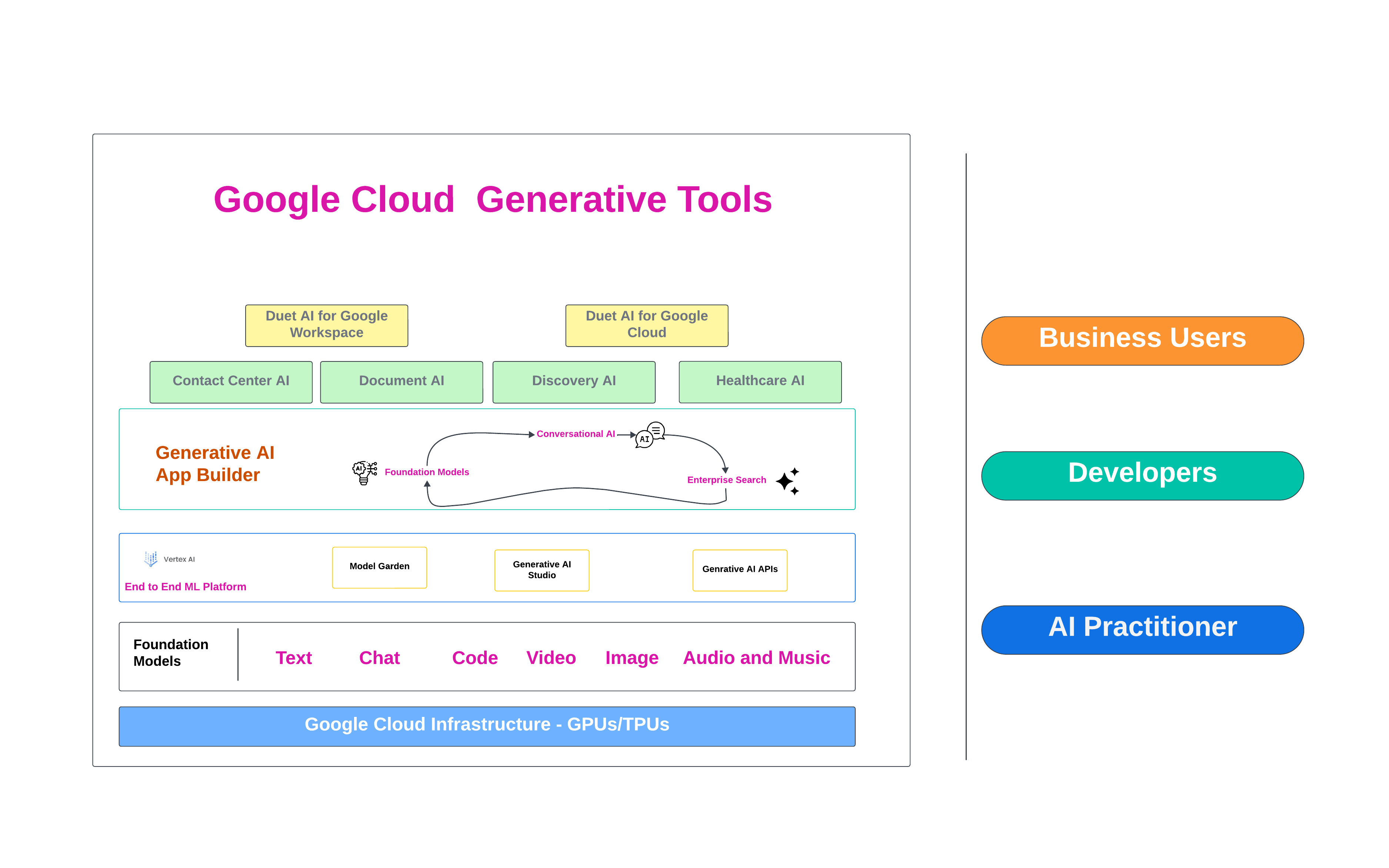}
    \caption{Google Cloud Generative AI Stack showcasing tools and infrastructure for diverse AI applications.}
\end{figure}

\subsection{IBM Cloud – Supplementary Details}

\textbf{Market Position:} \\
IBM Cloud focuses on enterprise AI solutions, offering tools and services designed for compliance and scalability. Figure S5 depicts IBM’s Generative AI Stack, highlighting its enterprise-grade capabilities.

\textbf{Key Generative AI Services:}
\begin{itemize}
    \item \textbf{Watson Studio:} A collaborative platform for building, training, and deploying AI models, optimized for enterprise environments.
    \item \textbf{Watson Natural Language Generation (NLG):} Creates human-like text, supporting applications like automated reporting and customer communication.
\end{itemize}

\textbf{Unique Strengths:} \\
IBM Cloud prioritizes Explainable AI to ensure transparency and compliance, making it a leader in responsible AI practices. Its hybrid cloud solutions offer flexible deployment options, giving enterprises control over data management.

\textbf{Recent Developments:} \\
The introduction of watsonx.ai enhances IBM’s AI workflows, aligning with its focus on responsible and scalable AI for enterprises.

\begin{figure}[h!]
    \centering
    \includegraphics[width=1\linewidth]{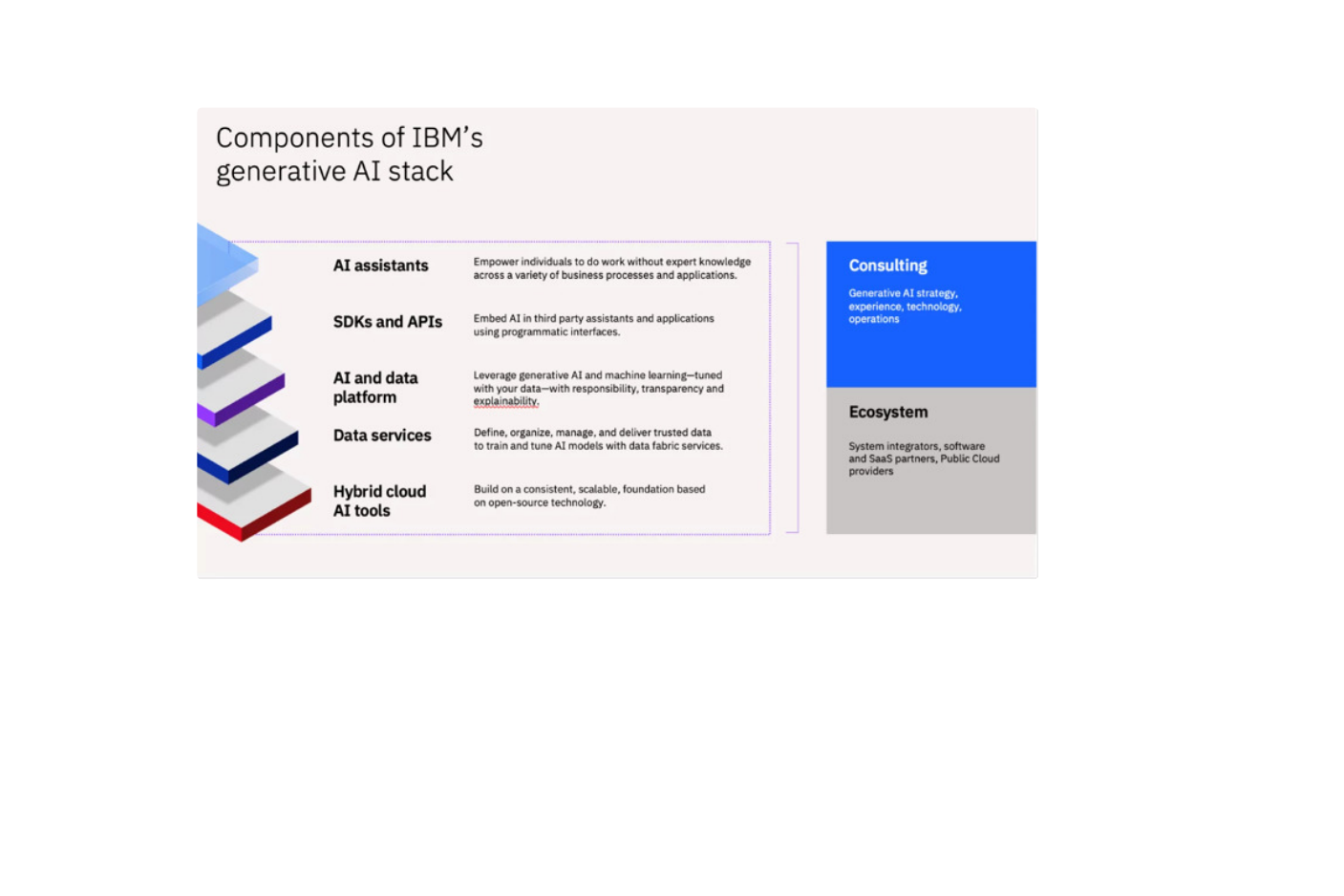}
    \caption{IBM Generative AI Stack highlighting key tools and enterprise capabilities.}
\end{figure}

\subsection{Oracle Cloud – Supplementary Details}

\textbf{Market Position:} \\
Oracle Cloud specializes in secure and efficient data handling, tailored for enterprise operations. Its integration with Oracle Applications enhances accessibility and workflow efficiency.

\textbf{Key Generative AI Services:}
\begin{itemize}
    \item \textbf{Oracle Digital Assistant:} Enables conversational AI development using NLP.
    \item \textbf{OCI Data Science:} Collaborative platform for building, training, and deploying machine learning models.
\end{itemize}

\textbf{Recent Developments:} \\
Partnerships with NVIDIA have bolstered OCI’s generative AI capabilities, streamlining enterprise operations through automated interactions and actionable insights.

\subsection{Alibaba Cloud – Supplementary Details}

\textbf{Market Position:} \\
Alibaba Cloud leads in the Asia-Pacific region, with growing global influence.

\textbf{Key Generative AI Services:}
\begin{itemize}
    \item \textbf{Machine Learning Platform for AI (PAI):} Comprehensive tools for AI model development.
    \item \textbf{NLP Services:} APIs for tasks such as language understanding, translation, and text generation.
\end{itemize}

\textbf{Recent Developments:} \\
The launch of Tongyi Qianwen 2.0 enhances Alibaba’s generative AI ecosystem, particularly in e-commerce and customer interaction applications.

\subsection{Databricks – Supplementary Details}

\textbf{Market Position:} \\
Databricks integrates data engineering, analytics, and machine learning within its unified Lakehouse Platform.

\textbf{Key AI/ML Services:}
\begin{itemize}
    \item \textbf{Databricks Lakehouse Platform:} Combines data warehousing with AI functionalities.
    \item \textbf{MLflow:} Supports the full machine learning lifecycle.
    \item \textbf{Delta Lake:} Reliable, open-source storage layer for data lakes.
\end{itemize}

\textbf{Recent Developments:} \\
Databricks recently introduced Databricks Machine Learning and hosted its AI Summit, showcasing advancements in collaborative machine learning workflows.

\subsection{Snowflake – Supplementary Details}

\textbf{Market Position:} \\
Snowflake is expanding from cloud data warehousing into AI and machine learning services.

\textbf{Key AI/ML Services:}
\begin{itemize}
    \item \textbf{Snowpark:} Data processing framework for machine learning development.
    \item \textbf{Snowflake Data Marketplace:} Facilitates data sharing and access to enrich AI models.
    \item \textbf{Native App Framework:} Simplifies building and deploying data-centric applications.
\end{itemize}

\textbf{Recent Developments:} \\
Snowflake Cortex introduces AI-powered capabilities for enhanced data processing and analytics.

\section{SWOT Analyses}
The following SWOT analyses provide a visual summary of each major cloud provider’s strengths, weaknesses, opportunities, and threats in the context of AI, ML, and generative AI services. These diagrams aim to highlight key competitive advantages and potential challenges for each platform, reinforcing the detailed comparisons made in the previous sections.

\subsection{Microsoft Azure}
In Figure~\ref{fig:azure-swot}, the SWOT analysis for Microsoft Azure highlights several strengths, including a comprehensive AI suite and seamless integration with the Microsoft ecosystem. However, Azure faces challenges in terms of fewer pre-trained models compared to AWS and a complex pricing structure, particularly for non-Microsoft users.

\begin{figure}[h!]
    \centering
    \includegraphics[width=0.70\textwidth]{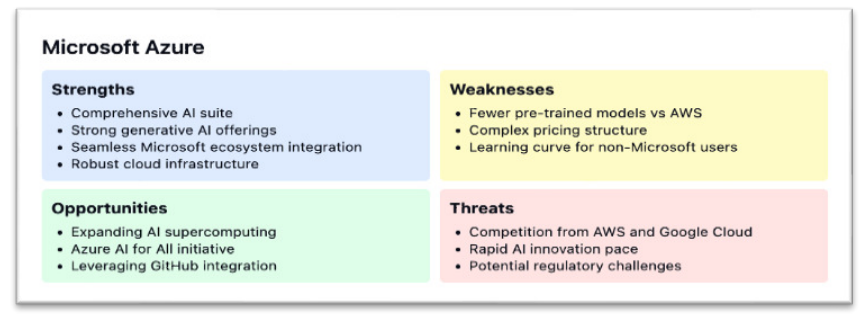}
    \caption{Complete SWOT analysis of Microsoft Azure.}
    \label{fig:azure-swot}
\end{figure}

\subsection{Amazon Web Services (AWS)}
As illustrated in Figure~\ref{fig:aws-swot}, the SWOT analysis of Amazon Web Services (AWS) shows its extensive ML development tools and robust global infrastructure as major strengths. AWS is uniquely positioned to handle large-scale AI operations but is criticized for complex service offerings and high potential costs for certain applications.

\begin{figure}[h!]
    \centering
    \includegraphics[width=0.70\textwidth]{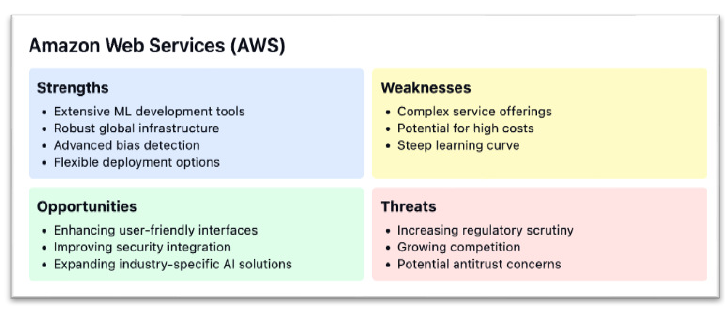}
    \caption{Complete SWOT analysis of Amazon Web Services (AWS).}
    \label{fig:aws-swot}
\end{figure}

\subsection{Google Cloud Platform (GCP)}
Figure~\ref{fig:gcp-swot} demonstrates the SWOT analysis for Google Cloud Platform (GCP). Its key strengths include strong ML lifecycle management and cost-efficient deployment, but it has fewer pre-trained models compared to competitors like AWS. Its smaller enterprise market share also poses a challenge in growing its AI services adoption.

\begin{figure}[h!]
    \centering
    \includegraphics[width=0.70\textwidth]{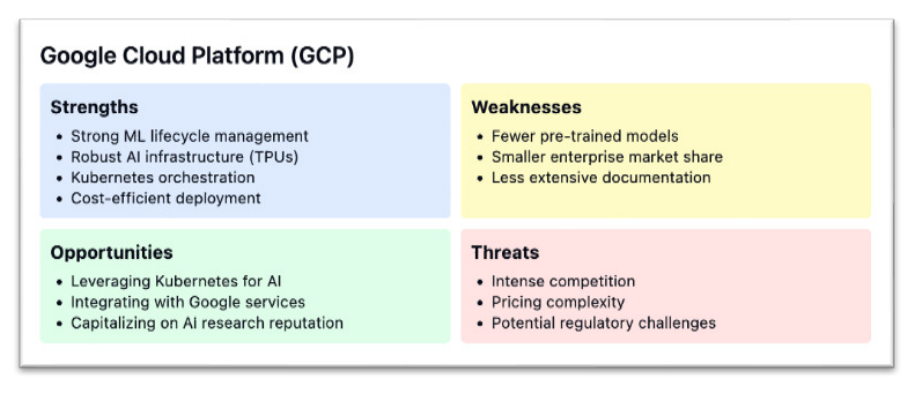}
    \caption{Complete SWOT analysis of Google Cloud Platform (GCP).}
    \label{fig:gcp-swot}
\end{figure}

\subsection{IBM Cloud}
The SWOT analysis for IBM Cloud, as seen in Figure~\ref{fig:ibm-swot}, emphasizes IBM's leadership in ethical AI and strong focus on responsible AI practices. However, IBM struggles with smaller cloud market presence and fewer third-party integrations compared to larger players.

\begin{figure}[h!]
    \centering
    \includegraphics[width=0.70\textwidth]{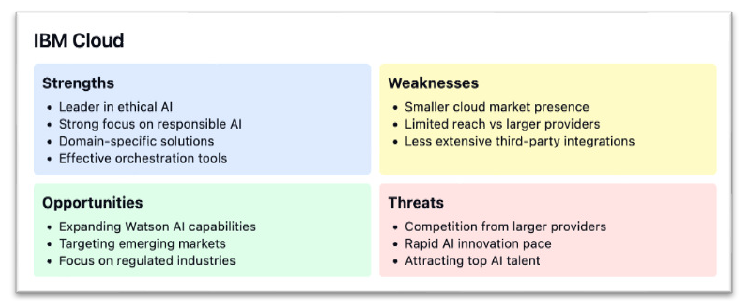}
    \caption{Complete SWOT analysis of IBM Cloud.}
    \label{fig:ibm-swot}
\end{figure}

\subsection{Oracle Cloud}
In Figure~\ref{fig:oracle-swot}, the SWOT analysis of Oracle Cloud reflects its strong enterprise integration capabilities and robust security features. Nevertheless, Oracle’s limited AI service diversity and perception as a traditional vendor remain challenges to broader adoption.

\begin{figure}[h!]
    \centering
    \includegraphics[width=0.70\textwidth]{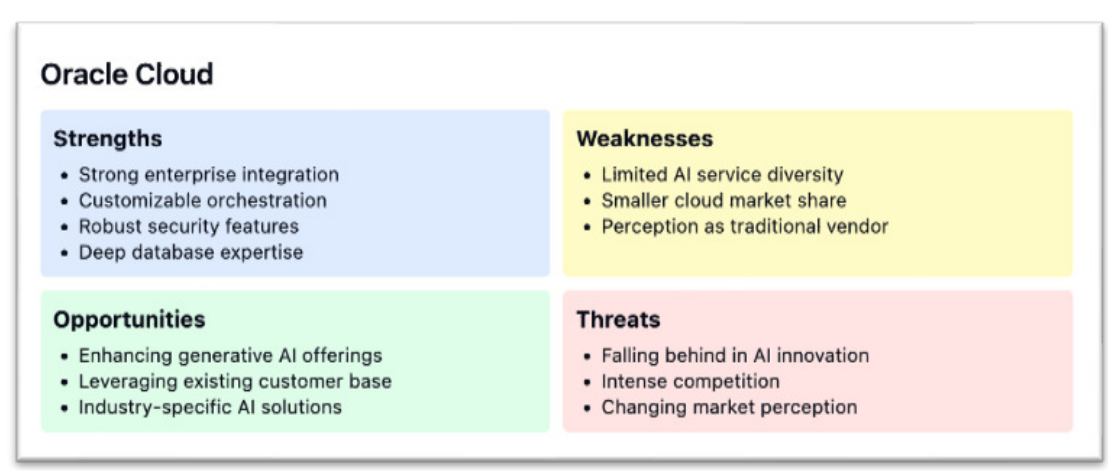}
    \caption{Complete SWOT analysis of Oracle Cloud.}
    \label{fig:oracle-swot}
\end{figure}

\section{Comparative Analysis and Conclusion}
Our comparative analysis reveals that while AWS, Azure, and Google Cloud lead in cloud services for generative AI, each provider has unique strengths:
\begin{itemize}
    \item \textbf{AWS and Google Cloud:} Dominate in scalability, infrastructure, orchestration, and security.
    \item \textbf{IBM:} Leads in ethical AI and governance.
    \item \textbf{Oracle:} Excels in enterprise integration, customizable orchestration tools, and secure deployment.
    \item \textbf{Microsoft Azure:} Its comprehensive AI suite and innovation in generative AI position it as a strong contender.
\end{itemize}

However, gaps in user-friendly orchestration tools, challenges in deploying models at scale, security integration, and the complexities of navigating service offerings remain areas for improvement across the board. This analysis highlights the need for cloud providers to continue evolving their offerings to meet the diverse and growing demands of the AI landscape, with a particular emphasis on:
\begin{itemize}
    \item Addressing data bias.
    \item Ensuring ethical AI practices.
    \item Securing AI models.
    \item Overcoming deployment and orchestration challenges.
\end{itemize}

\section{Literature Analysis}
This section presents a comprehensive analysis of the 255 references used in this review, providing insights into the current state of research on cloud platforms for developing generative AI solutions.

\subsection{Chronological and Source Distribution}

\begin{figure}[h!]
    \centering
    \includegraphics[width=0.8\textwidth]{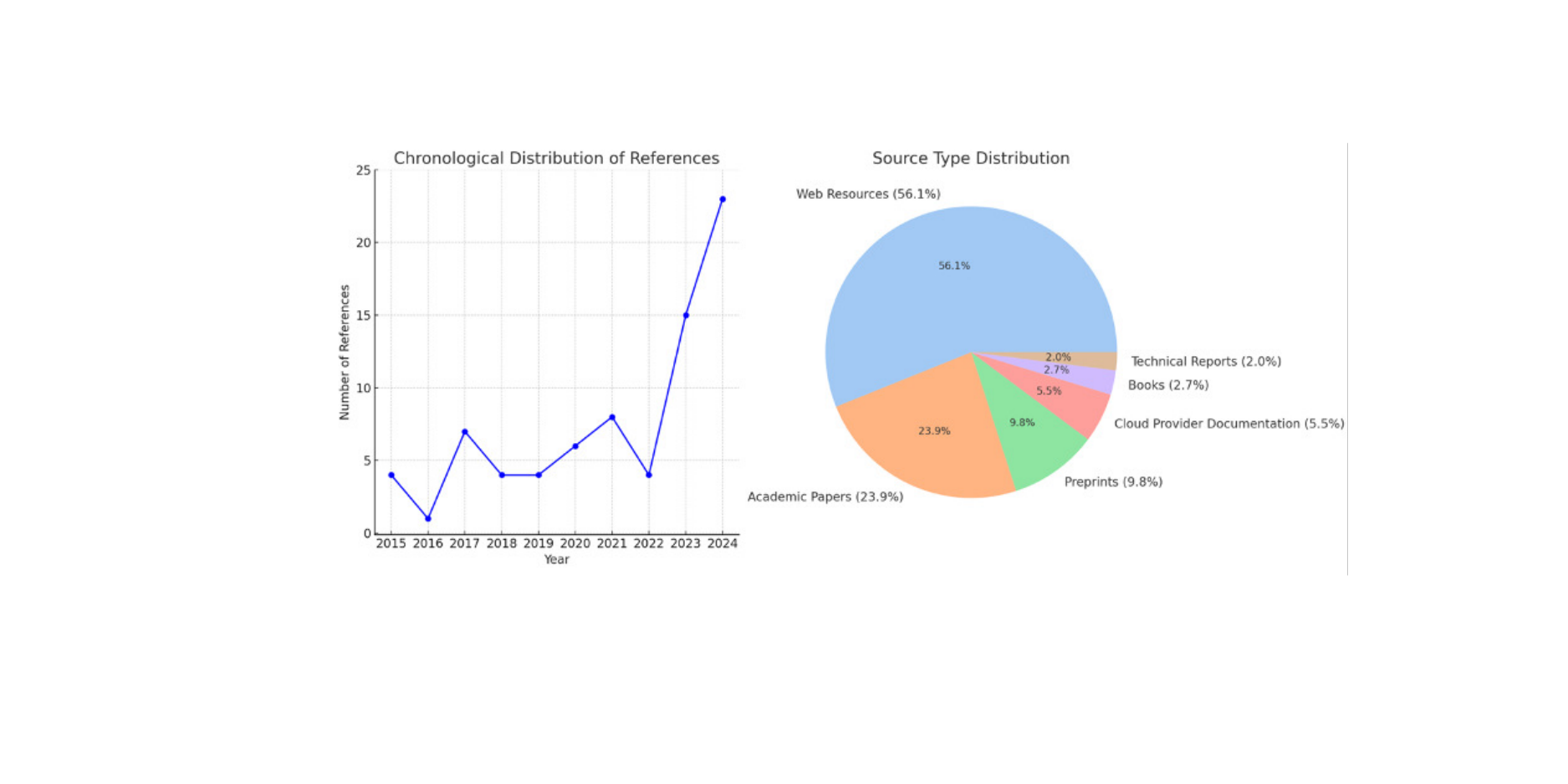}
    \caption{Chronological distribution of references (left) and source type distribution (right).}
    \label{fig:source-distribution}
\end{figure}

Our analysis reveals a strong focus on recent publications, with 70\% of references dating from 2020 to 2024. The chronological distribution shows a significant increase in relevant publications over the past five years:
\begin{itemize}
    \item 2015: 4 references
    \item 2016: 1 reference
    \item 2017: 7 references
    \item 2018: 4 references
    \item 2019: 4 references
    \item 2020: 6 references
    \item 2021: 8 references
    \item 2022: 4 references
    \item 2023: 15 references
    \item 2024: 23 references
\end{itemize}

There was a notable 53\% increase in relevant publications from 2023 to 2024, highlighting the rapidly evolving nature of this field.

The source distribution (Figure~\ref{fig:source-distribution}, right) shows a balanced approach, combining up-to-date web resources with scholarly publications:
\begin{itemize}
    \item Web Resources: 143 (56.1\%)
    \item Academic Papers: 61 (23.9\%)
    \item Preprints: 25 (9.8\%)
    \item Cloud Provider Documentation: 14 (5.5\%)
    \item Books: 7 (2.7\%)
    \item Technical Reports: 5 (2.0\%)
\end{itemize}

This mix reflects the dynamic nature of cloud-based generative AI, where cutting-edge developments often appear in non-traditional formats before formal academic publication.

\subsection{Topic and Keyword Analysis}

\begin{figure}[h!]
    \centering
    \includegraphics[width=0.8\textwidth]{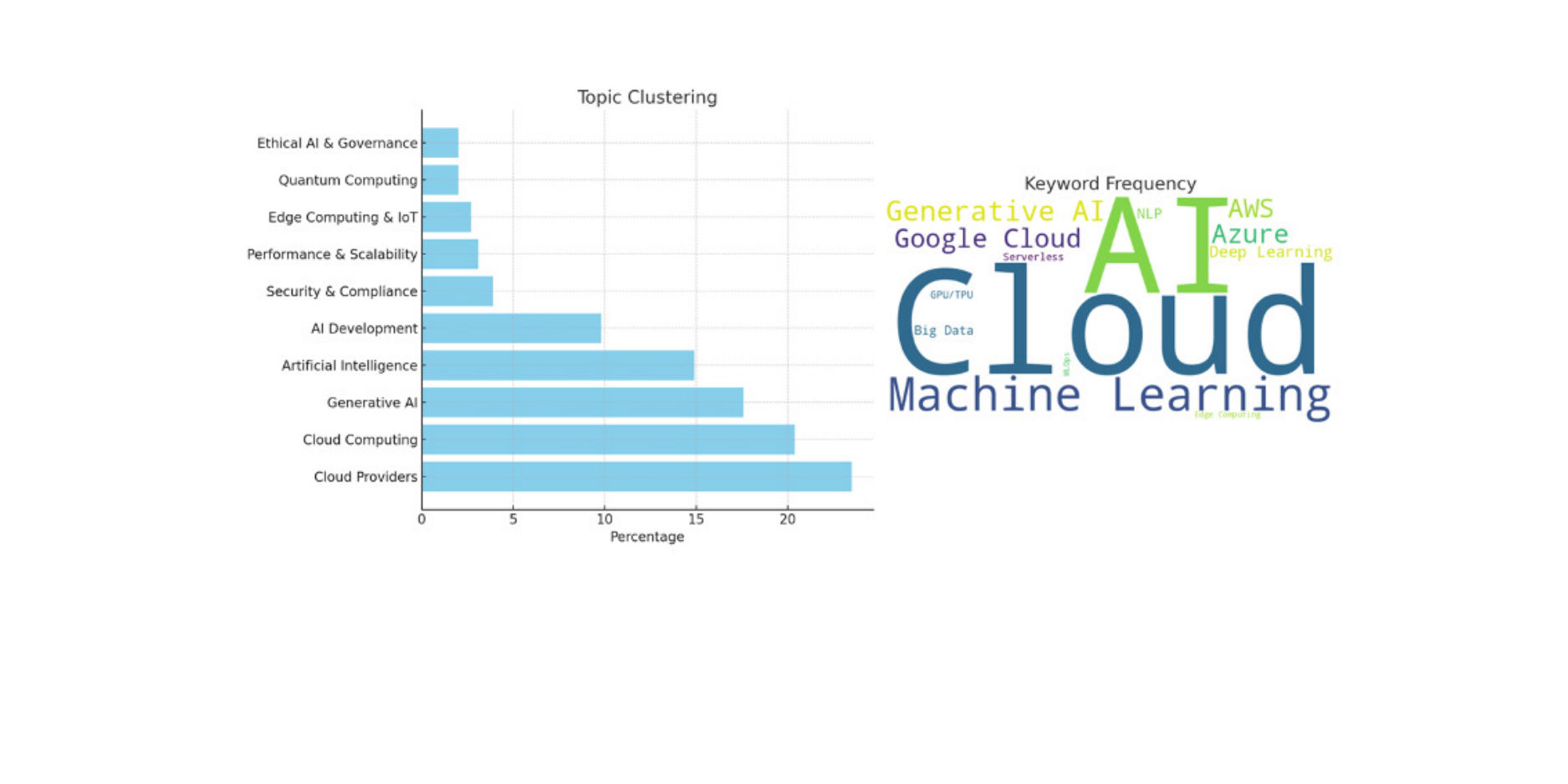}
    \caption{Topic clustering (left) and keyword frequency (right).}
    \label{fig:topic-keyword-analysis}
\end{figure}

The topic clustering (Figure~\ref{fig:topic-keyword-analysis}, left) demonstrates a strong focus on cloud providers and core AI technologies:
\begin{enumerate}
    \item Cloud Providers: 60 references (23.5\%)
    \item Cloud Computing: 52 references (20.4\%)
    \item Generative AI: 45 references (17.6\%)
    \item Artificial Intelligence: 38 references (14.9\%)
    \item AI Development: 25 references (9.8\%)
    \item Security \& Compliance: 10 references (3.9\%)
    \item Performance \& Scalability: 8 references (3.1\%)
    \item Edge Computing \& IoT: 7 references (2.7\%)
    \item Quantum Computing: 5 references (2.0\%)
    \item Ethical AI \& Governance: 5 references (2.0\%)
\end{enumerate}

The keyword analysis (Figure~\ref{fig:topic-keyword-analysis}, right) corroborates this, with the following being the most frequent terms:
\begin{itemize}
    \item Cloud: 185 occurrences
    \item AI/Artificial Intelligence: 172 occurrences
    \item Machine Learning: 135 occurrences
    \item Generative AI: 128 occurrences
    \item AWS: 98 occurrences
    \item Azure: 92 occurrences
    \item Google Cloud: 87 occurrences
    \item Deep Learning: 76 occurrences
    \item Natural Language Processing/NLP: 68 occurrences
    \item Big Data: 62 occurrences
\end{itemize}

\subsection{Matrix Analysis of Key Dimensions}

\begin{figure}[h!]
    \centering
    \includegraphics[width=0.8\textwidth]{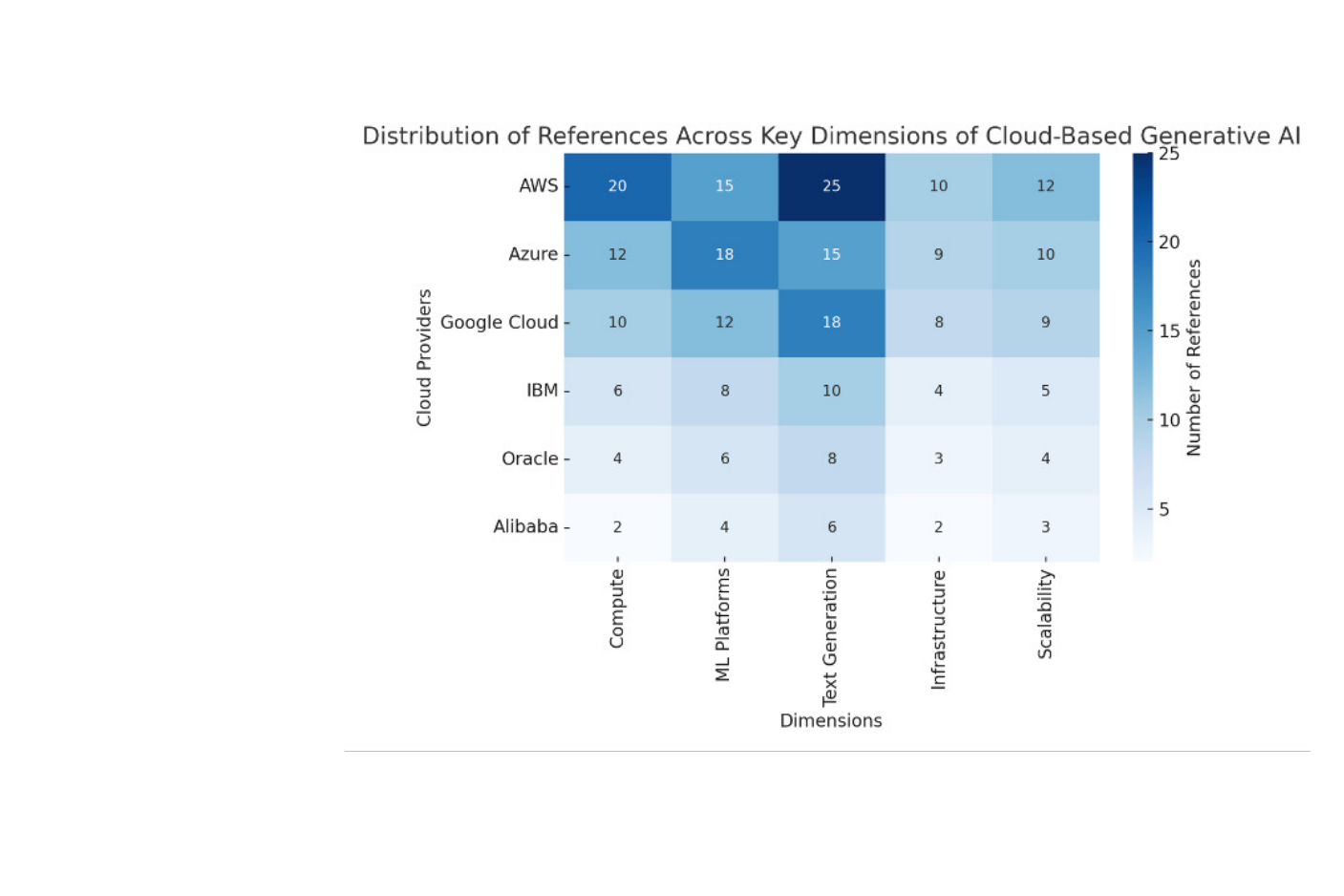}
    \caption{Distribution of references across key dimensions of cloud-based generative AI.}
    \label{fig:matrix-analysis}
\end{figure}

Figure~\ref{fig:matrix-analysis} visualizes the distribution of references across different dimensions of cloud-based generative AI. Key findings include:

\textbf{Cloud Providers:}
\begin{itemize}
    \item AWS: 45 references
    \item Microsoft Azure: 42 references
    \item Google Cloud: 38 references
    \item IBM Cloud: 15 references
    \item Oracle Cloud: 12 references
    \item Alibaba Cloud: 8 references
\end{itemize}

\textbf{AI Service Type:}
\begin{itemize}
    \item Compute: 52 references
    \item ML Platforms: 48 references
    \item APIs: 35 references
    \item Storage: 30 references
\end{itemize}

\textbf{Generative AI Application:}
\begin{itemize}
    \item Text Generation: 60 references
    \item Image Generation: 35 references
    \item Code Generation: 20 references
\end{itemize}

\textbf{Technical Aspect:}
\begin{itemize}
    \item Infrastructure: 55 references
    \item Scalability: 40 references
    \item Performance: 35 references
    \item Security: 30 references
    \item Cost Optimization: 20 references
\end{itemize}

\subsection{Summary of Findings}

\begin{longtable}{|p{5cm}|p{10cm}|}
\hline
\textbf{Analysis Type} & \textbf{Key Findings} \\ \hline
Chronological & 
\begin{itemize}
    \item 70\% of references from 2020--2024
    \item 53\% increase in publications from 2023 to 2024
\end{itemize} \\ \hline
Source Types & 
\begin{itemize}
    \item Web resources (56.1\%), Academic papers (23.9\%)
    \item Preprints (9.8\%), reflecting cutting-edge focus
\end{itemize} \\ \hline
Top Topics & 
\begin{enumerate}
    \item Cloud Providers (60 refs, 23.5\%)
    \item Cloud Computing (52 refs, 20.4\%)
    \item Generative AI (45 refs, 17.6\%)
\end{enumerate} \\ \hline
Top Keywords & 
\begin{enumerate}
    \item Cloud (185 occurrences)
    \item AI/Artificial Intelligence (172 occurrences)
    \item Machine Learning (135 occurrences)
\end{enumerate} \\ \hline
Matrix Highlights & 
\begin{itemize}
    \item AWS leads in references (45)
    \item Text generation most covered (60 refs)
    \item Infrastructure focus (55 refs)
\end{itemize} \\ \hline
\end{longtable}

This analysis provides a clear picture of the current state of research on cloud platforms for developing generative AI solutions. The focus is decidedly on recent developments, with a strong emphasis on practical, industry-oriented sources balanced by academic research. The bibliometric analysis not only provides a snapshot of the current research landscape but also highlights trends and potential gaps in the literature, offering valuable direction for future studies in this rapidly evolving field.

\section*{Glossary and Terminology}

\begin{longtable}{|p{3cm}|p{8cm}|p{4cm}|}
\hline
\textbf{Acronym/Term} & \textbf{Full Form/Description} & \textbf{Category} \\ \hline
AWS & Amazon Web Services, a leading cloud platform offering scalable computing services. & Cloud Providers \\ \hline
GCP & Google Cloud Platform, known for AI and ML innovations. & Cloud Providers \\ \hline
HPC & High-Performance Computing for intensive workloads. & Computing Infrastructure \\ \hline
IoT & Internet of Things, interconnected devices sharing data. & Networking \\ \hline
CDN & Content Delivery Network, for efficient content delivery. & Networking \\ \hline
AI & Artificial Intelligence, machines simulating human intelligence. & AI and ML Concepts \\ \hline
ML & Machine Learning, a subset of AI for data-driven learning. & AI and ML Concepts \\ \hline
IAM & Identity and Access Management for secure digital identities. & Security and Compliance \\ \hline
DLP & Data Loss Prevention for safeguarding sensitive information. & Security and Compliance \\ \hline
TPU & Tensor Processing Unit, hardware optimized for AI tasks. & Computing Infrastructure \\ \hline
EC2 & Elastic Compute Cloud, virtual server hosting by AWS. & Computing Infrastructure \\ \hline
S3 & Simple Storage Service, object storage by AWS. & Networking and Storage \\ \hline
NLP & Natural Language Processing for understanding human language. & AI and ML Concepts \\ \hline
GPU & Graphics Processing Unit for parallel computation. & Computing Infrastructure \\ \hline
LLM & Large Language Model, like GPT-4, for text generation. & AI and ML Concepts \\ \hline
GDPR & General Data Protection Regulation, EU data privacy rules. & Security and Compliance \\ \hline
HIPAA & Health Insurance Portability and Accountability Act, protecting medical data. & Security and Compliance \\ \hline
ISO & International Organization for Standardization for global standards. & Standards and Practices \\ \hline
API & Application Programming Interface for software interactions. & Development and Deployment \\ \hline
BERT & Bidirectional Encoder Representations from Transformers, an NLP model. & AI and ML Concepts \\ \hline
GAN & Generative Adversarial Network for generating realistic data. & AI and ML Concepts \\ \hline
DALL·E & AI for generating images from text descriptions. & Generative AI Applications \\ \hline
PaLM & Pathways Language Model, a versatile AI model by Google. & AI and ML Concepts \\ \hline
CLIP & Contrastive Language-Image Pre-training, a visual and language model. & AI and ML Concepts \\ \hline
SHAP & SHapley Additive exPlanations for explaining ML models. & AI and ML Concepts \\ \hline
LIME & Local Interpretable Model-agnostic Explanations, a tool for interpreting ML models. & AI and ML Concepts \\ \hline
IaaS & Infrastructure as a Service, on-demand cloud infrastructure. & Development and Deployment \\ \hline
PaaS & Platform as a Service for building and deploying apps. & Development and Deployment \\ \hline
AIaaS & AI as a Service for delivering AI solutions via the cloud. & Development and Deployment \\ \hline
AutoML & Automated Machine Learning for simplifying ML workflows. & Development and Deployment \\ \hline
CI/CD & Continuous Integration/Continuous Deployment for automation. & Development and Deployment \\ \hline
RDMA & Remote Direct Memory Access for high-speed, low-latency networking. & Networking \\ \hline
RoCE & RDMA over Converged Ethernet, enhancing data center performance. & Networking \\ \hline
NFV & Network Function Virtualization for decoupling network functions from hardware. & Networking \\ \hline
TCO & Total Cost of Ownership, the cost of maintaining a system. & Metrics and Performance \\ \hline
SWOT & Strengths, Weaknesses, Opportunities, Threats analysis framework. & Business Analysis \\ \hline
CCPA & California Consumer Privacy Act, protecting personal data in California. & Security and Compliance \\ \hline
POSIX & Portable Operating System Interface for cross-platform compatibility. & Standards and Practices \\ \hline
GPFS & General Parallel File System, IBM’s file system for HPC. & Storage \\ \hline
ONNX & Open Neural Network Exchange for AI model interoperability. & AI and ML Concepts \\ \hline
TTS & Text-to-Speech, converting text into natural-sounding speech. & AI Applications \\ \hline
RAG & Retrieval-Augmented Generation, enhancing generative AI outputs. & Generative AI Applications \\ \hline
ESB & Enterprise Service Bus for integrating various applications in enterprises. & Networking \\ \hline
NPU & Neural Processing Unit, a specialized chip for AI tasks. & Computing Infrastructure \\ \hline
FPGA & Field-Programmable Gate Array, reconfigurable hardware for AI tasks. & Computing Infrastructure \\ \hline
SDN & Software-Defined Networking for dynamic network management. & Networking \\ \hline
VPN & Virtual Private Network for secure connections. & Networking \\ \hline
MLOps & Machine Learning Operations for managing production ML. & Development and Deployment \\ \hline
IOPS & Input/Output Operations Per Second, for measuring storage device performance. & Metrics and Performance \\ \hline
PCI-DSS & Payment Card Industry Data Security Standard for secure card transactions. & Security and Compliance \\ \hline
SOX & Sarbanes-Oxley Act, financial regulation for data integrity. & Security and Compliance \\ \hline
FedRAMP & Federal Risk and Authorization Management Program for secure cloud deployments. & Security and Compliance \\ \hline
\end{longtable}

% Add further sections as per your document content.

\bibliographystyle{IEEEtran}

\end{document}